\documentclass[twocolumn]{aastex63}
\usepackage{amsmath}

    \defcitealias{Hurley2002}{H02}
    \defcitealias{Ivanova2003}{IT03}

\newcommand{\CIERA}{Center for Interdisciplinary Exploration and Research in Astrophysics (CIERA), Northwestern University, 1800 Sherman Ave, Evanston, IL 60201, USA}
\newcommand{\NUPA}{Department of Physics and Astronomy, Northwestern University, 2145 Sheridan Ave, Evanston, IL 60208, USA}
\newcommand{\Chicago}{Department of Astronomy and Astrophysics, University of Chicago, Chicago, IL 60637, USA}
\newcommand{\SUPA}{SUPA, School of Physics and Astronomy, University of Glasgow, Kelvin Building, University Ave, Glasgow G12 8QQ, UK}
\newcommand{\unige}{Departement d’Astronomie, Université de Genève, Chemin Pegasi 51, CH-1290 Versoix, Switzerland}
\newcommand{\CCA}{Center for Computational Astrophysics, Flatiron Institute, 162 Fifth Ave, New York, NY 10010, USA}
\newcommand{\ttech}{Department of Physics \& Astronomy, Texas Tech University, Box 41051, Lubbock, TX 79409, USA}
\newcommand{\UIUC}{Department of Physics, University of Illinois at Urbana-Champaign, 1110 W Green St, Urbana, IL 61801, USA}
\newcommand{\UFlorida}{Department of Physics, University of Florida, 2001 Museum Rd., Gainesville, FL 32611}
\newcommand{\ECE}{Electrical and Computer Engineering, Northwestern University, 2145 Sheridan Road, Evanston, IL 60208, USA}
\newcommand{\Princeton}{Department of Astrophysical Sciences, Princeton University, 4 Ivy
Lane, Princeton, NJ 08544, USA}
\begin{document}

\title{ Investigating the Lower Mass Gap with Low Mass X-ray Binary Population Synthesis }

\correspondingauthor{Jared Siegel}
\email{siegeljc@princeton.edu}

\author[0000-0002-9337-0902]{Jared~C.~Siegel}
\affiliation{\Chicago}
\affiliation{\CIERA}
\affiliation{\Princeton}

\author{Ilia~Kiato}
\affiliation{\UIUC}
\affiliation{\CIERA}

\author[0000-0001-9236-5469]{Vicky~Kalogera}
\affiliation{\NUPA}
\affiliation{\CIERA}

\author[0000-0003-3870-7215]{Christopher~P.~L.~Berry}
\affiliation{\SUPA}
\affiliation{\CIERA}

\author{Thomas~J.~Maccarone}
\affiliation{\ttech}

\author[0000-0001-5228-6598]{Katelyn~Breivik}
\affiliation{\CCA}

\author{Jeff J. Andrews}
\affiliation{\CIERA}
\affiliation{\UFlorida}

\author[0000-0002-3439-0321]{Simone~S.~Bavera}
\affiliation{\unige}

\author{Aaron Dotter}
\affiliation{\CIERA}

\author[0000-0003-1474-1523]{Tassos~Fragos}
\affiliation{\unige}

\author[0000-0003-3684-964X]{Konstantinos~Kovlakas}
\affiliation{\unige}

\author[0000-0003-4260-960X]{Devina~Misra}
\affiliation{\unige}

\author[0000-0003-4474-6528]{Kyle A. Rocha}
\affiliation{\NUPA}
\affiliation{\CIERA}

\author{Philipp M. Srivastava}
\affiliation{\ECE}
\affiliation{\CIERA}

\author[0000-0001-9037-6180]{Meng Sun}
\affiliation{\CIERA}

\author[0000-0002-0031-3029]{Zepei~Xing}
\affiliation{\unige}

\author{Emmanouil~Zapartas}
\affiliation{IAASARS, National Observatory of Athens, Penteli, 15236, Greece}


\begin{abstract}
Mass measurements from low-mass black hole X-ray binaries (LMXBs) and radio pulsars have been used to identify a gap between the most massive neutron stars (NSs) and the least massive black holes (BHs).  
BH mass measurements in LMXBs are typically only possible for transient systems: outburst periods enable detection via all-sky X-ray monitors, while quiescent periods enable radial-velocity measurements of the low-mass donor. 
We quantitatively study selection biases due to the requirement of transient behavior for BH mass measurements. 
Using rapid population synthesis simulations (\texttt{COSMIC}), detailed binary stellar-evolution models (\texttt{MESA}), and the disk instability model of transient behavior, we demonstrate that transient-LMXB selection effects introduce observational biases, and can suppress mass-gap BHs in the observed sample.
However, we find a population of transient LMXBs with mass-gap BHs form through accretion-induced collapse of a NS during the LMXB phase, which is inconsistent with observations. 
These results are robust against variations of binary evolution prescriptions. 
The significance of this accretion-induced collapse population depends upon the maximum NS birth mass $M_\mathrm{ NS, birth-max}$.
To reflect the observed dearth of low-mass BHs, \texttt{COSMIC} and \texttt{MESA} models favor $M_\mathrm{ NS, birth-max} \lesssim2M_{\odot}$.
In the absence of further observational biases against LMXBs with mass-gap BHs, our results indicate the need for additional physics connected to the modeling of LMXB formation and evolution.
\end{abstract}{} 

\section{Introduction}
\label{sec:intro}

Black hole X-ray binaries (XRBs) are composed of a black hole (BH) accreting mass from a non-degenerate donor star: either through Roche lobe overflow (RLO) or captured winds. 
There are currently $>50$ candidate BH XRB systems  \citep{Remillard2006,Corral-Santana2016}, broadly divided by donor mass into high-mass XRBs (HMXB) and low-mass XRBs (LMXB). 
HMXBs host donor stars of masses $M_\mathrm{donor} \gtrsim 5 M_{\odot}$, and predominately transfer mass to the compact object via strong winds, while LMXBs host donor stars of masses $M_\mathrm{donor} \lesssim 2 M_{\odot}$ and transfer mass via RLO. 
XRBs offer powerful insights into binary evolution, compact objects, and accretion disks. 
The XRB Cygnus X-1 provided the first evidence for the existence of a BH \citep{Bolton1972}, and since then X-ray and optical-infrared observations have allowed the identification of tens of transient BH systems \citep{Corral-Santana2016}.

In cases of RLO (most commonly in LMXBs) mass-transferred from the donor star forms an accretion disk around the BH. 
Depending on the structure and composition of the system, a thermal instability may form within the disk and generate transient X-ray emission, cycling through periods of bright outburst and dim quiescence \citep{Cannizzo1982, King1996, Lasota2008}.
Transient systems are prime targets for follow-up radial velocity measurements of the donor star: periods of bright outburst make the system discoverable in X-rays, while periods of quiescence allow for inference of the accretor's mass via uncontaminated radial velocity measurements of the donor star. 

Mass measurements provide key insights into the formation and evolution of BHs, and the majority of Galactic BHs with mass estimates are transient XRBs. 
Exceptions include the X-ray faint binary VFTS~243, which \citet{Shenar2022} proposed is composed of a $25M_{\odot}$ O-type star and a $>9M_{\odot}$ BH, as well as the proposed non-interacting low-mass binary companions to 2MASS~J05215658+4359220 \citep{Thompson2019} and V723~Mon \citep{Jayasinghe2021}; however, V723~Mon is consistent with a stripped-star companion instead of a compact object \citep{ElBadry2022}. 
These systems are outnumbered by the $\sim20$ transient LMXBs currently with dynamical mass measurements \citep{Corral-Santana2016}.
Using the mass measurements of the observed LMXB population, prior studies have attempted to constrain the underlying stellar-mass BH distribution \citep{Bailyn1998, Ozel2010, Farr2011}. 
Each of these investigations identified a gap between the lowest mass BHs ($\gtrsim 5 M_{\odot}$) and the maximum mass ($2$--$3 M_{\odot}$) for non-rotating neutron stars \citep[NSs;][]{Rhoades1974,Kalogera1996,Muller1996,ozel2016,Margalit2017,Ai2020,Shao2020,Raaijmakers2021}; this feature is commonly referred to as the \emph{lower mass gap}.

Understanding the nature of the purported lower mass gap offers key insights into the properties of degenerate matter and supernova explosion mechanisms \citep{Fryer2012, Mandel2020, Zevin2020, Liu2021, Patton2022}. 
If instability growth and launch of the supernova proceeds on rapid timescales ($\sim 10~\mathrm{ms}$), accretion onto proto-NSs is suppressed, and simulations predict a paucity of compact objects born with masses between $2$--$5 M_{\odot}$ \citep{Fryer2012, Belczynski2012, Fryer2022}; however, if instability growth is delayed (timescales greater than $\sim 200~\mathrm{ms}$), simulations predict a continuous compact object birth mass distribution. 
Motivated by the observed Galactic LMXB sample, prior studies have typically favored the rapid instability timescale, but an underlying assumption of these studies is that BH mass measurements are unbiased.

Recent detections of BH masses in systems other than LMXBs complicate the picture of the lower mass gap. 
The binary sources of gravitational waves GW190814 and GW200210\_092254 are each inferred to have a source containing a compact object lying in the mass gap:  $2.59^{+0.08}_{-0.09} M_{\odot}$ and $2.83^{+0.47}_{-0.42} M_{\odot}$, respectively \citep{abbott2020,GWTC3}.
The mass estimates for the compact-object companions to 2MASS~J05215658+4359220 and V723~Mon (assuming a compact object interpretation) also lie within the mass gap, $3.3_{-0.7}^{+2.8} M_{\odot}$ and $(3.04\pm 0.06) M_{\odot}$ respectively \citep{Thompson2019,Jayasinghe2021}. 
These measurements are in apparent contradiction with the existence of a lower mass gap, and must be reconciled with the Galactic LMXB BH mass distribution.

Understanding the nature of detection biases in the Galactic LMXB population is critical to interpreting the observed BH mass distribution. 
Through approximate scaling of the observed XRB sample, \citet{Ozel2010} argued detection biases due to outburst flux cannot account for the mass gap.
\citet{Jonker2021} noted that one potential selection effect comes from the fact that objects with high extinction are more difficult to obtain follow-up radial velocity measurements for. 
Assuming an anti-correlation between supernova kick magnitude and compact object mass \citep[e.g.,][]{Fryer2012, Atri2019}, this preference for LMXBs off the Galactic plane may bias against mass inference of higher mass BHs. 
While this effect alone would not produce a mass gap as an observational selection effect, it may still shape the observed BH mass distribution. 
\citet{Kreidberg2012} also demonstrated that assuming zero or constant emission from the accretion flow can lead to systematic overestimates of compact object masses. 
The role of detection biases in the Galactic LMXB population, particularly the requirement of transient behavior for BH mass measurement, remains unclear. 

Here we adopt a forward modeling approach to assess selection biases: generating samples of synthetic binary systems, evolving the systems forward in time, and inferring the detectable population using the disk instability model of LMXB transient behavior. 
With these methods, we investigate the impact of observational biases on the lower mass gap and the underlying BH mass distribution. 
Through our analysis, we uncover that the observed gap has implications for the maximum NS mass at birth ($M_\mathrm{ NS, birth-max}$). 

In Section~\ref{sec:pop_synth} and Section~\ref{sec:MESA}, we describe our population synthesis \citep[\texttt{COSMIC};][]{Breivik2020} and stellar evolution \citep[\texttt{MESA};][]{Paxton2011, Paxton2013, Paxton2015, Paxton2018, Paxton2019} methods, respectively; the population synthesis Milky Way model is outlined in Section~\ref{sec:MW}.
In Section~\ref{sec:transient} we outline our treatment of LMXB transient behavior, and in Section~\ref{sec:detect} we present the formulation of detection probabilities. 
In Section~\ref{sec:observing} we detail how we draw together a synthetic population.
We describe our results in Section~\ref{sec:results}, discuss their implications in Section~\ref{sec:discuss}, and summarize our conclusions in Section~\ref{sec:summary}.

\section{LMXB Simulations}

Here we outline the methods for simulating LMXB samples with rapid binary population synthesis (\texttt{COSMIC}; Section~\ref{sec:pop_synth}) and binary stellar-evolution models (\texttt{MESA}; Section~\ref{sec:MESA}). 
While \texttt{MESA} provides detailed modeling, the computational efficiency of \texttt{COSMIC} allows us to explore uncertain aspects of binary evolution physics. 
Taken together, the suite of proof-of-concept \texttt{COSMIC} simulations contextualize the results of improved mass-transfer modeling with \texttt{MESA}, by marginalizing over uncertain aspects of binary evolution physics. 

\subsection{Population models}
\label{sec:pop_synth}

Binary population synthesis offers valuable insights into binary evolution channels and the expected distributions of target populations. 
We employ the rapid binary population synthesis code \texttt{COSMIC} \citep{Breivik2020} to generate Milky Way LMXB populations. 
Originating from \texttt{BSE} \citep{Hurley2002}, \texttt{COSMIC} approximates single-star evolution using fitting formulae to detailed models and applies prescriptions for binary evolution to these formulae. 
To complement these simulations, we also consider a three-dimensional grid (in compact object mass $M_\mathrm{ CO}$, donor mass $M_\mathrm{ donor}$, and orbital period $P$ space) of LMXBs using \texttt{MESA} (Section~\ref{sec:MESA}). 
The computational efficiency of \texttt{COSMIC} allows us to explore uncertain aspects of binary evolution physics that would not be computationally feasible with \texttt{MESA}.

Independent of contextualizing \texttt{MESA}  simulations, verifying the extent to which binary population synthesis methods can replicate the observed LMXB sample is of interest. 
For example, \citet{Ivanova2006} proposed that some LMXBs could be fed by pre-main-sequence donor stars given the observed relation between LMXBs' orbital periods and the donors' effective temperatures, as well as the fact that only stars more massive than $\sim2M_{\odot}$ have a pre-main-sequence lifetime shorter than the main-sequence lifetime of a star that explodes and forms a BH ($\lesssim 10^7~\mathrm{yr}$). 
However, no rapid population synthesis codes simulate pre-main-sequence interactions.
These constraints on the binary population suppress the formation of LMXBs with massive BHs and shorter period binaries.
Any rapid population synthesis investigation of LMXBs that does not consider pre-main-sequence interactions is impacted by these limitations. 
While population synthesis methods likely cannot reproduce all aspects of the observed LMXB sample, quantifying these discrepancies is vital to developing improved population synthesis engines.
Moreover, we are employing population synthesis as a proof-of-concept survey of binary evolution prescriptions: rather than replicating all aspects of the observed LMXB distribution, we are interested in uncovering any significant dependencies between the synthetic observed LMXB sample and uncertain aspects of binary evolution physics. 

\begin{deluxetable*}{clcccl}\label{tab:MW}
\tablecaption{Milky Way Component Parameters. }
\tablehead{
\colhead{Component} &
\colhead{Age\tablenotemark{a}} &
\colhead{Metallicity} &
\colhead{Star formation history} &
\colhead{Stellar mass} &
\colhead{Spatial distribution parameters}\\
\colhead{} & \colhead{(Gyr)} & \colhead{ $[\mathrm{ Fe}/\mathrm{ H}]$ } & & \colhead{ ($10^{10} M_{\odot}$)} & 
}
\startdata
Thin disk & $\mathcal{A}_\mathrm{ MW}-1$~Gyr & 0.0 & Continuous & 4.32 & $z_\mathrm{ disk}=0.3$~kpc, $R_\mathrm{ disk}=2.90$~kpc\\
Thick disk & $\mathcal{A}_\mathrm{ MW}$ & $-0.8$ & $1$~Gyr burst & 1.44 & $z_\mathrm{ disk}=0.9$~kpc, $R_\mathrm{ disk}=3.31$kpc\\
 Bulge & $\mathcal{A}_\mathrm{ MW}-1$~Gyr & 0.0 & $1$~Gyr burst & 0.89 &  $a = 1.8$, $r_0 = 0.075$~kpc, $r_\mathrm{ cut} = 2.1$~kpc, $q = 0.5$
\enddata
\tablenotetext{a}{The adopted Milky Way age is $\mathcal{A}_\mathrm{ MW}=11$~Gyr.}
\tablecomments{ Component masses and spatial distributions follow \cite{McMillan2011}. Star formation histories and metallicities are adopted from \cite{Robin2003}. }
\end{deluxetable*}

To generate Galactic LMXB population synthesis samples, we simulate Milky Way stellar populations for a range of binary evolution prescriptions. For each \texttt{COSMIC} population:
\begin{enumerate}
    
    \item The maximum NS mass is set to $3M_{\odot}$ \citep[Section~\ref{sec:intro};][]{Rhoades1974,Kalogera1996}.
    \item Compact object birth masses from core-collapse supernovae are assigned following the \emph{Delayed} prescription of \citet{Fryer2012}; this differs from the alternative \emph{Rapid} prescription in the assumed timescale of instability growth and launch of the supernova. 
    The Delayed remnant prescription produces a continuous mass distribution, while the Rapid prescription produces a gap between the most massive NSs and the least massive BHs.
    The Delayed prescription is adopted to explore whether LMXB selection biases can hide an underlying population of low-mass BHs. 
    \item Compact object natal kicks are drawn from a bimodal distribution: standard iron core-collapse supernova kicks are drawn from a Maxwellian distribution with a dispersion of $265$~$\mathrm{km\,s^{-1}}$ \citep{Hobbs2005}, while kicks for electron-capture supernovae and ultra-stripped supernovae are drawn according to a Maxwellian distribution with a dispersion of $20~\mathrm{km\,s^{-1}}$ \citep{Giacobbo2019}. 
    Natal kicks are reduced by a factor of $1-f_\mathrm{fb}$, where $f_\mathrm{fb}$ is the fraction of the ejected supernova mass that will fall back onto the proto-compact object \citep{Fryer2012}.
    \item Initial conditions (e.g., eccentricity, orbital period, primary mass, and mass ratio) are independently drawn following \cite{Sana2012}.
\end{enumerate}
We investigate model dependence on six uncertain factors of binary evolution:
\begin{enumerate}
    \item Stars without a distinct core--envelope boundary (e.g., stars on the Hertzsprung gap) that instigate a common envelope (CE) event are either assumed to survive the CE (\emph{Optimistic}) or are assumed to merge  \citep[\emph{Pessimistic};][]{Belczynski2008}.
    \item The CE efficiency $\alpha$ parameterizes the transfer of orbital energy to the envelope, i.e., how easily the CE is unbound from the system \citep{Webbink1984,deKool1990}; higher values of $\alpha$ lead to longer period binaries post-CE. 
    Prior studies using one-dimensional simulations \citep{Fragos2019} and comparison to gravitation-wave observations \citep{Giacobbo2018,Santoliquido2021,Zevin2021} have favored higher CE efficiencies ($\alpha \sim 5$). 
    To explore these higher values, $\alpha$ is set to $\alpha=1.0$ or $\alpha=5.0$.
    \item The minimum ZAMS mass ratio between binary members is set to $q>0.01$, where $q \equiv M_\mathrm{donor}/M_\mathrm{primary}$, or is defined as a function of primary mass (\emph{Lifetime limited}). 
    For the Lifetime limited case, $q$ is restricted such that the pre-main-sequence lifetime of the secondary is shorter than the main-sequence lifetime of the primary. 
    \item In addition to gravitational radiation, close-in binaries (e.g., cataclysmic variables and LMXBs) can efficiently lose orbital angular momentum via magnetic braking of a tidally coupled magnetic wind. 
    Magnetic braking is implemented following \citet[][hereafter \citetalias{Hurley2002}]{Hurley2002} or \citet[][hereafter \citetalias{Ivanova2003}]{Ivanova2003}.
    \item The amount of mass accreted during Roche-lobe overflow is either fixed to $50\%$ efficiency \citep{Belczynski2008} or is a function of the accretor's type (\emph{State dependent}).
    For the State dependent prescription, the accretion efficiency during Roche-lobe overflow is either (i) limited to ten-times the thermal rate if the accretor is a main-sequence, Hertzsprung gap or core helium burning star, or (ii) unlimited if the accretor is a giant branch, early asymptotic giant branch or asymptotic giant branch star. 
    Compact objects also experience conservative mass transfer, provided the mass-transfer rate is sub-Eddington.
    Accretion onto a NS is treated independently of whether the system is transient. 
    While outburst events during accretion onto a white dwarf can lead to a loss of mass, only a small of fraction of accreting NSs show evidence of mass loss during bursts \citep{Degenaar2018}; this difference between white dwarf and NS accretors is likely attributable to the deeper potential wells of NSs.  
    \item The maximum NS birth mass from core-collapse supernova $M_\mathrm{NS,birth-max}$ is of particular interest to the formation and evolution of LMXBs; lowering $M_\mathrm{ NS, birth-max}$ suppresses the formation of LMXBs through accretion-induced collapse (AIC) of a NS. 
    To vary $M_\mathrm{ NS, birth-max}$ while preserving computational feasibility, $M_\mathrm{ NS, birth-max}$ is fixed to $3M_{\odot}$ within \texttt{COSMIC} and is lowered in post-processing by removing systems with a NS birth mass above a given threshold.
    This method of imposing a maximum NS birth mass is a non-physical toy model.
    The goal is to demonstrate how a truncated NS birth mass distribution impacts LMXB demographics.
    
\end{enumerate}

To simulate stellar populations analogous to the Milky Way, we assign birth times, spatial positions, and metallicities following a three-component Milky Way model (Section~\ref{sec:MW}).

\subsection{Milky Way model}
\label{sec:MW}

For a given set of binary evolution prescriptions, we generate a synthetic Galactic sample of LMXBs. 
To do so, we resample a given \texttt{COSMIC} population and assign system properties in accordance with the Milky Way (e.g., star-formation history and spatial distribution). 
The Milky Way is approximated as a composite of three sub-populations: thin disk, thick disk, and bulge.

The star-formation history and metallicity of each component is adopted from \citet{Robin2003}. 
For the thick disk and bulge, star formation is modeled as a $1~\mathrm{Gyr}$ continuous burst, beginning at the birth of the Milky Way and $1~\mathrm{Gyr}$ after the birth of the Milky Way, respectively. 
Star formation for the thin disk is continuous from $1~\mathrm{Gyr}$ after the birth of the Milky Way to the present. 
Stars in the thin disk and bulge have solar metallicity, while stars in the thick disk have $15\%$ of solar metallicity. 
System birth times are relative to a Milky Way age of $\mathcal{A}_\mathrm{ MW} = 11~\mathrm{Gyr}$. 
LMXBs' positions within the galaxy are fixed to their birthplaces; we do not include the effects of SN kicks on systemic velocity.

The component masses and spatial distributions for the Milky Way model follow \cite{McMillan2011}. 
The thin and thick disk spatial distributions are treated as an exponential,
\begin{align}
    \rho_\mathrm{disk}(R,z) \propto \exp \left( -\frac{\mid z \mid }{z_\mathrm{disk}} - \frac{R}{R_\mathrm{disk}}\right),
\end{align}
where $z_\mathrm{ disk}$ is the scale-height and $R_\mathrm{disk}$ is the scale length; for the thin disk, $z_\mathrm{ disk}=0.3~\mathrm{kpc}$ and $R_\mathrm{ disk}=2.9~\mathrm{kpc}$; for the thick disk, $z_\mathrm{ disk}=0.9~\mathrm{kpc}$ and $R_\mathrm{ disk}=3.31~\mathrm{kpc}$ \citep{McMillan2011}. 
The bulge spatial distribution is
\begin{align}
    \rho_\mathrm{ bulge}(R,z) &\propto \frac{ \exp \left[-(r^{\prime}/r_\mathrm{ cut})^2 \right] }{ (1+r^{\prime}/r_0)^{a} },\\
    r^{\prime} &= \sqrt{R^2 + (z/q)^2},
\end{align}
where $a = 1.8$, $r_0 = 0.075~\mathrm{kpc}$, $r_\mathrm{ cut} = 2.1~\mathrm{kpc}$, and the axial ratio $q = 0.5$. 
Each component is treated as axisymmetric. 

The parameters of the Milky Way model are summarized in Table~\ref{tab:MW}.

\subsection{\texttt{MESA} models}
\label{sec:MESA}

Rapid binary population synthesis enables the efficient modeling of statistically significant stellar samples, but such methods have drawbacks. 
Rapid binary population synthesis codes, including \texttt{COSMIC}, assume binary members have the same properties as a single star with matching mass and metallicity at thermal equilibrium; this can lead to systematic biases in the binaries' evolution \citep[][]{GallegosGarcia2021,Fragos2022}. 
As outlined above, the \texttt{COSMIC} simulations serve as a proof-of-concept survey over uncertain aspects of binary evolution physics and contextualize a smaller grid of binaries modeled in greater detail with \texttt{MESA}. 
Here we outline the LMXB grid evolved using \texttt{MESA}.

We consider a three-dimensional grid of binaries consisting of a compact object and a hydrogen-rich main-sequence star. 
Adopted from \citet{Fragos2022}, the grid consists of $1375$ binaries, initialized uniformly in $\log_{10}  (M_\mathrm{CO}/M_{\odot})$, $\log_{10} (M_\mathrm{donor}/M_{\odot})$, and $\log_{10}( P/\mathrm{day})$. 
Compact object masses span the range $1.1$--$10M_{\odot}$, donor star masses span $0.5$--$3M_{\odot}$, and orbital periods span $1.26$--$3162$~days. 
The grid includes LMXBs with mass-gap BH, which allows us to investigate the observability of such sources.
The maximum NS mass is set to $3 M_{\odot}$ (as in the \texttt{COSMIC} simulations). 
All systems are initialized at solar metallicity, and the donor stars are seeded with a rotational period equal to their orbital period \citep{Fragos2022}. 

To facilitate comparison between the population synthesis and \texttt{MESA} methods, we initialize and evolve grids of binaries with \texttt{COSMIC} at the same initial conditions as the \texttt{MESA} grid. 
We consider the same combinations of binary evolution prescriptions as Section~\ref{sec:pop_synth}, with the exception of CE efficiency and minimum ZAMS mass ratio (which do not apply to this phase of evolution).

\begin{figure*}
\gridline{\fig{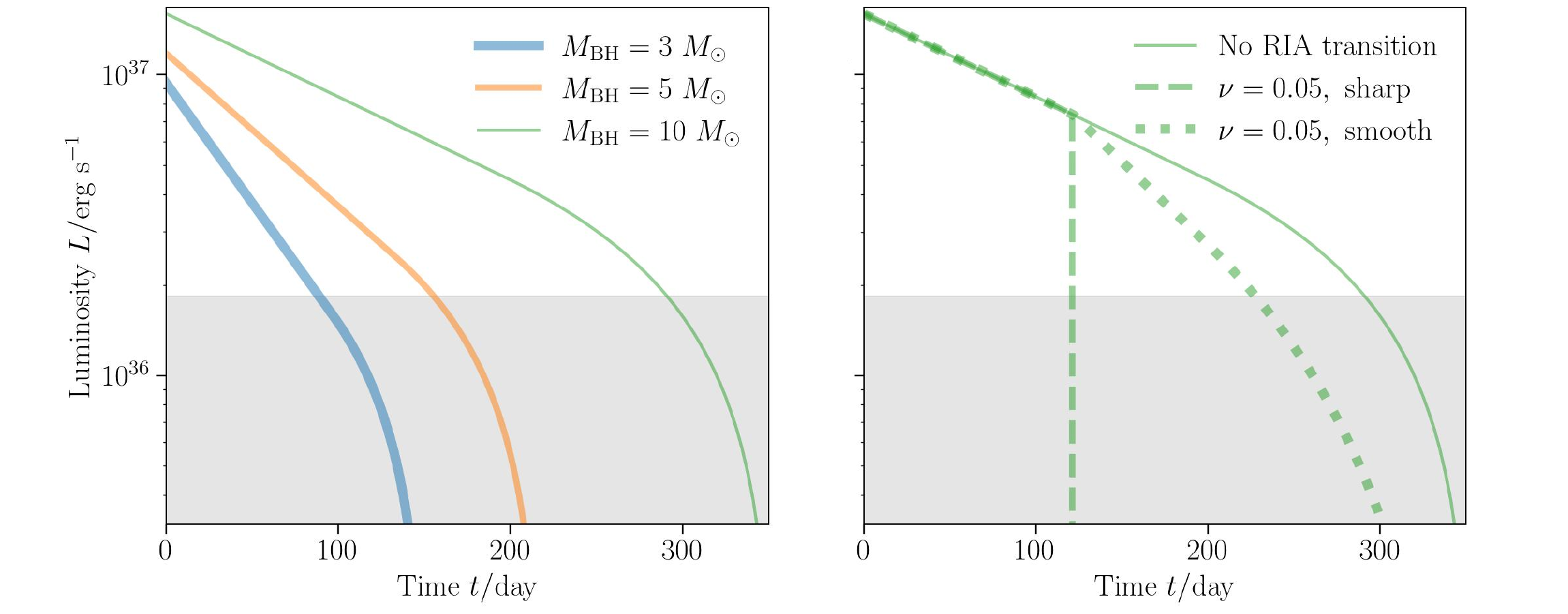}{\textwidth}{}}
\caption{X-ray outburst light curves for a typical LMXB system for a range of BH masses and transitions to RIA. 
On the \emph{left}, we assume no transition to RIA and vary the BH mass with fixed orbital period and companion mass. 
On the \emph{right}, we present the sharp and smooth transitions to RIA for fixed BH mass. 
Both panels assume $M_\mathrm{donor}=0.7M_{\odot}$, $P=10~\mathrm{hour}$, and $d=8~\mathrm{kpc}$. 
The shaded gray region highlights were the X-ray luminosity drops below the detectable limit ($10~\mathrm{mCrab}$). 
All luminosities have been converted to the $2$--$10~\mathrm{keV}$ range by dividing the bolometric luminosity by $f_\mathrm{ corr}=4$. }
\label{fig:lightcurve}
\end{figure*}

\subsection{Transient LMXB Behavior}
\label{sec:transient}

\subsubsection{Instability Cycle}

LMXB accretion disks are complex rotating, turbulent, irradiated systems. Despite this complexity, the outburst cycles of LMXBs are reasonably well approximated by the disk instability model \citep{King1996,Dubus1999,Lasota2001,Lasota2016}. 
This simplified model of accretion physics assumes (i) a constant mass-transfer rate, determined by the secular mass-transfer rate from the binary evolution calculations, (ii) that each outburst is identical, a consequence of the assumed accretion of the entire disk during an outburst, and (iii) that the mass transfer is conservative when sub-Eddington.  
Observational work shows these assumptions are imperfect \citep[e.g., the observed bimodality of outburst durations;][Section~4.1]{Lasota2001}. 
The disk instability model is adopted for this work given that it is quantitatively tractable, and efforts to improve the model are beyond the scope of this paper.
    
Under the disk instability model, a disk is thermally stable provided radiative cooling varies faster with temperature than viscous heating \citep{Frank1992,Dubus1999}. 
As a result, a rapid change in opacity instigates a thermal instability in the disk (e.g., if a disk is composed of ionized hydrogen, and the temperature falls low enough for hydrogen to recombine). 
If thermally unstable, a disk will cycle through hot outburst states and cold quiescent states. 

For a given LMXB, the conditions for thermal equilibrium of the accretion disk (typically presented in the disk surface density $\Sigma$ and effective temperature $T_\mathrm{eff}$ plane) can be numerically calculated as a function of disk radius. 
Equilibrium solutions in the $\Sigma$--$T_\mathrm{eff}$ plane (\emph{S-curves}) are composed of a stable hot branch and a stable cold branch connected by an instability strip \citep[][Section~3]{Lasota2001}. 
LMXBs where the mass-transfer rate results in a $T_\mathrm{eff}$ on the instability strip will be thermally unstable, leading the disk to cycle between the hot and cold branches. 
The inflection points of the S-curve correspond to critical accretion rates that bound the instability strip. 
By calculating S-curves for a range of disk radii and binary system parameters (e.g., orbital period, donor mass, and accretor mass), the cold branch and hot branch bounding critical accretion rates can be interpolated for a given LMXB. 

In principle, stable LMXB accretion disks can occupy the cold branch or the hot branch of the S-curve. 
However, binary evolution favors mass-transfer rates above the cold branch critical mass-transfer rate \citep{Lasota2001}. 
Systems may occupy the cold branch late in their evolution, when their donor stars become degenerate around the point that orbital periods begin to increase \citep[][]{Knigge2011}; in such cases, the outbursts would not contribute significantly to the detectable population of sources, because the outbursts would be short, faint, and infrequent. 
In this study, a LMXB is considered a transient source if its mass-transfer rate falls below the hot branch critical accretion rate $\dot{M}_\mathrm{donor} < \dot{M}_\mathrm{crit}$.

For systems with hydrogen-dominated donor stars, the hot branch critical mass-transfer rate for X-ray irradiated disks is \citep{Dubus1999}
\begin{align}
    \dot{ M}_\mathrm{ crit} = 1.5 \times 10^{15} & \left( \frac{ M_\mathrm{ BH}}{ M_{\odot}} \right)^{-0.4} \left( \frac{R_\mathrm{ disk}}{10^{10}~\mathrm{cm}} \right)^{2.1} \nonumber \\
    & \times \left( \frac{C}{5\times10^{-4}} \right) ~\mathrm{g~s^{-1}},
\end{align}
where $M_\mathrm{ BH}$ is the BH mass, $R_\mathrm{ disk}$ is the outer radius of the accretion disk, and $C$ is a disk structure constant \citep[which we fix to $C=5\times 10^{-4}$;][]{Dubus1999}.
For helium-dominated donors, where the opacity of the inflow and accretion disk are markedly different from the hydrogen-dominated case, the hot branch irradiated disk critical mass-transfer rate is \citep{Lasota2008}
\begin{align}
    \dot{ M }_\mathrm{crit} &= 2.1 \times 10^{16} \left( \frac{C_\mathrm{ He}}{10^{-3}} \right)^{-0.22} \nonumber \\
    & \times \left( \frac{\alpha}{0.1} \right)^{-0.03-0.01\log(C_\mathrm{ He}/10^{-3})}  \nonumber \\
    & \times \left( \frac{R_\mathrm{disk}}{10^{10}~\mathrm{cm}} \right)^{2.51-0.05\log(C_\mathrm{ He}/10^{-3})} \nonumber \\
    & \times \left( \frac{ M_\mathrm{ BH}}{ M_{\odot}} \right)^{-0.74+0.05\log(C_\mathrm{ He}/10^{-3})}~\mathrm{g\,s}^{-1}, 
\end{align}
where $C_\mathrm{ He}$ is a disk structure constant for helium-dominated donors and $\alpha$ is a viscosity parameter \citep[we fix  $C_\mathrm{ He}=10^{-3}$ and $\alpha = 0.1$;][]{Lasota2008}. We adopt the standard formulation of disk size: $R_\mathrm{ disk}=70\%$ of the BH's Roche lobe radius \citep{King1997a}.

If the mass-transfer rate is below the critical rate, the system will undergo a cycle of outburst and quiescent phases of duration $T_\mathrm{ outburst}$ and $T_\mathrm{ quiescence}$, respectively. 
If the mass-transfer rate is above the critical rate, the system will be a persistent X-ray source. 
For transient LMXBs, the period of the instability cycle is $T_\mathrm{ cycle} \equiv T_\mathrm{ quiescence} + T_\mathrm{ outburst}$. 
Assuming the entire disk is accreted onto the BH during an outburst event, the duration of the quiescent phase is
\begin{equation}
    \label{equ:t_q}
    T_\mathrm{ quiescence} = \frac{M_\mathrm{ disk-max}}{\dot{M}_\mathrm{ donor}},
\end{equation}
where $\dot{M}_\mathrm{ donor}$ is the mass-transfer rate from the donor star. 
Adopting the critical surface density $\Sigma_\mathrm{ crit}$ of \citet{Cannizzo1988}, and assuming the surface density of the quiescent disk prior to in-fall is comparable to $\Sigma_\mathrm{ crit}$ \citep{Dubus2001}, the maximum accretion disk mass is
\begin{equation}
    {M}_\mathrm{ disk-max} \approx \int_{R_\mathrm{ ISCO}}^{R_\mathrm{ disk}} 2 \pi \Sigma_\mathrm{ crit} (R)  R \,\mathrm{d}R,
\end{equation}
where $R_\mathrm{ ISCO}$ is the radius of the innermost stable circular orbit.

The duration of the outburst $T_\mathrm{ outburst}$ is the total time necessary for the disk to accrete onto the BH. 
Since the mass in-fall rate during the instability $\dot{ M}_\mathrm{ ins}(t)$ varies during the outburst, $T_\mathrm{ outburst}$ cannot be approximated analogously to $T_\mathrm{ quiescence}$ in Eq.~\eqref{equ:t_q}. 
Following \cite{King1998}, the mass in-fall rate $\dot{ M}_\mathrm{ ins}(t)$ is treated as an exponential decay followed by a linear decay. 

The bolometric outburst light curve is related to the in fall rate by
\begin{equation}
    L_\mathrm{ bol} (t) = \eta(t) c^2 \dot{ M}_\mathrm{ ins.} (t), 
\end{equation}
where $\eta(t)$ is the radiative efficiency and $c$ is the speed of light. 
Adopting the methods of \citet{Knevitt2014}, we consider two potential formulations of $\eta(t)$, corresponding to two different transitions to radiatively inefficient accretion (RIA): a \emph{sharp} transition,
\begin{equation}
\eta =  
\begin{cases}
      0, & L_\mathrm{ bol} < \nu L_\mathrm{ Edd} \\
      0.1, & L_\mathrm{ bol} \geq \nu L_\mathrm{ Edd}
\end{cases} ,
\end{equation}
or a \emph{smooth} transition,
\begin{equation}
 \eta =  
 \begin{cases}
      0.1\left( \dfrac{\dot{M}_\mathrm{ ins} }{ \nu \dot{ M}_\mathrm{ Edd} } \right), & L_\mathrm{ bol} < \nu L_\mathrm{ Edd} \\
      0.1, & L_\mathrm{ bol} \geq \nu L_\mathrm{ Edd}
\end{cases}  ,
\end{equation}
where we either fix $\nu$ to $0$ (no transition to RIA) or $0.05$; here $\dot{M}_\mathrm{ Edd}$ is the Eddington mass-transfer rate, and $L_\mathrm{ Edd}$ is the corresponding Eddington luminosity.

In this approximate outburst model, including a transition to RIA shortens the X-ray outburst lifetime without changing the peak luminosity of the outburst. 
To highlight this dependence, outburst light curves for a typical LMXB system under varying BH masses and transitions to RIA are presented in Figure~\ref{fig:lightcurve}.  

Since the outburst luminosity may drop below a detectable level before the full $T_\mathrm{outburst}$ has elapsed, we introduce the effective outburst time $T^{\prime}_\mathrm{outburst}$, the total amount of time that the X-ray outburst flux is above an observable threshold. 
For a limiting X-ray flux of $F_\mathrm{fiducial}$, the outburst is observable if
\begin{equation}
    F_\mathrm{fiducial} < \frac{ L_\mathrm{bol} (t)}{4\pi d^2 f_\mathrm{corr} },
\end{equation}
where $f_\mathrm{corr}$ converts the bolometric flux into the observable range and $d$ is the distance to the system. 
We adopt $f_\mathrm{corr}=4$ and $F_\mathrm{ fiducial}=10~\mathrm{mCrab}$ (Section~\ref{sec:detect}). 
For \texttt{COSMIC}, $d$ is drawn for each system from the spatial distributions described in Section~\ref{sec:MW}; for the \texttt{MESA} grid, $d$ is fixed to $8$~kpc (Section~\ref{sec:observing}). 

In Section~\ref{sec:detect}, LMXBs' effective outburst times $T^{\prime}_\mathrm{ outburst}$ and cycle lifetimes $T_\mathrm{ cycle}$ are mapped to probabilities of detection with an all-sky X-ray monitor.

\subsection{Detection Probabilities}
\label{sec:detect}

Transient X-ray binaries have predominately been discovered by all-sky monitors (ASMs). 
Although the observed sample of transient LMXBs has been built up by a myriad of instruments, we treat the ASM aboard the \textit{Rossi X-ray Timing Explorer} \citep[\textit{RXTE};][]{Bradt1993} as our benchmark detector. 
Active from 1996 to 2012, the \textit{RXTE} ASM observed in the $1.5$--$12~\mathrm{keV}$ band with a highly stochastic pointing pattern that typically scanned a given source 5--10 times per day \citep{Levine1996}. 
\textit{RXTE} discovered a significant portion of the transient LMXB sample \citep{Wen2006}. 
The detection requirements of the \textit{RXTE} ASM can be approximated as a flux limit of $F_\mathrm{ fiducial} = 10~\mathrm{mCrab}$ for a minimum emission time of $T_\mathrm{ fiducial} = 1$~day.

For a transient LMXB source to be identified by an all-sky survey and be eligible for mass inference, it must (i) be bright enough during outburst to be discovered, and (ii) transition from an outburst to a quiescent state during the survey. 
The latter requirement enables mass inference via radial-velocity follow-up of the low-mass donor. 
We neglect inclination effects and potential biases induced by the position of the Sun. 

If an outburst occurs during an ASM survey, the probability the outburst is detected by the ASM is \citep{Knevitt2014}
\begin{align}
    p_\mathrm{discover} =
    & \min \left\{1, \frac{T^{\prime}_\mathrm{outburst}}{T_\mathrm{fiducial}}\right\},
\end{align}
where  $T^{\prime}_\mathrm{ outburst}$ is the effective outburst time: the total amount of time that the X-ray outburst is brighter than $F_\mathrm{fiducial}$ (Section~\ref{sec:transient}).

In addition to the requirement on the system's outburst flux, a LMXB must also transition from outburst to quiescence during the survey in order to be classified as a transient. 
The probability of such a transition is
\begin{equation}
    p_\mathrm{transition} = \min \left\{1, \frac{T_\mathrm{survey}}{T_\mathrm{cycle}}\right\},
\end{equation}
where $T_\mathrm{cycle}$ is the period of the instability cycle, $T_\mathrm{cycle} \equiv T_\mathrm{quiescence} + T_\mathrm{outburst}$, and $T_\mathrm{survey}$ is the duration of the X-ray monitor's survey.

Considering both the probability of being seen in outburst and the probability of observing a transition, it follows that the probability of a transient LMXB source being identified is
\begin{align}
    \label{equ:prob_detect}
    p_\mathrm{detect}=
    p_\mathrm{discover}( T^{\prime}_\mathrm{ outburst}) \times
    p_\mathrm{transition}( T_\mathrm{cycle}).
\end{align}
Motivated by \textit{RXTE}, we adopt $T_\mathrm{ survey}=15$~yr. 
For persistent LMXBs, we set $p_\mathrm{detect}=0$. 

Using Eq.~\eqref{equ:prob_detect}, the probability of identifying a LMXB as a transient source can be inferred from the outburst time $T^{\prime}_\mathrm{outburst}$  and cycle lifetime $T_\mathrm{ cycle}$; $T^{\prime}_\mathrm{outburst}$ and $T_\mathrm{cycle}$ can in-turn be inferred from the evolutionary state of the LMXB, e.g., mass-transfer rate, BH mass, donor mass, and orbital period. 
In Section~\ref{sec:observing}, we leverage Eq.~\eqref{equ:prob_detect} to generate synthetic samples of observed LMXBs for both \texttt{COSMIC} and \texttt{MESA}.

\begin{deluxetable*}{ccccccc}[t!]\label{tab:pop_synth}
\tablecaption{Population Models}
\tablehead{
\colhead{CE survival} &
\colhead{$\alpha$} &
\colhead{$q$ limit} &
\colhead{Magnetic braking} & 
\colhead{Accretion limit} &
\colhead{$f_\mathrm{ low} (M_\mathrm{NS, birth} < 1.5 M_{\odot})$\tablenotemark{a} }  & 
\colhead{$f_\mathrm{ low} (M_\mathrm{NS, birth} < 3.0 M_{\odot})$ } 
}
\startdata
Optimistic & 1.0 & Lifetime limited & IT03 & State dependent & $0.39^{+0.04}_{-0.04}$ & $0.89^{+0.02}_{-0.03}$\\
Optimistic & 1.0 & Lifetime limited & IT03 & $50\%$ efficiency & $0.59^{+0.02}_{-0.04}$ & $0.67^{+0.02}_{-0.04}$\\
Optimistic & 1.0 & Lifetime limited & H02 & State dependent & $0.79^{+0.02}_{-0.02}$ & $0.955^{+0.005}_{-0.006}$\\
Optimistic & 1.0 & Lifetime limited & H02 & $50\%$ efficiency & $0.58^{+0.03}_{-0.03}$ & $0.72^{+0.02}_{-0.02}$\\
Optimistic & 1.0 & 0.01 & IT03 & State dependent & $0.37^{+0.02}_{-0.02}$ & $0.80^{+0.01}_{-0.02}$\\
Optimistic & 1.0 & 0.01 & IT03 & $50\%$ efficiency & $0.58^{+0.03}_{-0.02}$ & $0.65^{+0.04}_{-0.02}$\\
Optimistic & 1.0 & 0.01 & H02 & State dependent & $0.603^{+0.013}_{-0.009}$ & $0.880^{+0.006}_{-0.007}$\\
Optimistic & 1.0 & 0.01 & H02 & $50\%$ efficiency & $0.56^{+0.01}_{-0.02}$ & $0.67^{+0.01}_{-0.01}$\\
Optimistic & 5.0 & Lifetime limited & IT03 & State dependent & $0.44^{+0.02}_{-0.02}$ & $0.943^{+0.004}_{-0.004}$\\
Optimistic & 5.0 & Lifetime limited & IT03 & $50\%$ efficiency & $0.65^{+0.02}_{-0.03}$ & $0.80^{+0.01}_{-0.01}$\\
Optimistic & 5.0 & Lifetime limited & H02 & State dependent & $0.56^{+0.03}_{-0.03}$ & $0.965^{+0.002}_{-0.005}$\\
Optimistic & 5.0 & Lifetime limited & H02 & $50\%$ efficiency & $0.60^{+0.03}_{-0.04}$ & $0.84^{+0.01}_{-0.01}$\\
Optimistic & 5.0 & 0.01 & IT03 & State dependent & $0.39^{+0.02}_{-0.01}$ & $0.893^{+0.012}_{-0.008}$\\
Optimistic & 5.0 & 0.01 & IT03 & $50\%$ efficiency & $0.53^{+0.01}_{-0.02}$ & $0.70^{+0.02}_{-0.01}$\\
Optimistic & 5.0 & 0.01 & H02 & State dependent & $0.43^{+0.01}_{-0.02}$ & $0.914^{+0.005}_{-0.004}$\\
Optimistic & 5.0 & 0.01 & H02 & $50\%$ efficiency & $0.50^{+0.01}_{-0.02}$ & $0.741^{+0.009}_{-0.012}$\\
Pessimistic & 1.0 & Lifetime limited & IT03 & State dependent & $0.36^{+0.04}_{-0.05}$ & $0.88^{+0.01}_{-0.01}$\\
Pessimistic & 1.0 & Lifetime limited & IT03 & $50\%$ efficiency & $0.57^{+0.04}_{-0.05}$ & $0.64^{+0.03}_{-0.04}$\\
Pessimistic & 1.0 & Lifetime limited & H02 & State dependent & $0.63^{+0.03}_{-0.03}$ & $0.948^{+0.005}_{-0.004}$\\
Pessimistic & 1.0 & Lifetime limited & H02 & $50\%$ efficiency & $0.56^{+0.02}_{-0.03}$ & $0.71^{+0.04}_{-0.02}$\\
Pessimistic & 1.0 & 0.01 & IT03 & State dependent & $0.47^{+0.01}_{-0.04}$ & $0.837^{+0.008}_{-0.034}$\\
Pessimistic & 1.0 & 0.01 & IT03 & $50\%$ efficiency & $0.56^{+0.02}_{-0.02}$ & $0.64^{+0.04}_{-0.02}$\\
Pessimistic & 1.0 & 0.01 & H02 & State dependent & $0.48^{+0.01}_{-0.01}$ & $0.856^{+0.003}_{-0.003}$\\
Pessimistic & 1.0 & 0.01 & H02 & $50\%$ efficiency & $0.597^{+0.014}_{-0.009}$ & $0.69^{+0.02}_{-0.01}$\\
Pessimistic & 5.0 & Lifetime limited & IT03 & State dependent & $0.35^{+0.06}_{-0.05}$ & $0.988^{+0.002}_{-0.001}$\\
Pessimistic & 5.0 & Lifetime limited & IT03 & $50\%$ efficiency & $0.63^{+0.04}_{-0.05}$ & $0.83^{+0.02}_{-0.04}$\\
Pessimistic & 5.0 & Lifetime limited & H02 & State dependent & $0.70^{+0.02}_{-0.02}$ & $0.984^{+0.001}_{-0.002}$\\
Pessimistic & 5.0 & Lifetime limited & H02 & $50\%$ efficiency & $0.59^{+0.03}_{-0.02}$ & $0.930^{+0.006}_{-0.008}$\\
Pessimistic & 5.0 & 0.01 & IT03 & State dependent & $0.28^{+0.03}_{-0.03}$ & $0.928^{+0.014}_{-0.003}$\\
Pessimistic & 5.0 & 0.01 & IT03 & $50\%$ efficiency & $0.455^{+0.01}_{-0.0090}$ & $0.643^{+0.009}_{-0.013}$\\
Pessimistic & 5.0 & 0.01 & H02 & State dependent & $0.351^{+0.012}_{-0.007}$ & $0.911^{+0.002}_{-0.002}$\\
Pessimistic & 5.0 & 0.01 & H02 & $50\%$ efficiency & $0.40^{+0.03}_{-0.03}$ & $0.72^{+0.02}_{-0.01}$\\
\enddata
\tablenotetext{a}{The fraction $f_\mathrm{ low}(p_\mathrm{detect})$ is defined relative to $M_\mathrm{bound}=4.5 M_{\odot}$. }
\tablecomments{Using \texttt{COSMIC}, a set of population synthesis simulations spanning uncertain aspects of binary evolution physics reveals that rapid population synthesis consistently fills the mass gap. 
The detection-weighted fraction of LMXBs with a BH in the mass gap $f_\mathrm{ low}(p_\mathrm{detect})$ is presented for thirty-two exploratory \texttt{COSMIC} Milky Way populations (assuming no transition to RIA and either $M_\mathrm{ NS, birth-max} = 1.5 M_{\odot}$ or $3.0M_{\odot}$). 
For a given population, the average  and the $68\%$ uncertainty is reported from $5000$ snapshots of the evolution tracks between $10.5~\mathrm{Gyr}$ and $11.5~\mathrm{Gyr}$, to reflect the uncertainty in the age of the Milky Way. 
Each population adopts a different combination of binary evolution prescriptions, and each row of the table corresponds to a different \texttt{COSMIC} population. 
As outlined in Section~\ref{sec:pop_synth}, the prescriptions for uncertain aspects of binary evolution physics are varied between the \texttt{COSMIC} populations, including the conditions for merging during a CE event (Optimistic versus Pessimistic), the efficiency of CE ejection ($\alpha=1$ or $5$), the minimum mass ratio $q$ at ZAMS ($q>0.01$ or Lifetime limited, where $q$ is restricted such that the pre-main-sequence lifetime of the secondary is shorter than the main-sequence lifetime of the primary), the magnetic braking implementation (following either \citetalias{Hurley2002} or \citetalias{Ivanova2003}), the efficiency of accretion during RLO (either $50\%$ efficiency or State dependent efficiency, where compact objects have conservative mass accretion up to the Eddington limit). 
For each population, the compact object birth mass distribution is fixed to the Delayed prescription of \citet{Fryer2012}, the maximum NS mass is fixed to $3M_{\odot}$, and the binaries' initial conditions are independently drawn from \citet{Sana2012}.   } 

\end{deluxetable*}

\begin{figure*}
\gridline{\fig{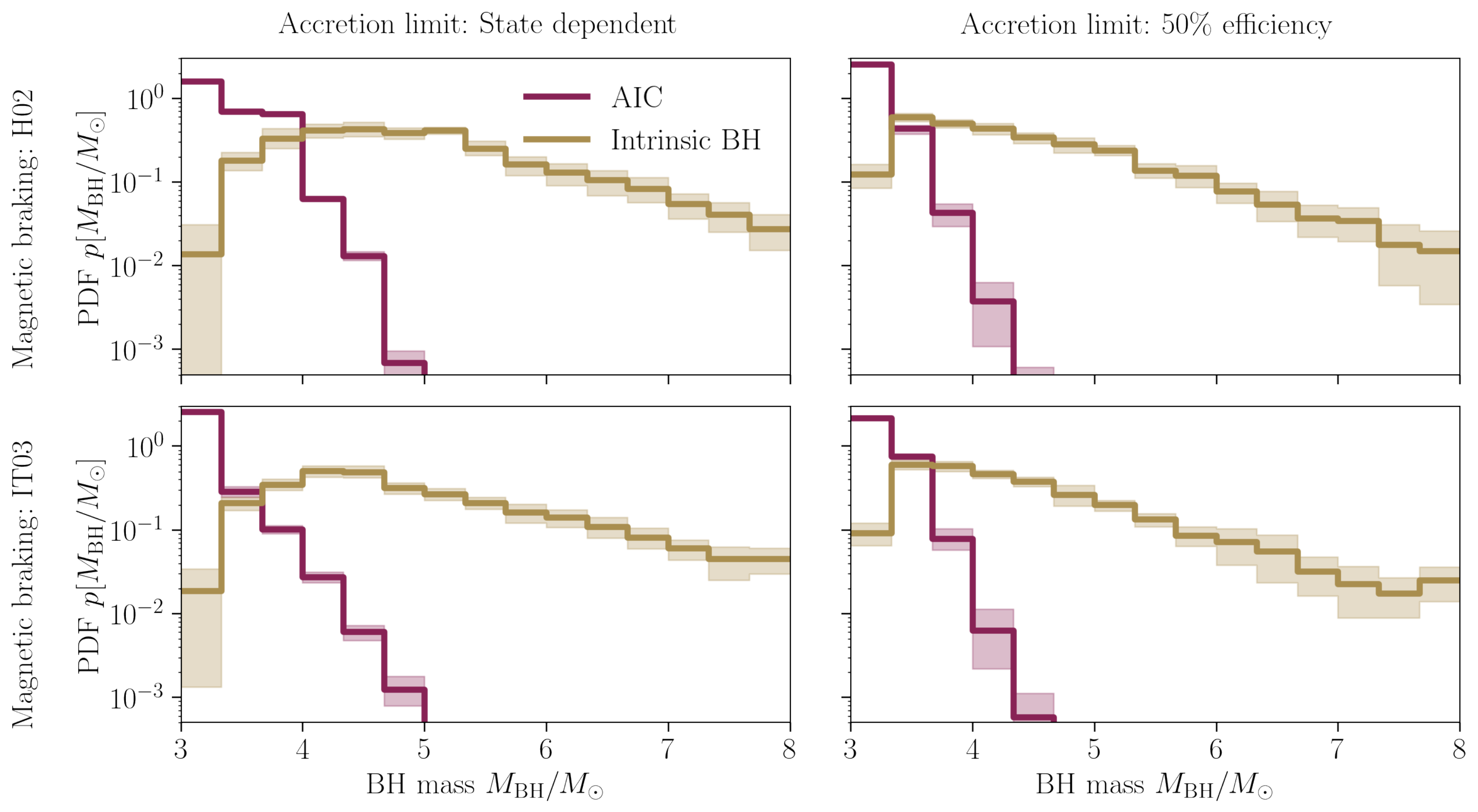}{\textwidth}{}}
\gridline{\fig{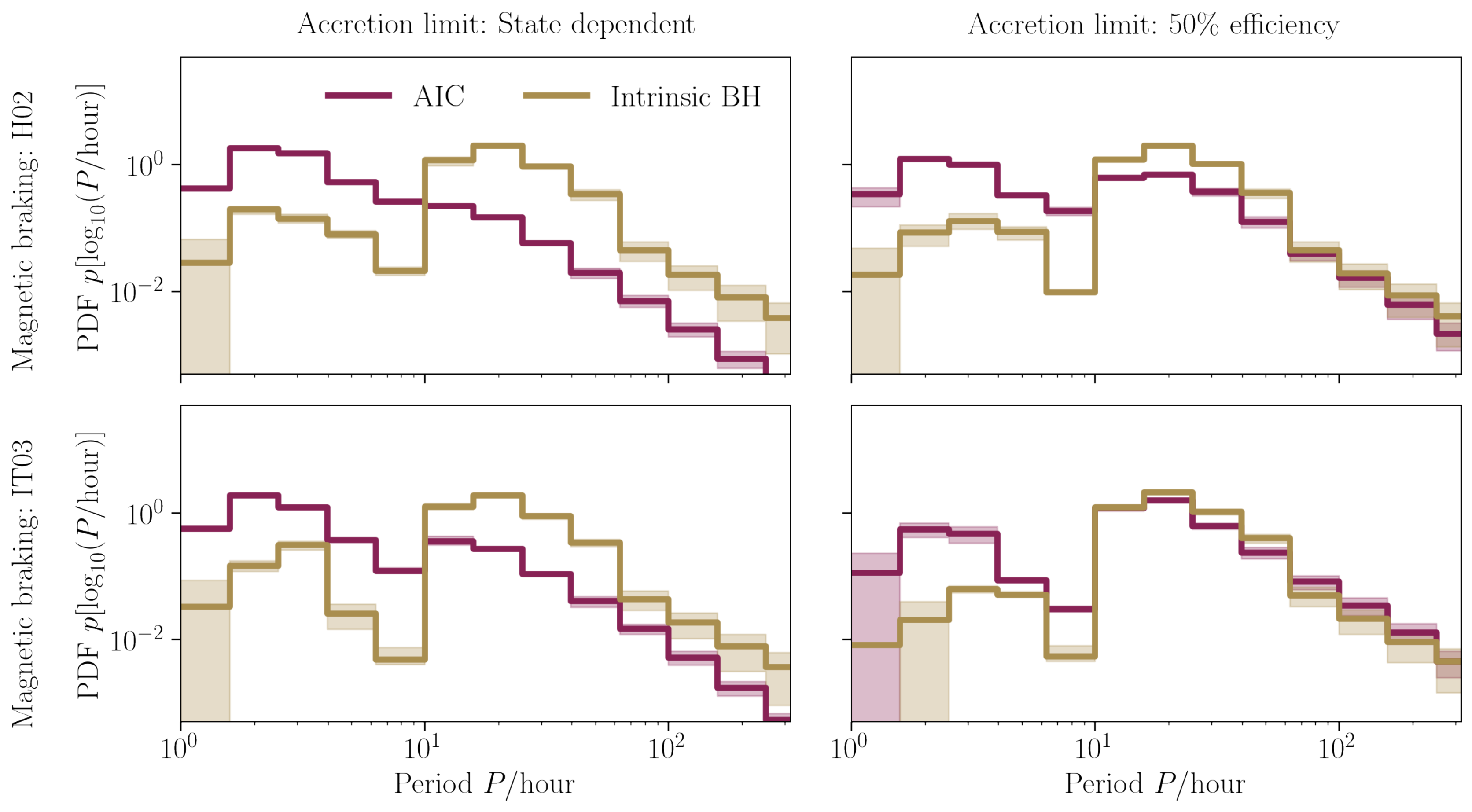}{\textwidth}{}}
\caption{
The BH mass (\emph{upper panels}) and orbital period (\emph{lower panels}) distributions for a selection of synthetic Galactic LMXB samples. 
Detection probabilities are calculated with no transition to RIA. 
Each panel corresponds to a different set of binary evolution prescriptions. 
For each presented population $\alpha=5$, the minimum $q$ at ZAMS follows the Lifetime limited prescription, and $M_\mathrm{NS,birth-max}=3.0M_\odot$.
The solid lines represent the average distribution across all $5000$ snapshots of the evolution tracks, and the shaded regions demarcate the $16$th--$84$th percentiles. 
AIC LMXBs are highlighted in purple. 
The detection-weighted BH mass and orbital period distributions of LMXBs from population synthesis are broadly insensitive to changes in the adopted binary evolution prescriptions.
}
\label{fig:channels}
\end{figure*}

\subsection{Synthetic observed samples}
\label{sec:observing}

To facilitate comparison between the observed LMXB sample and synthetic populations of LMXBs, a Monte Carlo observing procedure is employed. 
Samples are drawn from the evolution tracks of \texttt{COSMIC} and \texttt{MESA} based on the assumed age of the Milky Way, and are then compared with the Galactic LMXB population to investigate the nature of the lower mass gap.

While similar in philosophy, the synthetic observing process differs between the \texttt{COSMIC} Milky Way populations and the \texttt{MESA} grid; the differences are motivated by the far lower number of binaries simulated in the \texttt{MESA} grid. 
For a given \texttt{COSMIC} Milky Way population, we collect $5000 $ snapshots of the population within a time window of $10.5$--$11.5~\mathrm{Gyr}$, informed by our adopted Milky Way age of $\mathcal{A}_\mathrm{ MW} = (11\pm0.5)~\mathrm{Gyr}$. 
The time of each snapshot is independently and uniformly drawn within the window. 
At every snapshot, we infer the properties of the population via linear interpolation of the \texttt{COSMIC} time steps. 
Statistical properties of the synthetic observed sample can then be inferred by considering all $5000$ snapshots. 

As described in Section~\ref{sec:MESA}, we also consider a three-dimensional grid of compact object--main sequence binaries with \texttt{MESA}, and initialize \texttt{COSMIC} populations with matching initial conditions for a range of binary evolution prescriptions. 
Given the discrete distribution of initial conditions and the lower number of binaries, synthetic observed samples are drawn from the grids using a modified version of the Monte Carlo procedure applied to the \texttt{COSMIC} populations. 
For a given trial, we draw a random Milky Way age $\mathcal{A}_\mathrm{MW}$ from $\mathcal{U}(10.5,11.5)~\mathrm{Gyr}$ and draw each binary a birth time independently from $\mathcal{U}(0,\mathcal{A}_\mathrm{MW})$; this assignment of binary birth times is analogous to continuous star formation and is based on the thin disk component of the Milky Way model (Table~\ref{tab:MW}). 
Linear interpolation is then applied to infer each binary's properties at $\mathcal{A}_\mathrm{MW}$. While distances are assigned via spatial distributions for the \texttt{COSMIC} Milky Way populations, a nominal distance of $8~\mathrm{kpc}$ is adopted for each binary in the \texttt{MESA} grid, due to the low number of binaries. 
In total, $5000$ trails are drawn for each grid.

For a single snapshot, the number of observed transient LMXB systems is
\begin{equation}
     N_\mathrm{ transient} = \sum_{i=1}^{N_\mathrm{LMXB}} p_{\mathrm{detect}, i},
\end{equation}
where $N_\mathrm{LMXB}$ is the total number of LMXB systems in the population during the snapshot. Systems composed of a BH accreting mass from a non-degenerate donor star via RLO are considered LMXBs, which excludes degenerate donors, to mirror the observed sample of LMXBs with BH mass estimates.

To quantify the density of transient LMXB systems within the lower mass gap, we define the fraction 
\begin{equation}
 f_\mathrm{ low}(p_\mathrm{detect}) = \frac{1}{N_\mathrm{transient} }\sum_{i=1}^{N_\mathrm{LMXB}} 
\begin{cases}
      p_{\mathrm{detect}, i}, & M_\mathrm{BH}<M_\mathrm{bound}  \\
      0, & M_\mathrm{BH} \geq M_\mathrm{bound}
\end{cases} .
\end{equation} 
Here $M_\mathrm{ bound}$ is nominally fixed to $ 4.5M_{\odot}$.

The fraction $f_\mathrm{ low-intrinsic}$ is then defined as the unweighted fraction of LMXB systems with a BH mass less than $M_\mathrm{bound}$,
\begin{equation}
    f_\mathrm{ low-intrinsic} = \frac{1}{N_\mathrm{LMXB} }\sum_{i=1}^{N_\mathrm{LMXB}} 
    \begin{cases}
      1, & M_\mathrm{ BH} < M_\mathrm{ bound}  \\
      0, & M_\mathrm{ BH} \geq M_\mathrm{ bound}
\end{cases}.
\end{equation}  
This $f_\mathrm{ low-intrinsic}$ fraction characterizes the \emph{intrinsic} density of LMXBs with BHs within the lower mass gap, while $f_\mathrm{ low}(p_\mathrm{detect})$ characterizes the \emph{observed} density of LMXBs with a mass-gap BH. 
Considering $f_\mathrm{ low}(p_\mathrm{detect})$ and $f_\mathrm{ low-intrinsic}$ together highlights the influence of transient detection effects on the observed Galactic LMXB sample. 

Following the procedures above, we generate synthetic LMXB samples for each \texttt{COSMIC} population and \texttt{MESA} grid. 
We discuss these samples and their implications for the lower mass gap in Section~\ref{sec:results}.

\section{Results}
\label{sec:results}

Using rapid binary population synthesis (\texttt{COSMIC}) coupled with detailed stellar-structure and mass-transfer models (\texttt{MESA}), we investigate the impact of observational biases on the Galactic LMXB sample. 
We demonstrate that transient LMXB selection effects do introduce some biases into the observed LMXB sample, and that the observed lower mass gap has implications for the maximum NS birth-mass $M_\mathrm{ NS, birth-max}$.

To survey how well population synthesis methods match the observed LMXB sample, and to explore dependencies between the LMXB sample and uncertain aspects of binary evolution physics, our investigation begins with a suite of \texttt{COSMIC} rapid binary population synthesis simulations. 
The \texttt{COSMIC} results are discussed in Section~\ref{sec:results_COSMIC}.
For thirty-two combinations of binary evolution prescriptions, Galactic LMXB samples from \texttt{COSMIC} host mass-gap BHs, with a significant fraction of mass-gap BH forming through AIC of a NS. 
Regardless of binary evolution prescriptions, the \texttt{COSMIC} populations show minimal to moderate dependence on the adopted X-ray outburst selection effects and favor $M_\mathrm{ NS, birth-max} \lesssim 2 M_{\odot}$ (to suppress the formation of AIC LMXB).

Given the consistency of the \texttt{COSMIC} results, we next consider a three-dimensional grid of LMXBs evolved using \texttt{MESA}. These results are discussed in Section~\ref{sec:results_MESA}. 
While the \texttt{MESA} grid samples a narrower range of binary evolution prescriptions than the \texttt{COSMIC} populations, \texttt{MESA} employs a more detailed treatment of XRB mass-transfer.  
For $M_\mathrm{ NS, birth-max} = 3M_{\odot}$, \texttt{MESA} and \texttt{COSMIC} yield similar BH mass distributions, but if $M_\mathrm{NS, birth-max} \lesssim 2 M_{\odot}$ \texttt{MESA} produces a dearth of LMXBs with mass-gap BHs and reveals a greater dependence on the adopted transition to RIA in the disk instability model. 

\subsection{\texttt{COSMIC} results }
\label{sec:results_COSMIC}

Using \texttt{COSMIC}, we consider proof-of-concept Milky Way populations over a range of binary evolution prescriptions. 
In addition to evaluating the ability of population synthesis methods to replicate the observed LMXB sample, we leverage the simulation suite to investigate dependencies between the LMXB sample and uncertain aspects of binary evolution physics. 
Given the substantial level of uncertainty in several binary evolution prescriptions and the order-of-magnitude nature of our study, treating the \texttt{COSMIC} simulations in aggregate is preferred. 
Here, we first outline the principal results of the simulation suite, before investigating dependencies between BH mass-gap occupancy and different prescriptions of binary evolution physics.

For a given Milky Way \texttt{COSMIC} population, a synthetic observed sample of LMXBs is extracted from the evolution tracks through a Monte Carlo process (Section~\ref{sec:observing}). 
In Table~\ref{tab:pop_synth}, the $p_\mathrm{detect}$-weighted fraction of LMXBs in the synthetic observed sample hosting a BH in the mass gap $f_\mathrm{low}(p_\mathrm{detect})$ is reported for each population. 
Detection probabilities are assigned under the assumption that there is no transition to RIA. 
For a selection of representative populations, the mass and period distributions of the synthetic observed LMXBs are presented in Figure~\ref{fig:channels}. 
Every model predicts  $f_\mathrm{low}(p_\mathrm{detect})$ is greater than $40\%$ (using a benchmark $M_\mathrm{ NS, birth-max}=3M_{\odot}$ and $ M_\mathrm{bound}=4.5M_{\odot}$). 

For LMXBs hosting a BH within the mass gap, the BHs formed through one of two channels: AIC of a NS or intrinsic formation of a BH (directly post core-collapse of a star). 
While both channels are dependent on the maximum NS mass allowed by the equation of state (because it sets the transition between NS and BH in the compact object birth mass distribution), the AIC LMXBs also rely on $M_\mathrm{ NS, birth-max}$: if $M_\mathrm{ NS, birth-max}$ lies significantly below the maximum NS mass, the formation of AIC LMXBs will be suppressed relative to a population where $M_\mathrm{ NS, birth-max}$ lies near the maximum NS mass.

\begin{figure*}
\gridline{\fig{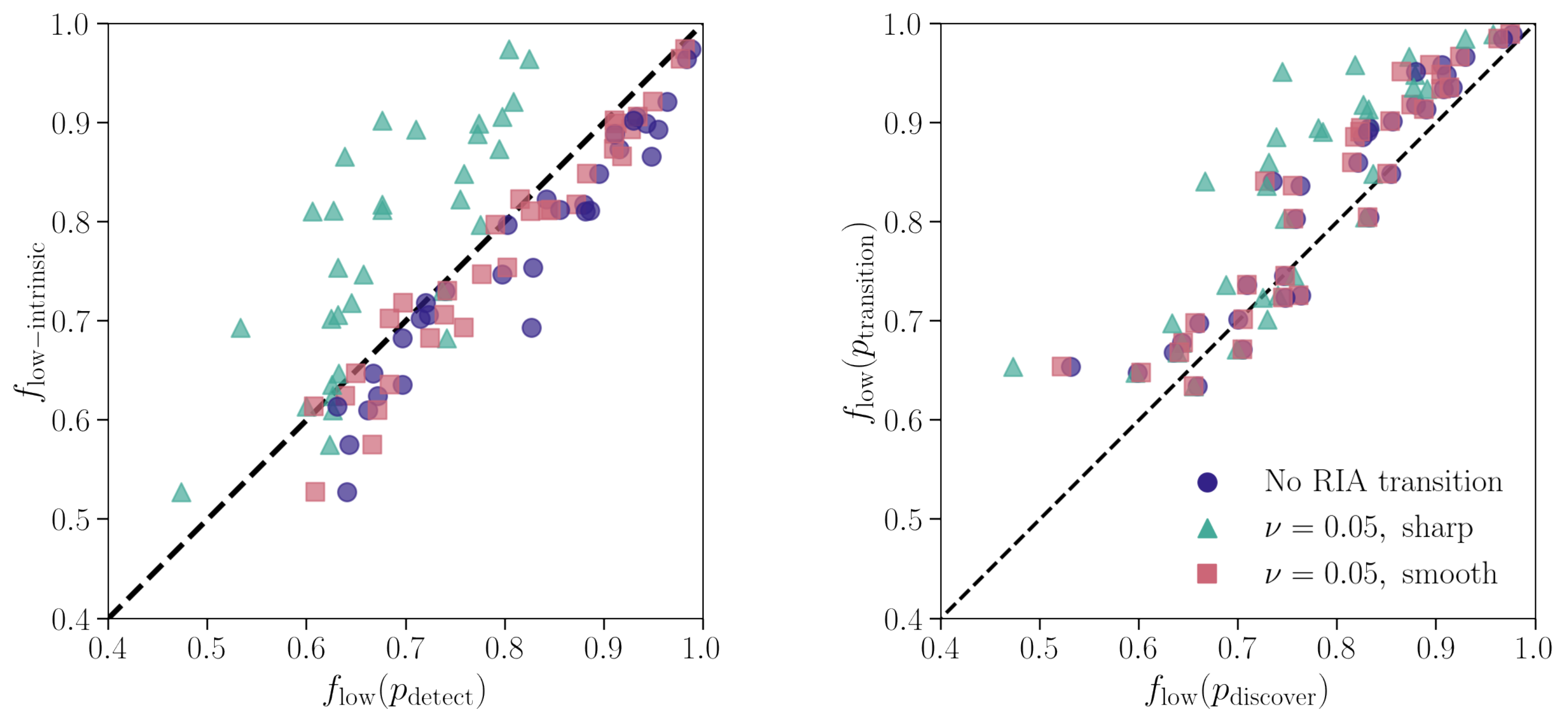}{\textwidth}{}}
\caption{
From population synthesis, each combination of binary evolution prescription and disk instability model yields an observed LMXB mass distribution with a substantial fraction of mass-gap BHs (for $M_\mathrm{ NS, birth-max}=3M_{\odot}$). 
\emph{Left}: The detection probability-weighted fraction of LMXBs with mass gap BHs, $f_\mathrm{ low}(p_\mathrm{detect})$, is presented against the unweighted fraction, $f_\mathrm{ low-intrinsic}$, for each model (using either no transition to RIA, a sharp transition, or a smooth transition). 
For points that fall above the dashed-black line, detection effects lead to a dearth of mass-gap BH detections relative to the intrinsic population.
\emph{Right}: The $p_\mathrm{discover}$ detection-weighted fraction of LMXBs with a mass-gap BH is presented against the $p_\mathrm{transition}$-weighted fraction. 
This panel breaks-down the two components of the LMXB detection probability: (i) being bright enough during outburst for discovery $p_\mathrm{discover}$, and (ii) transitioning from outburst to quiescence during the survey $p_\mathrm{transition}$. 
Relative to $p_\mathrm{transition}$, weighting by $p_\mathrm{discover}$ suppresses observation of mass-gap BHs. 
}
\label{fig:weighting}
\end{figure*}

\begin{figure*}
\gridline{\fig{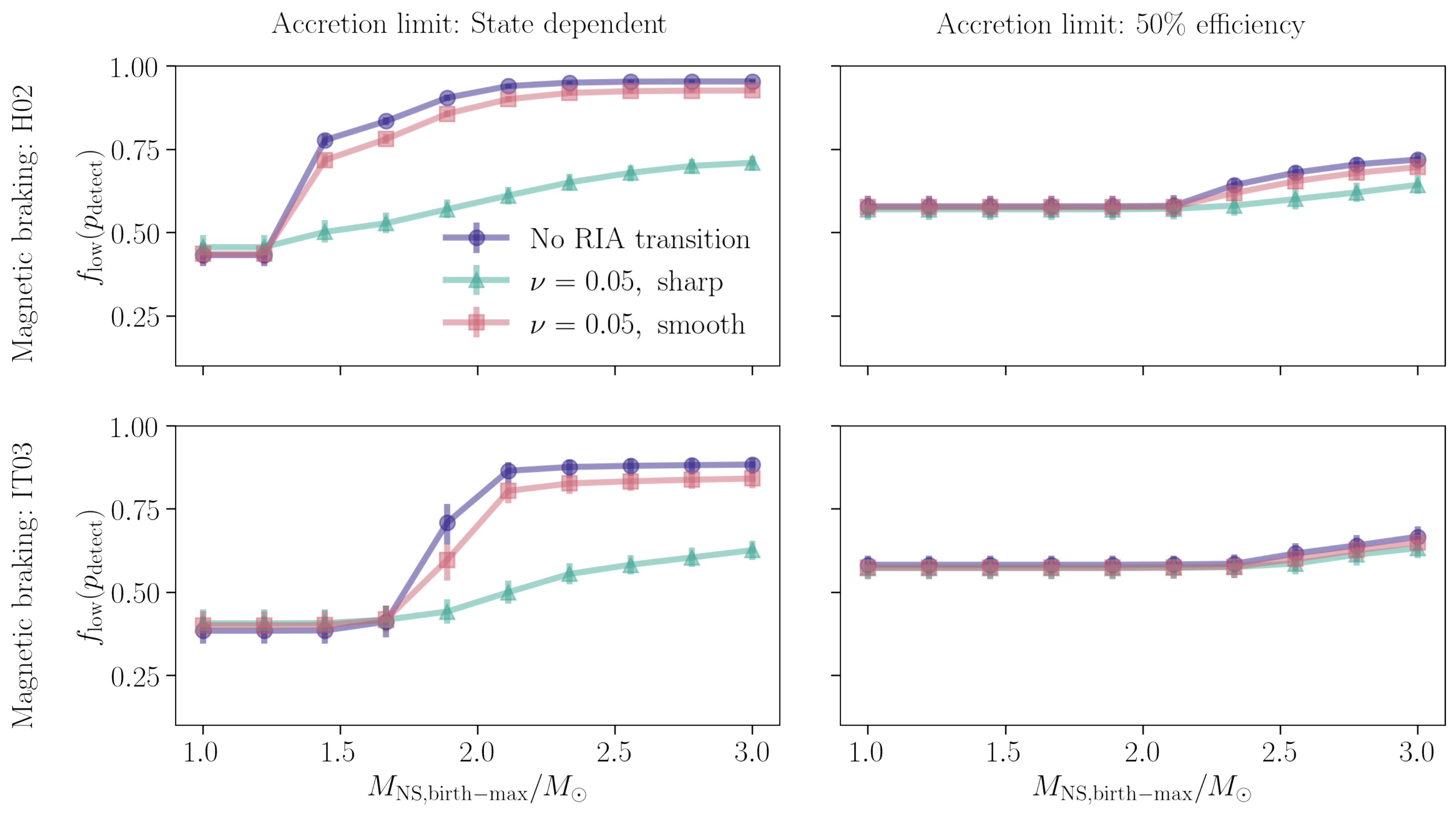}{\textwidth}{}}
\caption{
The detection-weighted fraction of LMXBs with a mass gap BH $f_\mathrm{low}(p_\mathrm{detect})$ as a function of $M_\mathrm{NS, birth-max}$ for a selection of populations.  
Fractions for each population are calculated by averaging over all $5000$ snapshots and the uncertainties are taken from the $68\%$ uncertainty across the snapshots. 
Lowering the maximum NS birth mass $M_\mathrm{ NS, birth-max}$ reduces the occurrence of LMXB systems with $M_\mathrm{BH} < 4.5M_{\odot}$. 
The magnitude of this trend depends on the adopted accretion prescription; changing the magnetic braking prescription moderately effects the fraction of mass-gap BH detections, while modifying the adopted accretion physics significantly affects $f_\mathrm{low}(p_\mathrm{detect})$.
}
\label{fig:f_vs_NS_birth}
\end{figure*}

AIC LMXBs account for a significant fraction ($>15\%$ and reaching as high as $90\%$, for $M_\mathrm{ NS, birth-max}=3M_{\odot}$ and assuming no transition to RIA) of the detection-weighted synthetic observed populations. 
The AIC LMXBs typically host BHs with $M_\mathrm{BH}<5M_\odot$.
The populations where the fraction of LMXBs formed via AIC is below $30\%$ all adopted a mass-accretion efficiency of $50\%$, while the populations where the fraction of LMXBs formed via AIC is above $90\%$ all adopted the State dependent mass-accretion prescription. 
The fraction of LMXBs formed via AIC is lowest for the population with $\alpha=1.0$, Lifetime limited $q$, Pessimistic CE survival, magnetic braking following \citetalias{Ivanova2003}, and $50\%$ accretion efficiency; the fraction of LMXBs formed via AIC is maximized for the population with $\alpha=5.0$, Lifetime limited minimum $q$, Pessimistic CE survival, magnetic braking following \citetalias{Ivanova2003}, and State dependent accretion efficiency. 

AIC LMXBs also dominate ($>20\%$ and reaching as high as $99\%$) the sub-population of systems with orbital periods shorter than $10$~hours. 
Consistent with the total fraction of LMXBs formed via AIC, the fraction of AIC LMXBs among systems with periods shorter than $10$~hours is reduced by adopting $50\%$ accretion efficiency and maximized by adopting State dependent accretion efficiency. 

We next consider how the principal results of the \texttt{COSMIC} simulation suite (i.e., population synthesis fails to replicate the observed LMXB population and consistently produces a population of AIC LMXBs) depends on the adopted binary evolution prescriptions, beginning with the details of the disk instability physics. 
From population synthesis, the detection of LMXBs hosting mass-gap BHs is largely insensitive to the adopted transition to RIA. 
In Figure~\ref{fig:weighting}, $f_\mathrm{ low}(p_\mathrm{detect})$ is presented against $f_\mathrm{ low-intrinsic}$ for every combination of population synthesis prescription and X-ray outburst prescription. 
Since $f_\mathrm{ low-intrinsic}$ characterizes the \emph{intrinsic} density of BHs within the lower mass gap, and $f_\mathrm{ low}(p_\mathrm{detect})$ characterizes the \emph{observed} density of LMXBs with a mass-gap BH, considering them together highlights the influence of transient detection effects on the observed Galactic LMXB sample. 

As shown in Figure~\ref{fig:weighting}, modeling the X-ray outburst with no RIA transition yields $f_\mathrm{ low}(p_\mathrm{detect})$ consistent with modeling the X-ray outburst with a smooth RIA transition. 
For both of these prescriptions, the detection-weighted mass distribution is only marginally different from the unweighted distribution: $f_\mathrm{ low}(p_\mathrm{detect})$ is $\sim10\%$ higher than $f_\mathrm{ low-intrinsic}$. 
Adopting a sharp transition to RIA lowers $f_\mathrm{ low}(p_\mathrm{detect})$ relative to the other RIA transitions; however, each population still yields an $f_\mathrm{ low}(p_\mathrm{detect})$ greater than $40\%$  (using a benchmark $M_\mathrm{ NS, birth-max}=3M_{\odot}$ and $ M_\mathrm{bound}=4.5M_{\odot}$). 
Since the outburst duration $T_\mathrm{ outburst}$ is positively correlated with orbital period, and the AIC LMXBs typically have orbital periods shorter than $10$~hours, the sensitivity of $f_\mathrm{ low}(p_\mathrm{detect})$ to the RIA transition prescription is expected. 
The sharp transition dramatically decreases the outburst duration (Figure~\ref{fig:lightcurve}), making the short-period LMXBs, which tend to be AIC LMXBs, no longer detectable. 

Discovering a transient LMXB source with an ASM relies on two independent probabilities: (i) the probability that the source is bright enough in outburst to be discovered ($p_\mathrm{ discover}$), and (ii) the probability that the source transitions from outburst to quiescence during the ASM's observing window ($p_\mathrm{transition}$). 
In Figure~\ref{fig:weighting}, the $p_\mathrm{ discover}$-weighted fraction of LMXB with $M_\mathrm{ BH} < 4.5M_{\odot}$ is presented against the $p_\mathrm{transition}$-weighted fraction. 
Similar to $f_\mathrm{ low-intrinsic}$, these fractions are not directly comparable to the observed LMXB sample; however, they outline the relative effects of $p_\mathrm{ discover}$ and $p_\mathrm{ transition}$ on the mass gap. 
As shown in Figure~\ref{fig:weighting}, weighting by $p_\mathrm{transition}$ fills the mass gap relative to $p_\mathrm{discover}$ weighting. 
Compared to short-period LMXBs, long-period LMXBs are disfavored by $p_\mathrm{ transition}$, because the quiescence time is positively related with orbital period. 
Since AIC LMXBs tend to have shorter periods than LMXBs with intrinsic BHs, $p_\mathrm{ transition}$ favors AIC LMXBs and raises the density of BHs in the mass gap relative to $p_\mathrm{discover}$ weighting.

The prevalence of LMXBs hosting a mass-gap BH in the \texttt{COSMIC} populations is discrepant with the Galactic LMXB sample. 
While AIC LMXBs comprise a substantial fraction of the \texttt{COSMIC} populations for a variety of disk instability models and binary evolution prescriptions, these systems are dependent on $M_\mathrm{ NS, birth-max}$. 
Lowering $M_\mathrm{NS, birth-max}$ suppresses the number of AIC LMXBs, so tuning $M_\mathrm{ NS, birth-max}$ could bring the population synthesis models into better agreement with observations. 
Below, we investigate the dependence between the AIC LMXB population and $M_\mathrm{ NS, birth-max}$.

In Figure~\ref{fig:f_vs_NS_birth}, $f_\mathrm{low}(p_\mathrm{detect})$ is presented as a function of $M_\mathrm{NS, birth-max}$ for a selection of \texttt{COSMIC} populations. 
While $f_\mathrm{ low}(p_\mathrm{detect})$ decreases for each population as $M_\mathrm{NS, birth-max}$ is lowered, the magnitude of the effect depends strongly on the accretion-limit prescription. 
This dependence reflects the higher occurrence rate of AIC LMXBs in populations with State dependent accretion limits. For the State dependent prescription, compact objects have conservative mass-transfer, provided the mass-transfer rate is sub-Eddington; accretors therefore retain a greater fraction of the transferred mass, which corresponds to more NSs undergoing AIC. 
The increased accretion rates from the State dependent prescription also translate to greater mass gain for the intrinsic BH LMXBs, which in turn reduces $f_\mathrm{low}(p_\mathrm{detect})$. 
Populations with $50\%$ accretion efficiency have both higher $f_\mathrm{low}(p_\mathrm{detect})$ for intrinsic BHs and weaker dependence on $M_\mathrm{NS, birth-max}$.

Over the range of uncertain binary evolution physics that we considered, LMXB populations modeled with \texttt{COSMIC} yield a substantial fraction of LMXB BHs with $M_\mathrm{BH}<4.5M_{\odot}$ (over $40\%$, assuming no transition to RIA) for $M_\mathrm{ NS, birth-max} = 3M_{\odot}$. 
The simulations are in tension with the observed sample of LMXBs, where there is a dearth of BHs with masses below $\approx 4.5M_\odot$ \citep{Bailyn1998, Ozel2010, Farr2011}.
The fraction of mass-gap BHs in the synthetic Galactic sample is minimized by adopting a CE efficiency of $\alpha=5$, $q>0.01$, magnetic braking following \citetalias{Ivanova2003}, State dependent accretion efficiency, pessimistic CE survival, no transition to RIA, and $M_\mathrm{ NS, birth-max}=1.5 M_{\odot}$; for this population, $28\%$ of LMXBs host a BH in the mass gap.
While lowering $M_\mathrm{ NS, birth-max}$ lowers the density of LMXB systems in the mass gap, $f_\mathrm{ low}(p_\mathrm{detect})$ remains greater than $25\%$ for every \texttt{COSMIC} population and is substantially higher for most, assuming no transition to RIA.

\begin{deluxetable*}{ccclccc}[t!]\label{tab:grid_comp}
\tablecaption{\texttt{COSMIC} and \texttt{MESA} grid comparison}
\tablehead{
\colhead{} & 
\colhead{Magnetic braking} & 
\colhead{Accretion limit} &
\colhead{RIA transition} &
\colhead{$f_\mathrm{low} (M_\mathrm{NS, birth} < 1.5 M_\odot)$\tablenotemark{a} }  & 
\colhead{$f_\mathrm{low} (M_\mathrm{NS, birth} < 3.0 M_\odot)$ } 
}
\startdata
\texttt{COSMIC} & H02 & $50\%$ efficiency &  No  transition & 0.345 & 0.406\\
 & H02 & $50\%$ efficiency &  $\nu=0.05$, sharp & 0.422 & 0.492\\
 & H02 & $50\%$ efficiency &  $\nu=0.05$, smooth & 0.347 & 0.407\\
 & IT03 & $50\%$ efficiency &  No  transition & 0.345 & 0.406\\
 & IT03 & $50\%$ efficiency &  $\nu=0.05$, sharp & 0.418 & 0.490\\
 & IT03 & $50\%$ efficiency &  $\nu=0.05$, smooth & 0.344 & 0.405\\
 & H02 & State dependent &  No  transition & 0.354 & 0.437\\
 & H02 & State dependent &  $\nu=0.05$, sharp & 0.439 & 0.516\\
 & H02 & State dependent &  $\nu=0.05$, smooth & 0.351 & 0.435\\
 & IT03 & State dependent &  No  transition & 0.352 & 0.435\\
 & IT03 & State dependent &  $\nu=0.05$, sharp & 0.439 & 0.515\\
 & IT03 & State dependent &  $\nu=0.05$, smooth & 0.353 & 0.437\\
\hline
\texttt{MESA} &  & &  No  transition & 0.105 & 0.428\\
 &  & &  $\nu=0.05$, sharp & 0.242 & 0.507\\
 &  & &  $\nu=0.05$, smooth & 0.105 & 0.429\\
\enddata
\tablenotetext{a}{The fraction $f_\mathrm{low}(p_\mathrm{detect})$  is defined relative to $M_\mathrm{bound}=4.5M_{\odot}$. }
\tablecomments{Relative to rapid population synthesis, LMXBs evolved using \texttt{MESA} under-populate the mass gap if $M_\mathrm{ NS, birth-max} \lesssim 2 M_{\odot}$, and show greater dependence on the adopted RIA transition. 
The detection-weighted fraction of LMXBs with a BH in the mass gap $f_\mathrm{low}(p_\mathrm{detect})$ is presented for a three-dimensional grid of binaries evolved with \texttt{MESA}, as well as for grids evolved using \texttt{COSMIC}; the grids evolved with \texttt{COSMIC} are initialized at the same initial conditions as the grid evolved using \texttt{MESA}. 
Values of the detection-weighted fraction $f_\mathrm{low}(p_\mathrm{detect})$ are reported for each combination of $M_\mathrm{ NS, birth-max} = 1.5M_{\odot}$ or $3.0M_{\odot}$ and smooth, sharp, or non-existent transition to RIA. 
To marginalize over uncertain aspects of binary evolution physics, different combinations of binary evolution prescriptions are adopted for the \texttt{COSMIC} grids, including the magnetic braking implementation (following either \citetalias{Hurley2002} or \citetalias{Ivanova2003}) and the efficiency of accretion during RLO (either $50\%$ efficiency or State dependent efficiency, where compact objects have conservative mass accretion up to the Eddington limit).  } 
\end{deluxetable*}

\begin{figure*}
\gridline{\fig{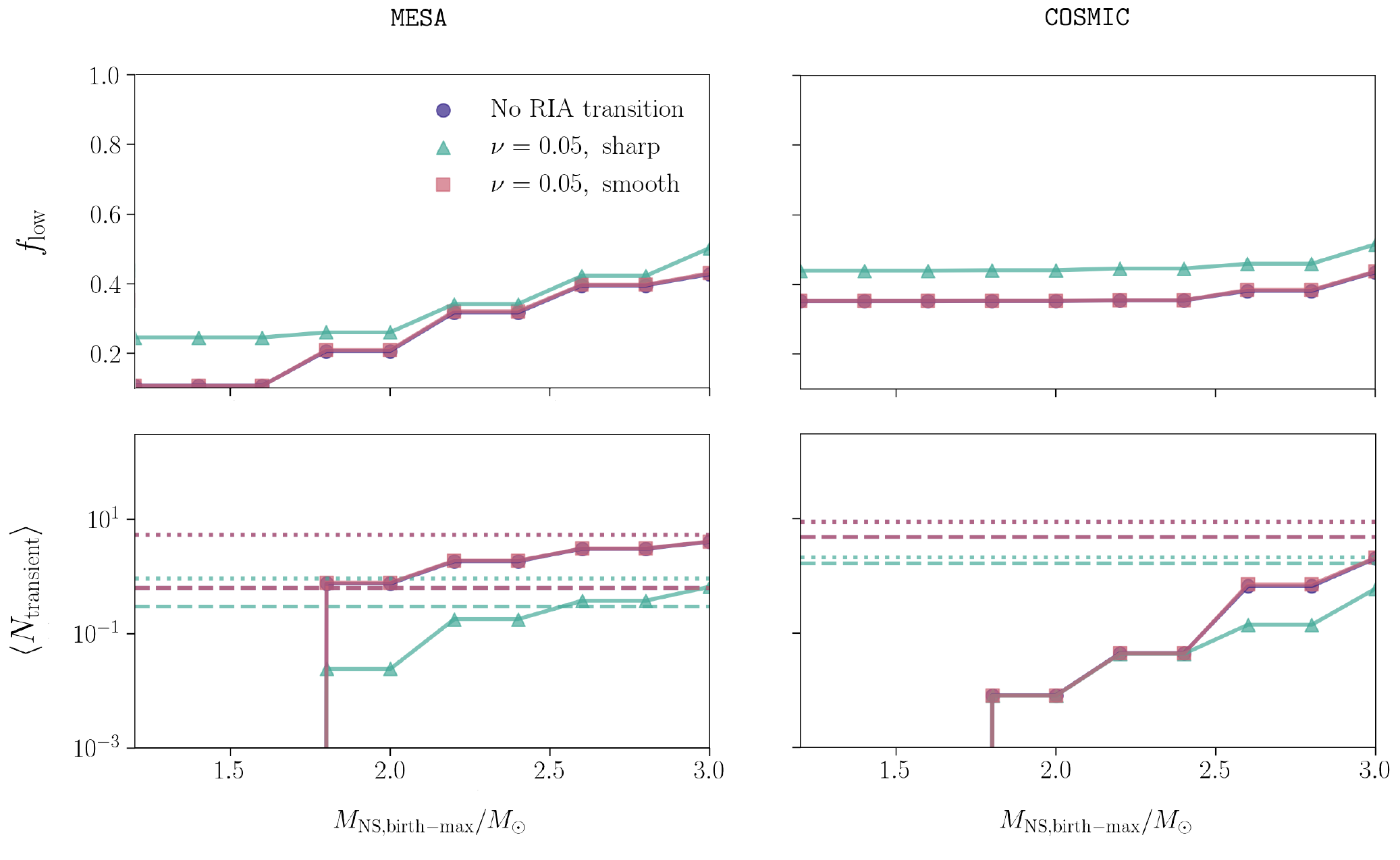}{\textwidth}{}}
\caption{
\emph{Top}: The detection-weighted fraction of LMXBs with a mass gap BH $f_\mathrm{ low}(p_\mathrm{detect})$ as a function of $M_\mathrm{ NS, birth-max}$. 
Each color corresponds to a different RIA transition. For each panel, the results for no transition to RIA overlap with the results for a smooth transition to RIA.
\emph{Bottom}: The average number of  detections of AIC LMXBs (over $N_\mathrm{trials}=5000$ snapshots of the evolutionary tracks) with $M_\mathrm{BH}<4.5M_{\odot}$ (as a solid line); the dashed horizontal line is the average number of detections of LMXB that formed with a BH and have $M_\mathrm{BH}<4.5M_{\odot}$; the dotted horizontal line is the average number of detections of LMXB (from any formation channel) with $M_\mathrm{BH}>4.5M_{\odot}$. 
The \emph{left column} presents the results from the \texttt{MESA} grid, while the \emph{right column} presents the results from the \texttt{COSMIC} grid. 
While \texttt{MESA} and \texttt{COSMIC} predict similar LMXB BH mass distributions when $M_\mathrm{NS, birth-max}= 3M_{\odot}$, \texttt{MESA} produces a dearth of LMXBs with mass gap BHs relative to \texttt{COSMIC} if $M_\mathrm{NS, birth-max} \lesssim 2M_{\odot}$. 
}
\label{fig:grid_f_for_NS}
\end{figure*}

\begin{figure*}
\gridline{\fig{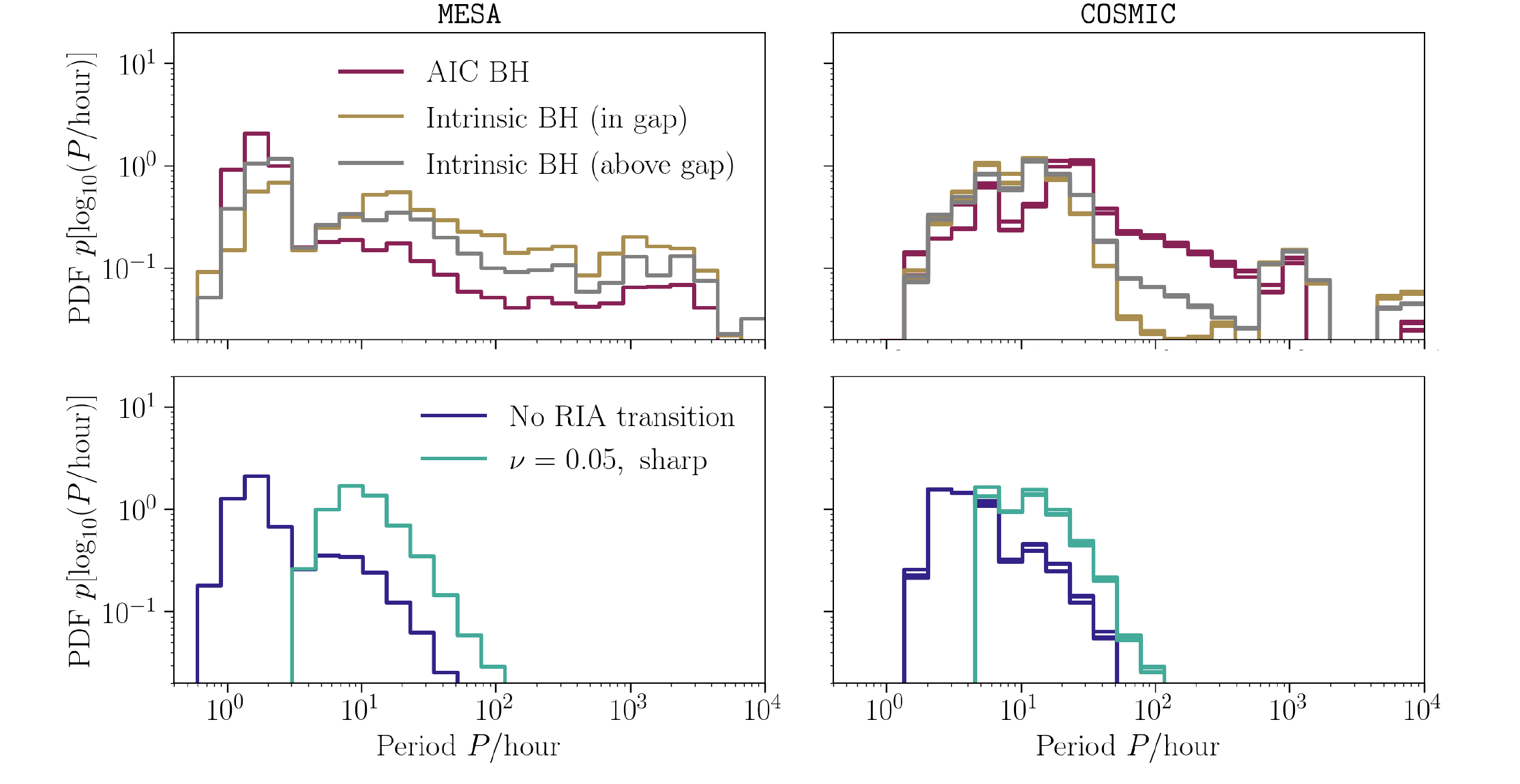}{\textwidth}{}}
\caption{ 
The LMXB orbital period distribution for the \texttt{MESA} grid is presented in the \emph{left} column, while the orbital period distributions of the \texttt{COSMIC} grids (which are generated with matching initial conditions) are presented on the \emph{right}. 
\emph{Top}: Monte Carlo sampled orbital period distributions are presented for three sub-populations of LMXBs: AIC LMXBs (purple), LMXBs with intrinsic BHs in the mass gap (gold), and LMXBs with intrinsic BHs above the mass gap (gray). 
\emph{Bottom}: Monte Carlo sampled and detection-weighted orbital period distributions for the entire population of LMXBs are presented for either no transition to RIA (blue) or a sharp transition (teal); here, applying a smooth transition to RIA is nearly indistinguishable from applying no transition. 
LMXBs evolved using \texttt{MESA} typically occupy shorter orbital periods than LMXBs evolved with \texttt{COSMIC}.
 }
\label{fig:grid_period}
\end{figure*}

\begin{figure*}
\gridline{\fig{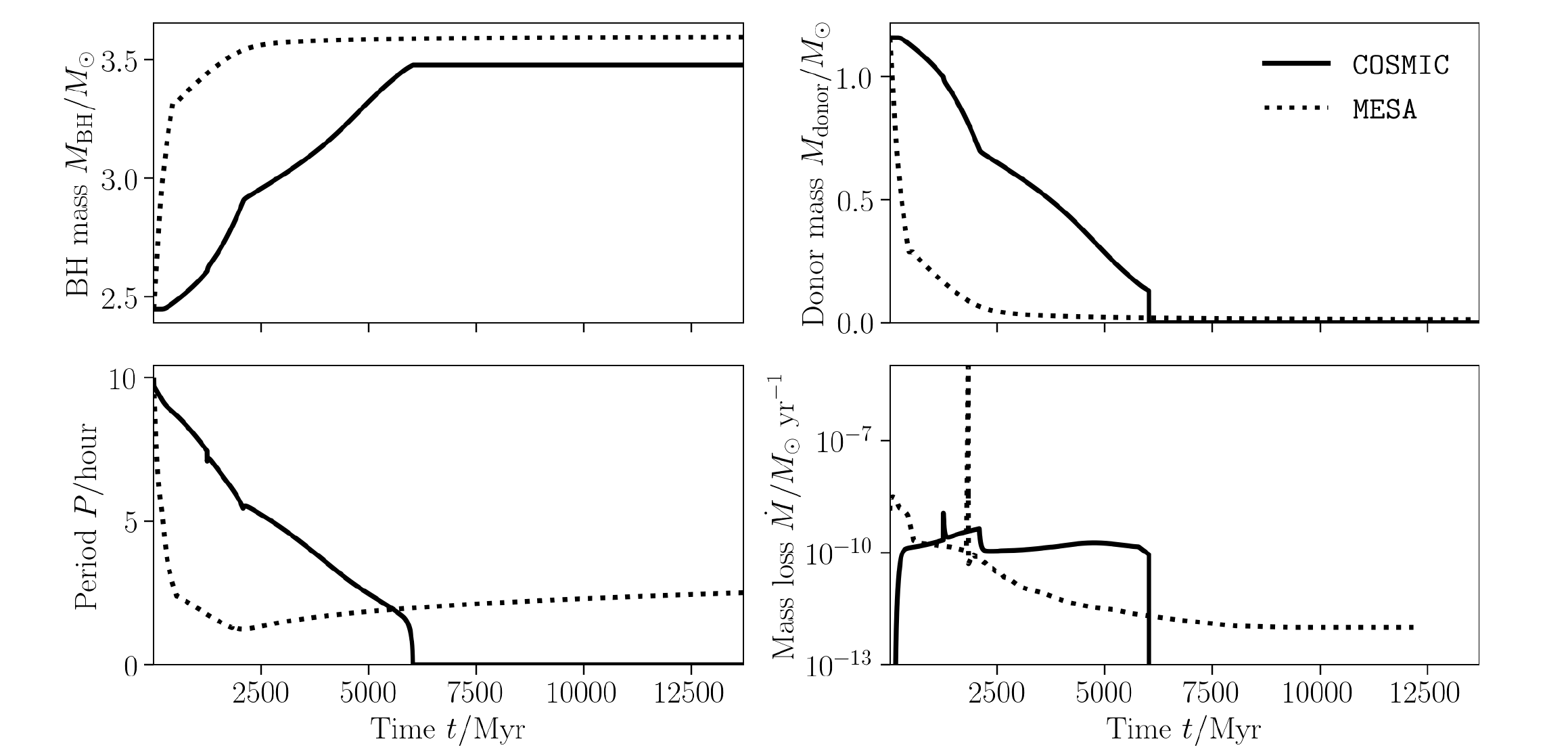}{\textwidth}{}}
\caption{The differences in the orbital period and mass distributions between \texttt{COSMIC} and \texttt{MESA} both originate from the treatment of RLO mass accretion. 
Relative to \texttt{COSMIC}, evolving LMXBs with \texttt{MESA} leads to greater mass accretion by the BHs, and by extension, shorter orbital periods. 
Here the evolution tracks of a single LMXB evolved with \texttt{MESA} (dashed lines) are presented against the evolution tracks when \texttt{COSMIC} is used (solid lines). 
The binary is initialized with $M_\mathrm{BH}=2.5M_{\odot}$, $M_\mathrm{donor}=1.16M_{\odot}$, and $P=0.413$~days. 
For the \texttt{COSMIC} evolution tracks, the magnetic braking formulation of \citetalias{Hurley2002} and the State dependent accretion efficiency were adopted. }
\label{fig:sequences}
\end{figure*}

\subsection{ \texttt{MESA} results }
\label{sec:results_MESA}

Our proof-of-concept rapid population synthesis Milky Way populations consistently populate the lower mass gap. These simulations also reveal a dependence between $M_\mathrm{ NS, birth-max}$ and the existence of the mass gap. 
Informed by the \texttt{COSMIC} simulations, here we investigate a grid of binary \texttt{MESA} models. 
We adopt a three-dimensional grid of binaries (in compact object mass, donor mass, and orbital period space), consisting of a compact object and a hydrogen-rich main-sequence star from \cite{Fragos2022}. 
A series of \texttt{COSMIC} populations are initialized with the same initial conditions as the \texttt{MESA} grid, to facilitate comparison. 

Using the Monte Carlo sampling procedure described in Section~\ref{sec:observing}, $f_\mathrm{ low}(p_\mathrm{detect})$ is reported for the \texttt{MESA} grid and each \texttt{COSMIC} grid in Table~\ref{tab:grid_comp}, for a range of accretion disk treatments and $M_\mathrm{ NS, birth-max}$. 
For $M_\mathrm{ NS, birth-max} = 3M_{\odot}$, the \texttt{MESA} and \texttt{COSMIC} grids yield generally consistent  $f_\mathrm{ low}(p_\mathrm{detect})$. 
However, when $M_\mathrm{ NS, birth-max}$ is lowered, \texttt{MESA} predicts a consistently lower $f_\mathrm{ low}(p_\mathrm{detect})$ than \texttt{COSMIC}, and the dependence on the assumed transition to RIA is exacerbated. 
When $M_\mathrm{ NS, birth-max}\lesssim2 M_{\odot}$, $f_\mathrm{ low}(p_\mathrm{detect})$ for a LMXB population modeled with \texttt{MESA} is approximately half that of a LMXB population modeled with \texttt{COSMIC}.

In Figure~\ref{fig:grid_f_for_NS}, $f_\mathrm{ low}(p_\mathrm{detect})$ is presented as a function of $M_\mathrm{ NS, birth-max}$ for the \texttt{MESA} grid and a representative \texttt{COSMIC} grid; for a more comprehensive view, the average number of LMXB detections (separated by formation channel and BH mass) is presented as well. 
For $M_\mathrm{ NS, birth-max} \lesssim 2 M_{\odot}$, the \texttt{MESA} grid yields $f_\mathrm{ low}(p_\mathrm{detect})$ a factor of two lower than the \texttt{COSMIC} grids. 
This disparity is enhanced when either a smooth or nonexistent transition to RIA is adopted. 
The sharp RIA transition actually reduces the number of detected AIC LMXBs for the \texttt{MESA} grid. 
However, the behavior of $f_\mathrm{ low}(p_\mathrm{detect})$ is dominated by the increased detection of LMXBs with $M_\mathrm{BH}>4.5M_{\odot}$ when either a smooth or nonexistent transition to RIA is adopted for the \texttt{MESA} grid and the less dramatic increase for the \texttt{COSMIC} grid (i.e., a smooth or nonexistent transition to RIA leads to slightly more mass-gap BH detections but a far greater increase in the number of higher mass LMXB detections). 

Similar to the BH mass distributions, the orbital period distributions from \texttt{MESA} and \texttt{COSMIC} also diverge. 
In Figure~\ref{fig:grid_period}, orbital period distributions are presented for three sub-populations of LMXBs (AIC LMXBs, LMXBs with intrinsic BH in the mass gap, and LMXBs with intrinsic BH above the mass gap) from the \texttt{MESA} grid and each \texttt{COSMIC} grid. 
To highlight the influence of $p_\mathrm{detect}$, Figure~\ref{fig:grid_period} includes both the underlying orbital period distributions (i.e., only  Monte Carlo sampled) and the synthetic observed orbital period distributions (i.e., Monte Carlo sampled and detection weighted).

While the observed orbital period distributions of the \texttt{COSMIC} grids are generally consistent with \texttt{MESA}, the underlying distributions show significant differences. 
Relative to the \texttt{COSMIC} grids, the underlying orbital period distribution of AIC LMXBs is skewed towards shorter orbital periods in the \texttt{MESA} grid; without $p_\mathrm{detect}$ weighting, $80\%$ of AIC LMXBs in the \texttt{MESA} grid have $P<10$~hour, but $<35\%$ of AIC LMXBs in the \texttt{COSMIC} grids have $P<10$~hour. 
For LMXBs with intrinsic BHs and $P<10$~hour, the density of LMXBs with mass-gap BHs is lower in the \texttt{MESA} grid, relative to the \texttt{COSMIC} grids; in the \texttt{MESA} grid, $f_\mathrm{low-intrinsic}=7\%$ for LMXBs with intrinsic BHs and $P<10$~hour, while the fraction is $\sim30\%$ in the \texttt{COSMIC} grids. 
For $P>10$~hour, this trend continues, but the magnitude is lessened; in the \texttt{MESA} grid, $f_\mathrm{low-intrinsic}=15\%$ for LMXBs with intrinsic BH and $P>10$~hour, while the fraction is $\sim20\%$ for the \texttt{COSMIC} grids. 

The differences in the orbital period and mass distributions between \texttt{COSMIC} and \texttt{MESA} are both consequences of how RLO mass transfer is treated in each code. 
\texttt{COSMIC} treats RLO mass transfer following \cite{Hurley2002}, in which stars are (nonphysically) approximated as in thermal equilibrium throughout mass loss. 
For main-sequence donors in the \texttt{MESA} models, mass is removed for stars overfilling their Roche lobes at the beginning of each time-step, such that the star's radius remains within its Roche lobe, while for giant stars, the methods of \citet{Kolb1990} are adopted; stars' radii are allowed to extend beyond their Roche lobes, and mass transfer is self-consistently calculated based on the local fluid conditions \citep{Fragos2022}. 
Differences in mass-transfer rate naturally affect a LMXB's BH mass and orbital evolution. 

Within the first $100~\mathrm{Myr}$ of evolution, every compact object in the grid accretes more mass when \texttt{MESA} is used than when \texttt{COSMIC} used. 
Similarly, every donor star in the grid loses more mass when \texttt{MESA} is used than when \texttt{COSMIC} is used. 
Of LMXBs in the grid that do not merge by $100~\mathrm{Myr}$, the compact objects of LMXBs evolved using \texttt{MESA} accrete on average $0.7M_{\odot}$ more mass by $100~\mathrm{Myr}$ than compact objects in LMXBs evolved using \texttt{COSMIC}; weighting by detection probability (assuming no transition to RIA), the compact objects of LMXBs evolved using \texttt{MESA} accrete on average $0.4M_{\odot}$ more mass by $100~\mathrm{Myr}$ than the compact objects of LMXBs evolved using \texttt{COSMIC}.
The greater mass accretion by compact objects in \texttt{MESA} results in both more massive compact objects and shorter orbital periods.

To highlight these differences, the evolution tracks of a single LMXB evolved using \texttt{MESA} are presented against the evolution tracks from \texttt{COSMIC} (for a LMXB with the same initial conditions) in Figure~\ref{fig:sequences}. 
The majority of both the orbital period evolution and mass gain for the compact object occurs in the first few hundred million years in the \texttt{MESA} grid. 
During this period, the mass-loss rate from the donor star is larger in \texttt{MESA} than in \texttt{COSMIC} by at least two orders of magnitude. 
Since the donor mass loss rate is still sub-Eddington, the mass exchange is conservative for both codes. 
In contrast, the timescales for orbital period evolution and mass gain in the \texttt{COSMIC} grid extend to several billion years. 
This leads to more dramatic orbital period shrinking and more accretion onto the compact object and thus results in the overall higher-mass BHs and shorter periods in the \texttt{MESA} grid when compared to \texttt{COSMIC}.

In summary, a \texttt{COSMIC} population study of uncertain aspects of binary evolution and disk instability models consistently found a substantial fraction of LMXBs hosting BHs with $M_\mathrm{BH}<4.5M_{\odot}$ (over $40\%$) for $M_\mathrm{ NS, birth-max} = 3M_{\odot}$. 
After lowering $M_\mathrm{ NS, birth-max}$ to $1.5M_{\odot}$, $f_\mathrm{ low}(p_\mathrm{detect}) > 25\%$ for every \texttt{COSMIC} population. 
For $M_\mathrm{ NS, birth-max} = 3M_{\odot}$, \texttt{MESA} and \texttt{COSMIC} yield similar BH mass distributions and both fill the lower mass gap. 
However, if $M_\mathrm{NS, birth-max}$ is lowered to below $2M_{\odot}$, the LMXBs produced with \texttt{MESA} have at least a factor of two fewer BHs with $M_\mathrm{BH}<4.5M_{\odot}$. 
This difference is exacerbated when a smooth or nonexistent transition to RIA is adopted. 
The results highlight the importance of accurately modeling mass transfer in understanding the formation of compact objects. 

\section{Discussion}
\label{sec:discuss}

While Galactic LMXBs have previously been used to constrain the underlying stellar-mass BH distribution \citep{Bailyn1998,Ozel2010,Farr2011}, the impact of detection biases on the observed LMXB sample has remained unclear \citep{Ozel2010,Kreidberg2012,Jonker2021}. 
Here, we presented a quantitative study of selection biases, motivated by the fact that BH mass measurement is possible only for \emph{transient} XRB systems. 
We examined whether the requirement for transient behavior can hide an underlying population of low-mass BHs and create an \emph{apparent} mass gap in the observed sample.

We followed a forward-modeling approach. Using both rapid binary population synthesis (\texttt{COSMIC}) and detailed stellar evolution models (\texttt{MESA}), we generated initial synthetic binary systems, evolved the systems forward in time, and inferred the detectable populations using the disk instability model of LMXB transient behavior.
To reflect the criteria for LMXB mass inference, synthetic samples were weighted by the probability a \textit{RXTE}-like ASM \citep{Levine1996} would detect the source as transient.
From these results, we demonstrated that transient LMXB selection effects do introduce some biases into the observed sample, but this is not sufficient to completely hide mass-gap BHs. 
Additionally, observations of the lower mass gap have implications for $M_\mathrm{ NS, birth-max}$, as this impacts the number of BHs (or lack thereof) formed through AIC. 
We conclude that while selection effects are important in understanding XRB observations, the lack of LMXB BH seen in the lower mass gap places constraints upon viable models of binary stellar evolution.  

Using \texttt{COSMIC}, we generated Milky Way binary populations for an array of binary evolution prescriptions and initial conditions. 
While current rapid population synthesis methods likely cannot reproduce all properties of the observed LMXB sample, they enable exploratory studies to uncover any significant dependencies between the synthetic observed LMXB sample and the binary evolution prescriptions. 
In concert with the suite of population synthesis simulations,  we considered a grid of \texttt{MESA} binary models. 
We adopted a three-dimensional grid of binaries consisting of a compact object and a hydrogen-rich main-sequence star from \citet{Fragos2022}. 
While the \texttt{MESA} grid does not consider the full range of binary evolution prescriptions investigated with \texttt{COSMIC}, here we explored an improved physical treatment of XRB mass-transfer.

Our key results are:
\begin{enumerate}
    \item Our rapid population synthesis results cannot replicate the observed mass distribution of Galactic LMXBs.
    The \texttt{COSMIC} results indicate that detection of LMXBs containing mass-gap BHs is a robust prediction of the Delayed explosion mechanism.
    For thirty-two combinations of binary evolution prescriptions and three disk instability models (smooth, sharp or nonexistent transition to RIA) over $40\%$ of the LMXBs in the synthetic Galactic samples have $M_\mathrm{BH}<4.5M_{\odot}$ (with $M_\mathrm{ NS, birth-max}=3M_{\odot}$).
    
    \item LMXBs with a mass-gap BH form through one of two channels: AIC of a NS or directly post core-collapse of a star. 
    Unlike LMXBs with intrinsic BHs, the formation of AIC LMXBs relies upon $M_\mathrm{ NS, birth-max}$ (i.e., lowering $M_\mathrm{ NS, birth-max}$ suppresses the formation of AIC LMXBs). 
    Since $M_\mathrm{BH} \lesssim 5M_{\odot}$ for the AIC LMXBs, the existence of a lower mass gap in the Galactic LMXB sample can potentially constrain $M_\mathrm{ NS, birth-max}$. 
    
    \item While rapid population synthesis and \texttt{MESA} models are consistent for high values of the maximum NS birth mass (e.g., $M_\mathrm{ NS, birth-max}=3M_{\odot}$), LMXBs evolved with \texttt{MESA} have a lower fraction of BHs in the mass gap relative to the \texttt{COSMIC} models, if $M_\mathrm{ NS, birth-max} \lesssim 2M_{\odot}$. 
    Among the synthetic LMXB samples, the \texttt{MESA} models produce a factor of two fewer LMXBs with a mass-gap BH than the analogous \texttt{COSMIC} systems.
    This trend is exacerbated when a smooth or nonexistent transition to RIA is adopted. 
    
    \item Relative to our rapid population synthesis models, BHs in LMXBs evolved using \texttt{MESA} generally undergo greater mass accretion. 
    This results in both more massive BHs and shorter period LMXBs in the \texttt{MESA} models. 
\end{enumerate}

Drawing population-level conclusions from the \texttt{MESA} grid alone presents considerable challenges.
Unlike the \texttt{COSMIC} Milky Way models, the initial conditions for the \texttt{MESA} grid are not drawn from informed distributions, and the \texttt{MESA} grid only considers the compact object--main sequence phase of evolution. 
While we find that the \texttt{COSMIC} simulated LMXB samples host significantly more BHs in the mass gap than samples evolved using \texttt{MESA}, we cannot predict the occurrence rate of mass-gap BH for the Milky Way using the \texttt{MESA} grid.

To suppress the formation of AIC LMXBs, both the \texttt{COSMIC} Milky Way populations and the \texttt{MESA} grid favor $M_\mathrm{ NS, birth-max} \lesssim 2M_{\odot}$. This constraint on $M_\mathrm{ NS, birth-max}$ is consistent with the observed NS mass distribution. 
Masses have been inferred for $>70$ NSs via pulsar timing, eclipsing X-ray binaries, and optical radial velocity measurements of a NS's companion \citep{Alsing2018, Lattimer2019}. 
From these measurements, the NS mass distribution displays a strong peak near $1.4M_{\odot}$ with a few objects heavier than $\sim1.6M_{\odot}$ and a maximum around $\sim2.3 M_{\odot}$ \citep{Shao2020}. 
However, the most massive NSs are in binaries with low-mass companions (typically white dwarfs) and likely underwent accretion growth. 
With the exceptions of GW190814 and GW200210\_092254, which are consistent with containing a compact object lying in the mass gap \citep{abbott2020,GWTC-2.1,GWTC3}, there are few gravitational-wave sources with significant support for a mass-gap component, and observations are consistent with a dip in the NS population above $\sim2.1M_{\odot}$ \citep{GWTC-3-Pops}. 
The preference for $M_\mathrm{ NS, birth-max} \lesssim 2 M_{\odot}$ in both the \texttt{COSMIC} and \texttt{MESA} simulations is approximately consistent with these observed samples. 

Over a wide range of binary evolution prescriptions, $M_\mathrm{ NS, birth-max}$, and disk instability models, \texttt{COSMIC} Milky Way populations fill the mass gap. 
In isolation, the robustness of this result (and its contradiction with observation) would favor a supernova explosion mechanism that inherently creates a mass gap \citep[e.g., the Rapid prescription of][]{Fryer2012}. 
However, gravitational-wave observations still favor mechanisms that (at least partially) populate the gap \citep[e.g., the Delayed prescription of][]{Fryer2012}. 
Therefore, there is still physics regarding the formation of mass-gap BHs (either from AIC of a NS or core collapse) that remains to be understood. 
Given the overproduction of mass-gap BHs with \texttt{COSMIC} relative to \texttt{MESA}, greater study of LMXB formation and observably using detailed stellar evolution models \citep[e.g., \texttt{POSYDON};][]{Fragos2022} including up-to-date supernova and stellar-collapse prescriptions \citep[e.g.,][]{Ertl2020,Fryer2022} is vital to untangling the mystery of the lower mass gap. 
Additionally, the presence of observational selection biases should motivate the building of more sensitive ASMs than \textit{RXTE}, such as those proposed for missions like \textit{eXTP} and \textit{STROBE-X} \citep[][]{extp,STROBEX}, as well as the use of sufficient optical and infrared resources to make mass estimates for the fainter X-ray transients which will tend to have fainter optical counterparts.

\section{Conclusions}
\label{sec:summary}

For a supernova engine that fills the lower BH mass gap, transient LMXB selection effects do introduce significant biases into the observed LMXB sample. 
However, unless there are further (unaccounted for) observational biases against finding LMXBs with mass-gap BHs, population synthesis models fail to reproduce this aspect of the observed LMXB population. 
This result is robust against variations of uncertain aspects of binary evolution physics, e.g., CE efficiency, CE survivability, minimum ZAMS mass ratio, magnetic braking, and accretion efficiency.
This points to the need for additional physics not currently included in our \texttt{COSMIC} and \texttt{MESA} simulations, such as a supernova mechanism that suppresses formation of mass-gap objects.

Regardless of whether the low-mass BHs form from core collapse, the results of our \texttt{COSMIC} and \texttt{MESA} models lead to the robust implication that the NS birth masses must be suppressed above $\sim2M_{\odot}$. 
Otherwise the mass gap would be filled by low-mass BHs formed through NS AIC. 
This constraint on the maximum NS birth mass is independent of whether the supernova engine forms mass-gap BHs.
This result alone motivates a reexamination of the physics included in the \texttt{COSMIC} and \texttt{MESA} simulations, such as a supernova engine that limits the maximum birth mass of NS. 

\section*{Acknowledgments}
The authors thank Michael Zevin and Simon Stevenson for useful discussions, and the referee for useful suggestions. 
JS and IK were supported as CIERA REU students by the National Science Foundation (NSF) under Grant Number 1757792. Any opinions, findings, and conclusions or recommendations expressed in this material are those of the author(s) and do not necessarily reflect the views of the NSF. 
JS was also partially supported by CIFAR (through VK's Senior Fellowship). 
VK was partially supported through a CIFAR Senior Fellowship, a Guggenheim Fellowship, and the Gordon and Betty Moore Foundation (grant award GBMF8477). 
CPLB was supported by the CIERA Board of Visitors Research Professorship. 
JJA was supported by Northwestern University through a CIERA Postdoctoral Fellowship. 
KR was supported by the Gordon and Betty Moore Foundation (PI Kalogera, grant award GBMF8477) and the Riedel Family Graduate Fellowship in CIERA. 
AD, PS, MS were supported by the Gordon and Betty Moore Foundation (grant award GBMF8477). 
SBB, TF, KK, DM, and ZX were supported from the Swiss National Science Foundation Professorship Grant (PP00P2\_176868; PI Fragos). 
KK acknowledges support from the Federal Commission for Scholarships for Foreign Students for the Swiss Government Excellence Scholarship (ESKAS No.\ 2021.0277). 
ZX acknowledges support from the Chinese Scholarship Council (CSC). 
EZ acknowledges funding support from the European Research Council (ERC) under the European Union’s Horizon 2020 research and innovation program (Grant agreement No.\ 772086). 
The Flatiron Institute is funded by the Simons Foundation.
This research was supported in part through the computational resources and staff contributions provided for the Quest high performance computing facility at Northwestern University which is jointly supported by the Office of the Provost, the Office for Research, and Northwestern University Information Technology. 

The data behind Table~\ref{tab:pop_synth} is openly available from the Zenodo repository at \href{https://zenodo.org/record/8155601}{10.5281/zenodo.8155601}.

\software{\texttt{astropy} \citep{astropy:2013, astropy:2018};
          \texttt{COSMIC} \citep{Breivik2020};
          \texttt{matplotlib}\ \citep{matplotlib}; 
          \texttt{MESA}\ \citep{Paxton2011, Paxton2013, Paxton2015, Paxton2018, Paxton2019};
          \texttt{numpy}\ \citep{numpy}; 
          \texttt{pandas}\ \citep{mckinney-proc-scipy-2010, reback2020pandas}; 
          \texttt{POSYDON} \citep{Fragos2022};
          \texttt{scipy}\ \citep{2020NatMe..17..261V}
          }
          
\bibliography{citations}

\begin{thebibliography}{}
\expandafter\ifx\csname natexlab\endcsname\relax\def\natexlab#1{#1}\fi
\providecommand{\url}[1]{\href{#1}{#1}}

\bibitem[{{Abbott} {et~al.}(2021{\natexlab{a}}){Abbott}, {Abbott}, {Acernese},
  {Ackley}, {Adams}, {Adhikari}, {Adhikari}, \& et~al.}]{GWTC-3-Pops}
{Abbott}, R., {Abbott}, T.~D., {Acernese}, F., {et~al.} 2021{\natexlab{a}},
  arXiv e-prints, arXiv:2111.03634

\bibitem[{{Abbott} {et~al.}(2020){Abbott}, {Abbott}, {Abraham}, {Acernese},
  {Ackley}, {Adams}, {Adhikari}, {Adya}, {Affeldt}, {Agathos}, {Agatsuma},
  {Aggarwal}, {Aguiar}, {Aich}, {Aiello}, {Ain}, {Ajith}, {Akcay}, {Allen},
  {Allocca}, {Altin}, {Amato}, {Anand}, {Ananyeva}, {Anderson}, {Anderson},
  {Angelova}, {Ansoldi}, {Antier}, {Appert}, {Arai}, {Araya}, {Areeda},
  {Ar{\`e}ne}, {Arnaud}, {Aronson}, {Arun}, {Asali}, {Ascenzi}, {Ashton},
  {Aston}, {Astone}, {Aubin}, {Aufmuth}, {AultONeal}, {Austin}, {Avendano},
  {Babak}, {Bacon}, {Badaracco}, {Bader}, {Bae}, {Baer}, {Baird}, {Baldaccini},
  {Ballardin}, {Ballmer}, {Bals}, {Balsamo}, {Baltus}, {Banagiri}, {Bankar},
  {Bankar}, {Barayoga}, {Barbieri}, {Barish}, {Barker}, {Barkett}, {Barneo},
  {Barone}, {Barr}, {Barsotti}, {Barsuglia}, {Barta}, {Bartlett}, {Bartos},
  {Bassiri}, {Basti}, {Bawaj}, {Bayley}, {Bazzan}, {B{\'e}csy}, {Bejger},
  {Belahcene}, {Bell}, {Beniwal}, {Benjamin}, {Benkel}, {Bentley}, {Bergamin},
  {Berger}, {Bergmann}, {Bernuzzi}, {Berry}, {Bersanetti}, {Bertolini},
  {Betzwieser}, {Bhandare}, {Bhandari}, {Bidler}, {Biggs}, {Bilenko},
  {Billingsley}, {Birney}, {Birnholtz}, {Biscans}, {Bischi}, {Biscoveanu},
  {Bisht}, {Bissenbayeva}, {Bitossi}, {Bizouard}, {Blackburn}, {Blackman},
  {Blair}, {Blair}, {Blair}, {Bobba}, {Bode}, {Boer}, {Boetzel}, {Bogaert},
  {Bondu}, {Bonilla}, {Bonnand}, {Booker}, {Boom}, {Bork}, {Boschi}, {Bose},
  {Bossilkov}, {Bosveld}, {Bouffanais}, {Bozzi}, {Bradaschia}, {Brady},
  {Bramley}, {Branchesi}, {Brau}, {Breschi}, {Briant}, {Briggs}, {Brighenti},
  {Brillet}, {Brinkmann}, {Brito}, {Brockill}, {Brooks}, {Brooks}, {Brown},
  {Brunett}, {Bruno}, {Bruntz}, {Buikema}, {Bulik}, {Bulten}, {Buonanno},
  {Buskulic}, {Byer}, {Cabero}, {Cadonati}, {Cagnoli}, {Cahillane}, {Bustillo},
  {Callaghan}, {Callister}, {Calloni}, {Camp}, {Canepa}, {Cannon}, {Cao},
  {Cao}, {Carapella}, {Carbognani}, {Caride}, {Carney}, {Carullo}, {Diaz},
  {Casentini}, {Casta{\~n}eda}, {Caudill}, {Cavagli{\`a}}, {Cavalier},
  {Cavalieri}, {Cella}, {Cerd{\'a}-Dur{\'a}n}, {Cesarini}, {Chaibi},
  {Chakravarti}, {Chan}, {Chan}, {Chao}, {Charlton}, {Chase},
  {Chassande-Mottin}, {Chatterjee}, {Chaturvedi}, {Chatziioannou}, {Chen},
  {Chen}, {Chen}, {Cheng}, {Cheong}, {Chia}, {Chiadini}, {Chierici},
  {Chincarini}, {Chiummo}, {Cho}, {Cho}, {Cho}, {Christensen}, {Chu}, {Chua},
  {Chung}, {Chung}, {Ciani}, {Ciecielag}, {Cie{\'s}lar}, {Ciobanu}, {Ciolfi},
  {Cipriano}, {Cirone}, {Clara}, {Clark}, {Clearwater}, {Clesse}, {Cleva},
  {Coccia}, {Cohadon}, {Cohen}, {Colleoni}, {Collette}, {Collins}, {Colpi},
  {Constancio}, {Conti}, {Cooper}, {Corban}, {Corbitt}, {Cordero-Carri{\'o}n},
  {Corezzi}, {Corley}, {Cornish}, {Corre}, {Corsi}, {Cortese}, {Costa},
  {Cotesta}, {Coughlin}, {Coughlin}, {Coulon}, {Countryman}, {Couvares},
  {Covas}, {Coward}, {Cowart}, {Coyne}, {Coyne}, {Creighton}, {Creighton},
  {Cripe}, {Croquette}, {Crowder}, {Cudell}, {Cullen}, {Cumming}, {Cummings},
  {Cunningham}, {Cuoco}, {Curylo}, {Canton}, {D{\'a}lya}, {Dana},
  {Daneshgaran-Bajastani}, {D'Angelo}, {Danilishin}, {D'Antonio}, {Danzmann},
  {Darsow-Fromm}, {Dasgupta}, {Datrier}, {Dattilo}, {Dave}, {Davier}, {Davies},
  {Davis}, {Daw}, {DeBra}, {Deenadayalan}, {Degallaix}, {De Laurentis},
  {Del{\'e}glise}, {Delfavero}, {De Lillo}, {Del Pozzo}, {DeMarchi},
  {D'Emilio}, {Demos}, {Dent}, {De Pietri}, {De Rosa}, {De Rossi}, {DeSalvo},
  {de Varona}, {Dhurandhar}, {D{\'\i}az}, {Diaz-Ortiz}, {Dietrich}, {Di Fiore},
  {Di Fronzo}, {Di Giorgio}, {Di Giovanni}, {Di Giovanni}, {Di Girolamo}, {Di
  Lieto}, {Ding}, {Di Pace}, {Di Palma}, {Di Renzo}, {Divakarla}, {Dmitriev},
  {Doctor}, {Donovan}, {Dooley}, {Doravari}, {Dorrington}, {Downes}, {Drago},
  {Driggers}, {Du}, {Ducoin}, {Dupej}, {Durante}, {D'Urso}, {Dwyer}, {Easter},
  {Eddolls}, {Edelman}, {Edo}, {Edy}, {Effler}, {Ehrens}, {Eichholz},
  {Eikenberry}, {Eisenmann}, {Eisenstein}, {Ejlli}, {Errico}, {Essick},
  {Estelles}, {Estevez}, {Etienne}, {Etzel}, {Evans}, {Evans}, {Ewing},
  {Fafone}, {Fairhurst}, {Fan}, {Farinon}, {Farr}, {Farr}, {Fauchon-Jones},
  {Favata}, {Fays}, {Fazio}, {Feicht}, {Fejer}, {Feng}, {Fenyvesi}, {Ferguson},
  {Fernandez-Galiana}, {Ferrante}, {Ferreira}, {Ferreira}, {Fidecaro}, {Fiori},
  {Fiorucci}, {Fishbach}, {Fisher}, {Fittipaldi}, {Fitz-Axen}, {Fiumara},
  {Flaminio}, {Floden}, {Flynn}, {Fong}, {Font}, {Forsyth}, {Fournier},
  {Frasca}, {Frasconi}, {Frei}, {Freise}, {Frey}, {Frey}, {Fritschel},
  {Frolov}, {Fronz{\`e}}, {Fulda}, {Fyffe}, {Gabbard}, {Gadre}, {Gaebel},
  {Gair}, {Galaudage}, {Ganapathy}, {Ganguly}, {Gaonkar},
  {Garc{\'\i}a-Quir{\'o}s}, {Garufi}, {Gateley}, {Gaudio}, {Gayathri}, {Gemme},
  {Genin}, {Gennai}, {George}, {George}, {Gergely}, {Ghonge}, {Ghosh}, {Ghosh},
  {Ghosh}, {Giacomazzo}, {Giaime}, {Giardina}, {Gibson}, {Gier}, {Gill},
  {Glanzer}, {Gniesmer}, {Godwin}, {Goetz}, {Goetz}, {Gohlke}, {Goncharov},
  {Gonz{\'a}lez}, {Gopakumar}, {Gossan}, {Gosselin}, {Gouaty}, {Grace},
  {Grado}, {Granata}, {Grant}, {Gras}, {Grassia}, {Gray}, {Gray}, {Greco},
  {Green}, {Green}, {Gretarsson}, {Griggs}, {Grignani}, {Grimaldi}, {Grimm},
  {Grote}, {Grunewald}, {Gruning}, {Guidi}, {Guimaraes}, {Guix{\'e}}, {Gulati},
  {Guo}, {Gupta}, {Gupta}, {Gupta}, {Gustafson}, {Gustafson}, {Haegel},
  {Halim}, {Hall}, {Hamilton}, {Hammond}, {Haney}, {Hanke}, {Hanks}, {Hanna},
  {Hannam}, {Hannuksela}, {Hansen}, {Hanson}, {Harder}, {Hardwick}, {Haris},
  {Harms}, {Harry}, {Harry}, {Hasskew}, {Haster}, {Haughian}, {Hayes}, {Healy},
  {Heidmann}, {Heintze}, {Heinze}, {Heitmann}, {Hellman}, {Hello}, {Hemming},
  {Hendry}, {Heng}, {Hennes}, {Hennig}, {Heurs}, {Hild}, {Hinderer}, {Hoback},
  {Hochheim}, {Hofgard}, {Hofman}, {Holgado}, {Holland}, {Holt}, {Holz},
  {Hopkins}, {Horst}, {Hough}, {Howell}, {Hoy}, {Huang}, {H{\"u}bner},
  {Huerta}, {Huet}, {Hughey}, {Hui}, {Husa}, {Huttner}, {Huxford},
  {Huynh-Dinh}, {Idzkowski}, {Iess}, {Inchauspe}, {Ingram}, {Intini}, {Isac},
  {Isi}, {Iyer}, {Jacqmin}, {Jadhav}, {Jadhav}, {James}, {Jani}, {Janthalur},
  {Jaranowski}, {Jariwala}, {Jaume}, {Jenkins}, {Jiang}, {Johns},
  {Johnson-McDaniel}, {Jones}, {Jones}, {Jones}, {Jones}, {Jones}, {Jonker},
  {Ju}, {Junker}, {Kalaghatgi}, {Kalogera}, {Kamai}, {Kandhasamy}, {Kang},
  {Kanner}, {Kapadia}, {Karki}, {Kashyap}, {Kasprzack}, {Kastaun},
  {Katsanevas}, {Katsavounidis}, {Katzman}, {Kaufer}, {Kawabe},
  {K{\'e}f{\'e}lian}, {Keitel}, {Keivani}, {Kennedy}, {Key}, {Khadka},
  {Khalili}, {Khan}, {Khan}, {Khan}, {Khazanov}, {Khetan}, {Khursheed},
  {Kijbunchoo}, {Kim}, {Kim}, {Kim}, {Kim}, {Kim}, {Kim}, {Kim}, {Kimball},
  {King}, {Kinley-Hanlon}, {Kirchhoff}, {Kissel}, {Kleybolte}, {Klimenko},
  {Knowles}, {Knyazev}, {Koch}, {Koehlenbeck}, {Koekoek}, {Koley},
  {Kondrashov}, {Kontos}, {Koper}, {Korobko}, {Korth}, {Kovalam}, {Kozak},
  {Kringel}, {Krishnendu}, {Kr{\'o}lak}, {Krupinski}, {Kuehn}, {Kumar},
  {Kumar}, {Kumar}, {Kumar}, {Kumar}, {Kuo}, {Kutynia}, {Lackey}, {Laghi},
  {Lalande}, {Lam}, {Lamberts}, {Landry}, {Landry}, {Lane}, {Lang}, {Lange},
  {Lantz}, {Lanza}, {La Rosa}, {Lartaux-Vollard}, {Lasky}, {Laxen},
  {Lazzarini}, {Lazzaro}, {Leaci}, {Leavey}, {Lecoeuche}, {Lee}, {Lee}, {Lee},
  {Lee}, {Lee}, {Lehmann}, {Leroy}, {Letendre}, {Levin}, {Li}, {Li}, {li},
  {Li}, {Li}, {Linde}, {Linker}, {Linley}, {Littenberg}, {Liu}, {Liu},
  {Llorens-Monteagudo}, {Lo}, {Lockwood}, {London}, {Longo}, {Lorenzini},
  {Loriette}, {Lormand}, {Losurdo}, {Lough}, {Lousto}, {Lovelace}, {L{\"u}ck},
  {Lumaca}, {Lundgren}, {Ma}, {Macas}, {Macfoy}, {MacInnis}, {Macleod},
  {MacMillan}, {Macquet}, {Hernandez}, {Maga{\~n}a-Sandoval}, {Magee},
  {Majorana}, {Maksimovic}, {Malik}, {Man}, {Mandic}, {Mangano}, {Mansell},
  {Manske}, {Mantovani}, {Mapelli}, {Marchesoni}, {Marion}, {M{\'a}rka},
  {M{\'a}rka}, {Markakis}, {Markosyan}, {Markowitz}, {Maros}, {Marquina},
  {Marsat}, {Martelli}, {Martin}, {Martin}, {Martinez}, {Martynov},
  {Masalehdan}, {Mason}, {Massera}, {Masserot}, {Massinger}, {Masso-Reid},
  {Mastrogiovanni}, {Matas}, {Matichard}, {Mavalvala}, {Maynard}, {McCann},
  {McCarthy}, {McClelland}, {McCormick}, {McCuller}, {McGuire}, {McIsaac},
  {McIver}, {McManus}, {McRae}, {McWilliams}, {Meacher}, {Meadors}, {Mehmet},
  {Mehta}, {Villa}, {Melatos}, {Mendell}, {Mercer}, {Mereni}, {Merfeld},
  {Merilh}, {Merritt}, {Merzougui}, {Meshkov}, {Messenger}, {Messick},
  {Metzdorff}, {Meyers}, {Meylahn}, {Mhaske}, {Miani}, {Miao}, {Michaloliakos},
  {Michel}, {Middleton}, {Milano}, {Miller}, {Millhouse}, {Mills}, {Milotti},
  {Milovich-Goff}, {Minazzoli}, {Minenkov}, {Mishkin}, {Mishra}, {Mistry},
  {Mitra}, {Mitrofanov}, {Mitselmakher}, {Mittleman}, {Mo}, {Mogushi},
  {Mohapatra}, {Mohite}, {Molina-Ruiz}, {Mondin}, {Montani}, {Moore}, {Moraru},
  {Morawski}, {Moreno}, {Morisaki}, {Mours}, {Mow-Lowry}, {Mozzon},
  {Muciaccia}, {Mukherjee}, {Mukherjee}, {Mukherjee}, {Mukherjee}, {Mukund},
  {Mullavey}, {Munch}, {Mu{\~n}iz}, {Murray}, {Nagar}, {Nardecchia},
  {Naticchioni}, {Nayak}, {Neil}, {Neilson}, {Nelemans}, {Nelson}, {Nery},
  {Neunzert}, {Ng}, {Ng}, {Nguyen}, {Nguyen}, {Nichols}, {Nichols}, {Nissanke},
  {Nocera}, {Noh}, {North}, {Nothard}, {Nuttall}, {Oberling}, {O'Brien},
  {Oganesyan}, {Ogin}, {Oh}, {Oh}, {Ohme}, {Ohta}, {Okada}, {Oliver},
  {Olivetto}, {Oppermann}, {Oram}, {O'Reilly}, {Ormiston}, {Ortega},
  {O'Shaughnessy}, {Ossokine}, {Osthelder}, {Ottaway}, {Overmier}, {Owen},
  {Pace}, {Pagano}, {Page}, {Pagliaroli}, {Pai}, {Pai}, {Palamos}, {Palashov},
  {Palomba}, {Pan}, {Panda}, {Pang}, {Pankow}, {Pannarale}, {Pant}, {Paoletti},
  {Paoli}, {Parida}, {Parker}, {Pascucci}, {Pasqualetti}, {Passaquieti},
  {Passuello}, {Patricelli}, {Payne}, {Pearlstone}, {Pechsiri}, {Pedersen},
  {Pedraza}, {Pele}, {Penn}, {Perego}, {Perez}, {P{\'e}rigois}, {Perreca},
  {Perri{\`e}s}, {Petermann}, {Pfeiffer}, {Phelps}, {Phukon}, {Piccinni},
  {Pichot}, {Piendibene}, {Piergiovanni}, {Pierro}, {Pillant}, {Pinard},
  {Pinto}, {Piotrzkowski}, {Pirello}, {Pitkin}, {Plastino}, {Poggiani}, {Pong},
  {Ponrathnam}, {Popolizio}, {Porter}, {Powell}, {Prajapati}, {Prasai},
  {Prasanna}, {Pratten}, {Prestegard}, {Principe}, {Prodi}, {Prokhorov},
  {Punturo}, {Puppo}, {P{\"u}rrer}, {Qi}, {Quetschke}, {Quinonez}, {Raab},
  {Raaijmakers}, {Radkins}, {Radulesco}, {Raffai}, {Rafferty}, {Raja}, {Rajan},
  {Rajbhandari}, {Rakhmanov}, {Ramirez}, {Ramos-Buades}, {Rana}, {Rao},
  {Rapagnani}, {Raymond}, {Razzano}, {Read}, {Regimbau}, {Rei}, {Reid},
  {Reitze}, {Rettegno}, {Ricci}, {Richardson}, {Richardson}, {Ricker},
  {Riemenschneider}, {Riles}, {Rizzo}, {Robertson}, {Robinet}, {Rocchi},
  {Rodriguez-Soto}, {Rolland}, {Rollins}, {Roma}, {Romanelli}, {Romano},
  {Romel}, {Romero-Shaw}, {Romie}, {Rose}, {Rose}, {Rose}, {Rosi{\'n}ska},
  {Rosofsky}, {Ross}, {Rowan}, {Rowlinson}, {Roy}, {Roy}, {Roy}, {Ruggi},
  {Rutins}, {Ryan}, {Sachdev}, {Sadecki}, {Sakellariadou}, {Salafia},
  {Salconi}, {Saleem}, {Salemi}, {Samajdar}, {Sanchez}, {Sanchez},
  {Sanchis-Gual}, {Sanders}, {Santiago}, {Santos}, {Sarin}, {Sassolas},
  {Sathyaprakash}, {Sauter}, {Savage}, {Savant}, {Sawant}, {Sayah}, {Schaetzl},
  {Schale}, {Scheel}, {Scheuer}, {Schmidt}, {Schnabel}, {Schofield},
  {Sch{\"o}nbeck}, {Schreiber}, {Schulte}, {Schutz}, {Schwarm}, {Schwartz},
  {Scott}, {Scott}, {Seidel}, {Sellers}, {Sengupta}, {Sennett}, {Sentenac},
  {Sequino}, {Sergeev}, {Setyawati}, {Shaddock}, {Shaffer}, {Shahriar},
  {Sharma}, {Sharma}, {Shawhan}, {Shen}, {Shikauchi}, {Shink}, {Shoemaker},
  {Shoemaker}, {Shukla}, {ShyamSundar}, {Siellez}, {Sieniawska}, {Sigg},
  {Singer}, {Singh}, {Singh}, {Singha}, {Singhal}, {Sintes}, {Sipala},
  {Skliris}, {Slagmolen}, {Slaven-Blair}, {Smetana}, {Smith}, {Smith},
  {Somala}, {Son}, {Soni}, {Sorazu}, {Sordini}, {Sorrentino}, {Souradeep},
  {Sowell}, {Spencer}, {Spera}, {Srivastava}, {Srivastava}, {Staats},
  {Stachie}, {Standke}, {Steer}, {Steinhoff}, {Steinke}, {Steinlechner},
  {Steinlechner}, {Steinmeyer}, {Stevenson}, {Stocks}, {Stops}, {Stover},
  {Strain}, {Stratta}, {Strunk}, {Sturani}, {Stuver}, {Sudhagar}, {Sudhir},
  {Summerscales}, {Sun}, {Sunil}, {Sur}, {Suresh}, {Sutton}, {Swinkels},
  {Szczepa{\'n}czyk}, {Tacca}, {Tait}, {Talbot}, {Tanasijczuk}, {Tanner},
  {Tao}, {T{\'a}pai}, {Tapia}, {San Martin}, {Tasson}, {Taylor}, {Tenorio},
  {Terkowski}, {Thirugnanasambandam}, {Thomas}, {Thomas}, {Thompson},
  {Thondapu}, {Thorne}, {Thrane}, {Tinsman}, {Saravanan}, {Tiwari}, {Tiwari},
  {Tiwari}, {Toland}, {Tonelli}, {Tornasi}, {Torres-Forn{\'e}}, {Torrie},
  {Tosta e Melo}, {T{\"o}yr{\"a}}, {Trail}, {Travasso}, {Traylor}, {Tringali},
  {Tripathee}, {Trovato}, {Trudeau}, {Tsang}, {Tse}, {Tso}, {Tsukada}, {Tsuna},
  {Tsutsui}, {Turconi}, {Ubhi}, {Ueno}, {Ugolini}, {Unnikrishnan}, {Urban},
  {Usman}, {Utina}, {Vahlbruch}, {Vajente}, {Valdes}, {Valentini}, {van Bakel},
  {van Beuzekom}, {van den Brand}, {Van Den Broeck}, {Vander-Hyde}, {van der
  Schaaf}, {Van Heijningen}, {van Veggel}, {Vardaro}, {Varma}, {Vass},
  {Vas{\'u}th}, {Vecchio}, {Vedovato}, {Veitch}, {Veitch}, {Venkateswara},
  {Venugopalan}, {Verkindt}, {Veske}, {Vetrano}, {Vicer{\'e}}, {Viets},
  {Vinciguerra}, {Vine}, {Vinet}, {Vitale}, {Vivanco}, {Vo}, {Vocca},
  {Vorvick}, {Vyatchanin}, {Wade}, {Wade}, {Wade}, {Walet}, {Walker},
  {Wallace}, {Wallace}, {Walsh}, {Wang}, {Wang}, {Wang}, {Ward}, {Warden},
  {Warner}, {Was}, {Watchi}, {Weaver}, {Wei}, {Weinert}, {Weinstein}, {Weiss},
  {Wellmann}, {Wen}, {We{\ss}els}, {Westhouse}, {Wette}, {Whelan}, {Whiting},
  {Whittle}, {Wilken}, {Williams}, {Willis}, {Willke}, {Winkler}, {Wipf},
  {Wittel}, {Woan}, {Woehler}, {Wofford}, {Wong}, {Wright}, {Wu}, {Wysocki},
  {Xiao}, {Yamamoto}, {Yang}, {Yang}, {Yang}, {Yap}, {Yazback}, {Yeeles}, {Yu},
  {Yu}, {Yuen}, {Zadro{\.z}ny}, {Zadro{\.z}ny}, {Zanolin}, {Zelenova},
  {Zendri}, {Zevin}, {Zhang}, {Zhang}, {Zhang}, {Zhao}, {Zhao}, {Zhou}, {Zhou},
  {Zhu}, {Zimmerman}, {Zucker}, {Zweizig}, {LIGO Scientific Collaboration}, \&
  {Virgo Collaboration}}]{abbott2020}
{Abbott}, R., {Abbott}, T.~D., {Abraham}, S., {et~al.} 2020, \apjl, 896, L44

\bibitem[{{Abbott} {et~al.}(2021{\natexlab{b}}){Abbott}, {Abbott}, {Acernese},
  {Ackley}, {Adams}, {Adhikari}, {Adhikari}, {Adya}, {Affeldt}, {Agarwal},
  {Agathos}, {Agatsuma}, {Aggarwal}, {Aguiar}, {Aiello}, {Ain}, {Ajith},
  {Akcay}, {Akutsu}, {Albanesi}, {Allocca}, {Altin}, {Amato}, {Anand}, {Anand},
  {Ananyeva}, {Anderson}, {Anderson}, {Ando}, {Andrade}, {Andres},
  {Andri{\'c}}, {Angelova}, {Ansoldi}, {Antelis}, {Antier}, {Appert}, {Arai},
  {Arai}, {Arai}, {Araki}, {Araya}, {Araya}, {Areeda}, {Ar{\`e}ne}, {Aritomi},
  {Arnaud}, {Arogeti}, {Aronson}, {Arun}, {Asada}, {Asali}, {Ashton}, {Aso},
  {Assiduo}, {Aston}, {Astone}, {Aubin}, {Austin}, {Babak}, {Badaracco},
  {Bader}, {Badger}, {Bae}, {Bae}, {Baer}, {Bagnasco}, {Bai}, {Baiotti},
  {Baird}, {Bajpai}, {Ball}, {Ballardin}, {Ballmer}, {Balsamo}, {Baltus},
  {Banagiri}, {Bankar}, {Barayoga}, {Barbieri}, {Barish}, {Barker}, {Barneo},
  {Barone}, {Barr}, {Barsotti}, {Barsuglia}, {Barta}, {Bartlett}, {Barton},
  {Bartos}, {Bassiri}, {Basti}, {Bawaj}, {Bayley}, {Baylor}, {Bazzan},
  {B{\'e}csy}, {Bedakihale}, {Bejger}, {Belahcene}, {Benedetto}, {Beniwal},
  {Bennett}, {Bentley}, {BenYaala}, {Bergamin}, {Berger}, {Bernuzzi}, {Berry},
  {Bersanetti}, {Bertolini}, {Betzwieser}, {Beveridge}, {Bhandare}, {Bhardwaj},
  {Bhattacharjee}, {Bhaumik}, {Bilenko}, {Billingsley}, {Bini}, {Birney},
  {Birnholtz}, {Biscans}, {Bischi}, {Biscoveanu}, {Bisht}, {Biswas}, {Bitossi},
  {Bizouard}, {Blackburn}, {Blair}, {Blair}, {Blair}, {Bobba}, {Bode}, {Boer},
  {Bogaert}, {Boldrini}, {Bonavena}, {Bondu}, {Bonilla}, {Bonnand}, {Booker},
  {Boom}, {Bork}, {Boschi}, {Bose}, {Bose}, {Bossilkov}, {Boudart},
  {Bouffanais}, {Bozzi}, {Bradaschia}, {Brady}, {Bramley}, {Branch},
  {Branchesi}, {Brandt}, {Brau}, {Breschi}, {Briant}, {Briggs}, {Brillet},
  {Brinkmann}, {Brockill}, {Brooks}, {Brooks}, {Brown}, {Brunett}, {Bruno},
  {Bruntz}, {Bryant}, {Bulik}, {Bulten}, {Buonanno}, {Buscicchio}, {Buskulic},
  {Buy}, {Byer}, {Cabourn Davies}, {Cadonati}, {Cagnoli}, {Cahillane},
  {Calder{\'o}n Bustillo}, {Callaghan}, {Callister}, {Calloni}, {Cameron},
  {Camp}, {Canepa}, {Canevarolo}, {Cannavacciuolo}, {Cannon}, {Cao}, {Cao},
  {Capocasa}, {Capote}, {Carapella}, {Carbognani}, {Carlin}, {Carney},
  {Carpinelli}, {Carrillo}, {Carullo}, {Carver}, {Casanueva Diaz}, {Casentini},
  {Castaldi}, {Caudill}, {Cavagli{\`a}}, {Cavalier}, {Cavalieri}, {Ceasar},
  {Cella}, {Cerd{\'a}-Dur{\'a}n}, {Cesarini}, {Chaibi}, {Chakravarti},
  {Chalathadka Subrahmanya}, {Champion}, {Chan}, {Chan}, {Chan}, {Chan},
  {Chan}, {Chandra}, {Chanial}, {Chao}, {Chapman-Bird}, {Charlton}, {Chase},
  {Chassande-Mottin}, {Chatterjee}, {Chatterjee}, {Chatterjee}, {Chaturvedi},
  {Chaty}, {Chatziioannou}, {Chen}, {Chen}, {Chen}, {Chen}, {Chen}, {Chen},
  {Chen}, {Chen}, {Cheng}, {Cheong}, {Cheung}, {Chia}, {Chiadini}, {Chiang},
  {Chiarini}, {Chierici}, {Chincarini}, {Chiofalo}, {Chiummo}, {Cho}, {Cho},
  {Choudhary}, {Choudhary}, {Christensen}, {Chu}, {Chu}, {Chu}, {Chua},
  {Chung}, {Ciani}, {Ciecielag}, {Cie{\'s}lar}, {Cifaldi}, {Ciobanu}, {Ciolfi},
  {Cipriano}, {Cirone}, {Clara}, {Clark}, {Clark}, {Clarke}, {Clearwater},
  {Clesse}, {Cleva}, {Coccia}, {Codazzo}, {Cohadon}, {Cohen}, {Cohen},
  {Colleoni}, {Collette}, {Colombo}, {Colpi}, {Compton}, {Constancio}, {Conti},
  {Cooper}, {Corban}, {Corbitt}, {Cordero-Carri{\'o}n}, {Corezzi}, {Corley},
  {Cornish}, {Corre}, {Corsi}, {Cortese}, {Costa}, {Cotesta}, {Coughlin},
  {Coulon}, {Countryman}, {Cousins}, {Couvares}, {Coward}, {Cowart}, {Coyne},
  {Coyne}, {Creighton}, {Creighton}, {Criswell}, {Croquette}, {Crowder},
  {Cudell}, {Cullen}, {Cumming}, {Cummings}, {Cunningham}, {Cuoco},
  {Cury{\l}o}, {Dabadie}, {Dal Canton}, {Dall'Osso}, {D{\'a}lya}, {Dana},
  {DaneshgaranBajastani}, {D'Angelo}, {Danila}, {Danilishin}, {D'Antonio},
  {Danzmann}, {Darsow-Fromm}, {Dasgupta}, {Datrier}, {Datta}, {Dattilo},
  {Dave}, {Davier}, {Davis}, {Davis}, {Daw}, {de Alarc{\'o}n}, {Dean}, {DeBra},
  {Deenadayalan}, {Degallaix}, {De Laurentis}, {Del{\'e}glise}, {Del Favero},
  {De Lillo}, {De Lillo}, {Del Pozzo}, {DeMarchi}, {De Matteis}, {D'Emilio},
  {Demos}, {Dent}, {Depasse}, {De Pietri}, {De Rosa}, {De Rossi}, {DeSalvo},
  {De Simone}, {Dhurandhar}, {D{\'\i}az}, {Diaz-Ortiz}, {Didio}, {Dietrich},
  {Di Fiore}, {Di Fronzo}, {Di Giorgio}, {Di Giovanni}, {Di Giovanni}, {Di
  Girolamo}, {Di Lieto}, {Ding}, {Di Pace}, {Di Palma}, {Di Renzo},
  {Divakarla}, {Dmitriev}, {Doctor}, {D'Onofrio}, {Donovan}, {Dooley},
  {Doravari}, {Dorrington}, {Drago}, {Driggers}, {Drori}, {Ducoin}, {Dupej},
  {Durante}, {D'Urso}, {Duverne}, {Dwyer}, {Eassa}, {Easter}, {Ebersold},
  {Eckhardt}, {Eddolls}, {Edelman}, {Edo}, {Edy}, {Effler}, {Eguchi},
  {Eichholz}, {Eikenberry}, {Eisenmann}, {Eisenstein}, {Ejlli}, {Engelby},
  {Enomoto}, {Errico}, {Essick}, {Estell{\'e}s}, {Estevez}, {Etienne}, {Etzel},
  {Evans}, {Evans}, {Ewing}, {Fafone}, {Fair}, {Fairhurst}, {Farah}, {Farinon},
  {Farr}, {Farr}, {Farrow}, {Fauchon-Jones}, {Favaro}, {Favata}, {Fays},
  {Fazio}, {Feicht}, {Fejer}, {Fenyvesi}, {Ferguson}, {Fernandez-Galiana},
  {Ferrante}, {Ferreira}, {Fidecaro}, {Figura}, {Fiori}, {Fishbach}, {Fisher},
  {Fittipaldi}, {Fiumara}, {Flaminio}, {Floden}, {Fong}, {Font}, {Fornal},
  {Forsyth}, {Franke}, {Frasca}, {Frasconi}, {Frederick}, {Freed}, {Frei},
  {Freise}, {Frey}, {Fritschel}, {Frolov}, {Fronz{\'e}}, {Fujii}, {Fujikawa},
  {Fukunaga}, {Fukushima}, {Fulda}, {Fyffe}, {Gabbard}, {Gabella}, {Gadre},
  {Gair}, {Gais}, {Galaudage}, {Gamba}, {Ganapathy}, {Ganguly}, {Gao},
  {Gaonkar}, {Garaventa}, {Garc{\'\i}a}, {Garc{\'\i}a-N{\'u}{\~n}ez},
  {Garc{\'\i}a-Quir{\'o}s}, {Garufi}, {Gateley}, {Gaudio}, {Gayathri}, {Ge},
  {Gemme}, {Gennai}, {George}, {George}, {Gerberding}, {Gergely}, {Gewecke},
  {Ghonge}, {Ghosh}, {Ghosh}, {Ghosh}, {Ghosh}, {Giacomazzo}, {Giacoppo},
  {Giaime}, {Giardina}, {Gibson}, {Gier}, {Giesler}, {Giri}, {Gissi},
  {Glanzer}, {Gleckl}, {Godwin}, {Goetz}, {Goetz}, {Gohlke}, {Golomb},
  {Goncharov}, {Gonz{\'a}lez}, {Gopakumar}, {Gosselin}, {Gouaty}, {Gould},
  {Grace}, {Grado}, {Granata}, {Granata}, {Grant}, {Gras}, {Grassia}, {Gray},
  {Gray}, {Greco}, {Green}, {Green}, {Gretarsson}, {Gretarsson}, {Griffith},
  {Griffiths}, {Griggs}, {Grignani}, {Grimaldi}, {Grimm}, {Grote}, {Grunewald},
  {Gruning}, {Guerra}, {Guidi}, {Guimaraes}, {Guix{\'e}}, {Gulati}, {Guo},
  {Guo}, {Gupta}, {Gupta}, {Gupta}, {Gustafson}, {Gustafson}, {Guzman}, {Ha},
  {Haegel}, {Hagiwara}, {Haino}, {Halim}, {Hall}, {Hamilton}, {Hammond}, {Han},
  {Haney}, {Hanks}, {Hanna}, {Hannam}, {Hannuksela}, {Hansen}, {Hansen},
  {Hanson}, {Harder}, {Hardwick}, {Haris}, {Harms}, {Harry}, {Harry},
  {Hartwig}, {Hasegawa}, {Haskell}, {Hasskew}, {Haster}, {Hattori}, {Haughian},
  {Hayakawa}, {Hayama}, {Hayes}, {Healy}, {Heidmann}, {Heidt}, {Heintze},
  {Heinze}, {Heinzel}, {Heitmann}, {Hellman}, {Hello}, {Helmling-Cornell},
  {Hemming}, {Hendry}, {Heng}, {Hennes}, {Hennig}, {Hennig}, {Hernandez},
  {Hernandez Vivanco}, {Heurs}, {Hild}, {Hill}, {Himemoto}, {Hines},
  {Hiranuma}, {Hirata}, {Hirose}, {Hochheim}, {Hofman}, {Hohmann}, {Holcomb},
  {Holland}, {Holley-Bockelmann}, {Hollows}, {Holmes}, {Holt}, {Holz}, {Hong},
  {Hopkins}, {Hough}, {Hourihane}, {Howell}, {Hoy}, {Hoyland}, {Hreibi},
  {Hsieh}, {Hsu}, {Huang}, {Huang}, {Huang}, {Huang}, {Huang}, {Huang},
  {H{\"u}bner}, {Huddart}, {Hughey}, {Hui}, {Hui}, {Husa}, {Huttner},
  {Huxford}, {Huynh-Dinh}, {Ide}, {Idzkowski}, {Iess}, {Ikenoue}, {Imam},
  {Inayoshi}, {Ingram}, {Inoue}, {Ioka}, {Isi}, {Isleif}, {Ito}, {Itoh},
  {Iyer}, {Izumi}, {JaberianHamedan}, {Jacqmin}, {Jadhav}, {Jadhav}, {James},
  {Jan}, {Jani}, {Janquart}, {Janssens}, {Janthalur}, {Jaranowski}, {Jariwala},
  {Jaume}, {Jenkins}, {Jenner}, {Jeon}, {Jeunon}, {Jia}, {Jin}, {Johns},
  {Johnson-McDaniel}, {Jones}, {Jones}, {Jones}, {Jones}, {Jones}, {Jonker},
  {Ju}, {Jung}, {Jung}, {Junker}, {Juste}, {Kaihotsu}, {Kajita}, {Kakizaki},
  {Kalaghatgi}, {Kalogera}, {Kamai}, {Kamiizumi}, {Kanda}, {Kandhasamy},
  {Kang}, {Kanner}, {Kao}, {Kapadia}, {Kapasi}, {Karat}, {Karathanasis},
  {Karki}, {Kashyap}, {Kasprzack}, {Kastaun}, {Katsanevas}, {Katsavounidis},
  {Katzman}, {Kaur}, {Kawabe}, {Kawaguchi}, {Kawai}, {Kawasaki},
  {K{\'e}f{\'e}lian}, {Keitel}, {Key}, {Khadka}, {Khalili}, {Khan}, {Khazanov},
  {Khetan}, {Khursheed}, {Kijbunchoo}, {Kim}, {Kim}, {Kim}, {Kim}, {Kim},
  {Kim}, {Kimball}, {Kimura}, {Kinley-Hanlon}, {Kirchhoff}, {Kissel}, {Kita},
  {Kitazawa}, {Kleybolte}, {Klimenko}, {Knee}, {Knowles}, {Knyazev}, {Koch},
  {Koekoek}, {Kojima}, {Kokeyama}, {Koley}, {Kolitsidou}, {Kolstein}, {Komori},
  {Kondrashov}, {Kong}, {Kontos}, {Koper}, {Korobko}, {Kotake}, {Kovalam},
  {Kozak}, {Kozakai}, {Kozu}, {Kringel}, {Krishnendu}, {Kr{\'o}lak}, {Kuehn},
  {Kuei}, {Kuijer}, {Kulkarni}, {Kumar}, {Kumar}, {Kumar}, {Kumar}, {Kume},
  {Kuns}, {Kuo}, {Kuo}, {Kuromiya}, {Kuroyanagi}, {Kusayanagi}, {Kuwahara},
  {Kwak}, {Lagabbe}, {Laghi}, {Lalande}, {Lam}, {Lamberts}, {Landry}, {Lane},
  {Lang}, {Lange}, {Lantz}, {La Rosa}, {Lartaux-Vollard}, {Lasky}, {Laxen},
  {Lazzarini}, {Lazzaro}, {Leaci}, {Leavey}, {Lecoeuche}, {Lee}, {Lee}, {Lee},
  {Lee}, {Lee}, {Lee}, {Lehmann}, {Lema{\^\i}tre}, {Leonardi}, {Leroy},
  {Letendre}, {Levesque}, {Levin}, {Leviton}, {Leyde}, {Li}, {Li}, {Li}, {Li},
  {Li}, {Li}, {Lin}, {Lin}, {Lin}, {Lin}, {Lin}, {Linde}, {Linker}, {Linley},
  {Littenberg}, {Liu}, {Liu}, {Liu}, {Liu}, {Llamas}, {Llorens-Monteagudo},
  {Lo}, {Lockwood}, {Loh}, {London}, {Longo}, {Lopez}, {Lopez Portilla},
  {Lorenzini}, {Loriette}, {Lormand}, {Losurdo}, {Lott}, {Lough}, {Lousto},
  {Lovelace}, {Lucaccioni}, {L{\"u}ck}, {Lumaca}, {Lundgren}, {Luo}, {Lynam},
  {Macas}, {MacInnis}, {Macleod}, {MacMillan}, {Macquet}, {Maga{\~n}a
  Hernandez}, {Magazz{\`u}}, {Magee}, {Maggiore}, {Magnozzi}, {Mahesh},
  {Majorana}, {Makarem}, {Maksimovic}, {Maliakal}, {Malik}, {Man}, {Mandic},
  {Mangano}, {Mango}, {Mansell}, {Manske}, {Mantovani}, {Mapelli},
  {Marchesoni}, {Marchio}, {Marion}, {Mark}, {M{\'a}rka}, {M{\'a}rka},
  {Markakis}, {Markosyan}, {Markowitz}, {Maros}, {Marquina}, {Marsat},
  {Martelli}, {Martin}, {Martin}, {Martinez}, {Martinez}, {Martinez},
  {Martinovic}, {Martynov}, {Marx}, {Masalehdan}, {Mason}, {Massera},
  {Masserot}, {Massinger}, {Masso-Reid}, {Mastrogiovanni}, {Matas},
  {Mateu-Lucena}, {Matichard}, {Matiushechkina}, {Mavalvala}, {McCann},
  {McCarthy}, {McClelland}, {McClincy}, {McCormick}, {McCuller}, {McGhee},
  {McGuire}, {McIsaac}, {McIver}, {McRae}, {McWilliams}, {Meacher}, {Mehmet},
  {Mehta}, {Meijer}, {Melatos}, {Melchor}, {Mendell}, {Menendez-Vazquez},
  {Menoni}, {Mercer}, {Mereni}, {Merfeld}, {Merilh}, {Merritt}, {Merzougui},
  {Meshkov}, {Messenger}, {Messick}, {Meyers}, {Meylahn}, {Mhaske}, {Miani},
  {Miao}, {Michaloliakos}, {Michel}, {Michimura}, {Middleton}, {Milano},
  {Miller}, {Miller}, {Miller}, {Millhouse}, {Mills}, {Milotti}, {Minazzoli},
  {Minenkov}, {Mio}, {Mir}, {Miravet-Ten{\'e}s}, {Mishra}, {Mishra}, {Mistry},
  {Mitra}, {Mitrofanov}, {Mitselmakher}, {Mittleman}, {Miyakawa}, {Miyamoto},
  {Miyazaki}, {Miyo}, {Miyoki}, {Mo}, {Modafferi}, {Moguel}, {Mogushi},
  {Mohapatra}, {Mohite}, {Molina}, {Molina-Ruiz}, {Mondin}, {Montani}, {Moore},
  {Moraru}, {Morawski}, {More}, {Moreno}, {Moreno}, {Mori}, {Morisaki},
  {Moriwaki}, {Morr{\'a}s}, {Mours}, {Mow-Lowry}, {Mozzon}, {Muciaccia},
  {Mukherjee}, {Mukherjee}, {Mukherjee}, {Mukherjee}, {Mukherjee}, {Mukund},
  {Mullavey}, {Munch}, {Mu{\~n}iz}, {Murray}, {Musenich}, {Muusse}, {Nadji},
  {Nagano}, {Nagano}, {Nagar}, {Nakamura}, {Nakano}, {Nakano}, {Nakashima},
  {Nakayama}, {Napolano}, {Nardecchia}, {Narikawa}, {Naticchioni}, {Nayak},
  {Nayak}, {Negishi}, {Neil}, {Neilson}, {Nelemans}, {Nelson}, {Nery},
  {Neubauer}, {Neunzert}, {Ng}, {Ng}, {Nguyen}, {Nguyen}, {Nguyen}, {Nguyen
  Quynh}, {Ni}, {Nichols}, {Nishizawa}, {Nissanke}, {Nitoglia}, {Nocera},
  {Norman}, {North}, {Nozaki}, {Nu{\~n}o Siles}, {Nuttall}, {Oberling},
  {O'Brien}, {Obuchi}, {O'Dell}, {Oelker}, {Ogaki}, {Oganesyan}, {Oh}, {Oh},
  {Oh}, {Ohashi}, {Ohishi}, {Ohkawa}, {Ohme}, {Ohta}, {Okada}, {Okutani},
  {Okutomi}, {Olivetto}, {Oohara}, {Ooi}, {Oram}, {O'Reilly}, {Ormiston},
  {Ormsby}, {Ortega}, {O'Shaughnessy}, {O'Shea}, {Oshino}, {Ossokine},
  {Osthelder}, {Otabe}, {Ottaway}, {Overmier}, {Pace}, {Pagano}, {Page},
  {Pagliaroli}, {Pai}, {Pai}, {Palamos}, {Palashov}, {Palomba}, {Pan}, {Pan},
  {Panda}, {Pang}, {Pang}, {Pankow}, {Pannarale}, {Pant}, {Panther},
  {Paoletti}, {Paoli}, {Paolone}, {Parisi}, {Park}, {Park}, {Parker},
  {Pascucci}, {Pasqualetti}, {Passaquieti}, {Passuello}, {Patel}, {Pathak},
  {Patricelli}, {Patron}, {Paul}, {Payne}, {Pedraza}, {Pegoraro}, {Pele},
  {Pe{\~n}a Arellano}, {Penn}, {Perego}, {Pereira}, {Pereira}, {Perez},
  {P{\'e}rigois}, {Perkins}, {Perreca}, {Perri{\`e}s}, {Petermann},
  {Petterson}, {Pfeiffer}, {Pham}, {Phukon}, {Piccinni}, {Pichot},
  {Piendibene}, {Piergiovanni}, {Pierini}, {Pierro}, {Pillant}, {Pillas},
  {Pilo}, {Pinard}, {Pinto}, {Pinto}, {Piotrzkowski}, {Piotrzkowski},
  {Pirello}, {Pitkin}, {Placidi}, {Planas}, {Plastino}, {Pluchar}, {Poggiani},
  {Polini}, {Pong}, {Ponrathnam}, {Popolizio}, {Porter}, {Poulton}, {Powell},
  {Pracchia}, {Pradier}, {Prajapati}, {Prasai}, {Prasanna}, {Pratten},
  {Principe}, {Prodi}, {Prokhorov}, {Prosposito}, {Prudenzi}, {Puecher},
  {Punturo}, {Puosi}, {Puppo}, {P{\"u}rrer}, {Qi}, {Quetschke},
  {Quitzow-James}, {Qutob}, {Raab}, {Raaijmakers}, {Radkins}, {Radulesco},
  {Raffai}, {Rail}, {Raja}, {Rajan}, {Ramirez}, {Ramirez}, {Ramos-Buades},
  {Rana}, {Rapagnani}, {Rapol}, {Ray}, {Raymond}, {Raza}, {Razzano}, {Read},
  {Rees}, {Regimbau}, {Rei}, {Reid}, {Reid}, {Reitze}, {Relton}, {Renzini},
  {Rettegno}, {Reza}, {Rezac}, {Ricci}, {Richards}, {Richardson}, {Richardson},
  {Riemenschneider}, {Riles}, {Rinaldi}, {Rink}, {Rizzo}, {Robertson}, {Robie},
  {Robinet}, {Rocchi}, {Rodriguez}, {Rolland}, {Rollins}, {Romanelli},
  {Romano}, {Romel}, {Romero-Rodr{\'\i}guez}, {Romero-Shaw}, {Romie},
  {Ronchini}, {Rosa}, {Rose}, {Rosi{\'n}ska}, {Ross}, {Rowan}, {Rowlinson},
  {Roy}, {Roy}, {Roy}, {Rozza}, {Ruggi}, {Ruiz-Rocha}, {Ryan}, {Sachdev},
  {Sadecki}, {Sadiq}, {Sago}, {Saito}, {Saito}, {Sakai}, {Sakai},
  {Sakellariadou}, {Sakuno}, {Salafia}, {Salconi}, {Saleem}, {Salemi},
  {Samajdar}, {Sanchez}, {Sanchez}, {Sanchez}, {Sanchis-Gual}, {Sanders},
  {Sanuy}, {Saravanan}, {Sarin}, {Sassolas}, {Satari}, {Sathyaprakash}, {Sato},
  {Sato}, {Sauter}, {Savage}, {Sawada}, {Sawant}, {Sawant}, {Sayah},
  {Schaetzl}, {Scheel}, {Scheuer}, {Schiworski}, {Schmidt}, {Schmidt},
  {Schnabel}, {Schneewind}, {Schofield}, {Sch{\"o}nbeck}, {Schulte}, {Schutz},
  {Schwartz}, {Scott}, {Scott}, {Seglar-Arroyo}, {Sekiguchi}, {Sekiguchi},
  {Sellers}, {Sengupta}, {Sentenac}, {Seo}, {Sequino}, {Sergeev}, {Setyawati},
  {Shaffer}, {Shahriar}, {Shams}, {Shao}, {Sharma}, {Sharma}, {Shawhan},
  {Shcheblanov}, {Shibagaki}, {Shikauchi}, {Shimizu}, {Shimoda}, {Shimode},
  {Shinkai}, {Shishido}, {Shoda}, {Shoemaker}, {Shoemaker}, {ShyamSundar},
  {Sieniawska}, {Sigg}, {Singer}, {Singh}, {Singh}, {Singha}, {Sintes},
  {Sipala}, {Skliris}, {Slagmolen}, {Slaven-Blair}, {Smetana}, {Smith},
  {Smith}, {Soldateschi}, {Somala}, {Somiya}, {Son}, {Soni}, {Soni}, {Sordini},
  {Sorrentino}, {Sorrentino}, {Sotani}, {Soulard}, {Souradeep}, {Sowell},
  {Spagnuolo}, {Spencer}, {Spera}, {Srinivasan}, {Srivastava}, {Srivastava},
  {Staats}, {Stachie}, {Steer}, {Steinhoff}, {Steinlechner}, {Steinlechner},
  {Stevenson}, {Stops}, {Stover}, {Strain}, {Strang}, {Stratta}, {Strunk},
  {Sturani}, {Stuver}, {Sudhagar}, {Sudhir}, {Sugimoto}, {Suh}, {Sullivan},
  {Sullivan}, {Summerscales}, {Sun}, {Sun}, {Sunil}, {Sur}, {Suresh}, {Sutton},
  {Suzuki}, {Suzuki}, {Swinkels}, {Szczepa{\'n}czyk}, {Szewczyk}, {Tacca},
  {Tagoshi}, {Tait}, {Takahashi}, {Takahashi}, {Takamori}, {Takano}, {Takeda},
  {Takeda}, {Talbot}, {Talbot}, {Tanaka}, {Tanaka}, {Tanaka}, {Tanaka},
  {Tanaka}, {Tanasijczuk}, {Tanioka}, {Tanner}, {Tao}, {Tao}, {Tapia San
  Mart{\'\i}n}, {Taranto}, {Tasson}, {Telada}, {Tenorio}, {Terhune},
  {Terkowski}, {Thirugnanasambandam}, {Thomas}, {Thomas}, {Thomas}, {Thompson},
  {Thondapu}, {Thorne}, {Thrane}, {Tiwari}, {Tiwari}, {Tiwari}, {Toivonen},
  {Toland}, {Tolley}, {Tomaru}, {Tomigami}, {Tomura}, {Tonelli},
  {Torres-Forn{\'e}}, {Torrie}, {Tosta e Melo}, {T{\"o}yr{\"a}}, {Trapananti},
  {Travasso}, {Traylor}, {Trevor}, {Tringali}, {Tripathee}, {Troiano},
  {Trovato}, {Trozzo}, {Trudeau}, {Tsai}, {Tsai}, {Tsang}, {Tsang}, {Tsao},
  {Tse}, {Tso}, {Tsubono}, {Tsuchida}, {Tsukada}, {Tsuna}, {Tsutsui},
  {Tsuzuki}, {Turbang}, {Turconi}, {Tuyenbayev}, {Ubhi}, {Uchikata},
  {Uchiyama}, {Udall}, {Ueda}, {Uehara}, {Ueno}, {Ueshima}, {Unnikrishnan},
  {Uraguchi}, {Urban}, {Ushiba}, {Utina}, {Vahlbruch}, {Vajente}, {Vajpeyi},
  {Valdes}, {Valentini}, {Valsan}, {van Bakel}, {van Beuzekom}, {van den
  Brand}, {Van Den Broeck}, {Vander-Hyde}, {van der Schaaf}, {van Heijningen},
  {Vanosky}, {van Putten}, {van Remortel}, {Vardaro}, {Vargas}, {Varma},
  {Vas{\'u}th}, {Vecchio}, {Vedovato}, {Veitch}, {Veitch}, {Venneberg},
  {Venugopalan}, {Verkindt}, {Verma}, {Verma}, {Veske}, {Vetrano},
  {Vicer{\'e}}, {Vidyant}, {Viets}, {Vijaykumar}, {Villa-Ortega}, {Vinet},
  {Virtuoso}, {Vitale}, {Vo}, {Vocca}, {von Reis}, {von Wrangel}, {Vorvick},
  {Vyatchanin}, {Wade}, {Wade}, {Wagner}, {Walet}, {Walker}, {Wallace},
  {Wallace}, {Walsh}, {Wang}, {Wang}, {Wang}, {Ward}, {Warner}, {Was},
  {Washimi}, {Washington}, {Watchi}, {Weaver}, {Webster}, {Weinert},
  {Weinstein}, {Weiss}, {Weller}, {Weller}, {Wellmann}, {Wen}, {We{\ss}els},
  {Wette}, {Whelan}, {White}, {Whiting}, {Whittle}, {Wilken}, {Williams},
  {Williams}, {Williams}, {Williamson}, {Willis}, {Willke}, {Wilson},
  {Winkler}, {Wipf}, {Wlodarczyk}, {Woan}, {Woehler}, {Wofford}, {Wong}, {Wu},
  {Wu}, {Wu}, {Wu}, {Wysocki}, {Xiao}, {Xu}, {Yamada}, {Yamamoto}, {Yamamoto},
  {Yamamoto}, {Yamamoto}, {Yamashita}, {Yamazaki}, {Yang}, {Yang}, {Yang},
  {Yang}, {Yang}, {Yap}, {Yeeles}, {Yelikar}, {Ying}, {Yokogawa}, {Yokoyama},
  {Yokozawa}, {Yoo}, {Yoshioka}, {Yu}, {Yu}, {Yuzurihara}, {Zadro{\.z}ny},
  {Zanolin}, {Zeidler}, {Zelenova}, {Zendri}, {Zevin}, {Zhan}, {Zhang},
  {Zhang}, {Zhang}, {Zhang}, {Zhang}, {Zhao}, {Zhao}, {Zhao}, {Zhao}, {Zheng},
  {Zhou}, {Zhou}, {Zhu}, {Zhu}, {Zimmerman}, {Zlochower}, {Zucker}, \&
  {Zweizig}}]{GWTC3}
{Abbott}, R., {Abbott}, T.~D., {Acernese}, F., {et~al.} 2021{\natexlab{b}},
  arXiv e-prints, arXiv:2111.03606

\bibitem[{{Abbott} {et~al.}(2021{\natexlab{c}}){Abbott}, {Abbott}, {Acernese},
  {Ackley}, {Adams}, {Adhikari}, {Adhikari}, {Adya}, \& et~al.}]{GWTC-2.1}
---. 2021{\natexlab{c}}, arXiv e-prints, arXiv:2108.01045

\bibitem[{{Ai} {et~al.}(2020){Ai}, {Gao}, \& {Zhang}}]{Ai2020}
{Ai}, S., {Gao}, H., \& {Zhang}, B. 2020, \apj, 893, 146

\bibitem[{{Alsing} {et~al.}(2018){Alsing}, {Silva}, \& {Berti}}]{Alsing2018}
{Alsing}, J., {Silva}, H.~O., \& {Berti}, E. 2018, \mnras, 478, 1377

\bibitem[{{Atri} {et~al.}(2019){Atri}, {Miller-Jones}, {Bahramian}, {Plotkin},
  {Jonker}, {Nelemans}, {Maccarone}, {Sivakoff}, {Deller}, {Chaty}, {Torres},
  {Horiuchi}, {McCallum}, {Natusch}, {Phillips}, {Stevens}, \&
  {Weston}}]{Atri2019}
{Atri}, P., {Miller-Jones}, J.~C.~A., {Bahramian}, A., {et~al.} 2019, \mnras,
  489, 3116

\bibitem[{{Bailyn} {et~al.}(1998){Bailyn}, {Jain}, {Coppi}, \&
  {Orosz}}]{Bailyn1998}
{Bailyn}, C.~D., {Jain}, R.~K., {Coppi}, P., \& {Orosz}, J.~A. 1998, \apj, 499,
  367

\bibitem[{{Belczynski} {et~al.}(2008){Belczynski}, {Kalogera}, {Rasio}, {Taam},
  {Zezas}, {Bulik}, {Maccarone}, \& {Ivanova}}]{Belczynski2008}
{Belczynski}, K., {Kalogera}, V., {Rasio}, F.~A., {et~al.} 2008, \apjs, 174,
  223

\bibitem[{{Belczynski} {et~al.}(2012){Belczynski}, {Wiktorowicz}, {Fryer},
  {Holz}, \& {Kalogera}}]{Belczynski2012}
{Belczynski}, K., {Wiktorowicz}, G., {Fryer}, C.~L., {Holz}, D.~E., \&
  {Kalogera}, V. 2012, \apj, 757, 91

\bibitem[{{Bolton}(1972)}]{Bolton1972}
{Bolton}, C.~T. 1972, \nat, 235, 271

\bibitem[{{Bradt} {et~al.}(1993){Bradt}, {Rothschild}, \& {Swank}}]{Bradt1993}
{Bradt}, H.~V., {Rothschild}, R.~E., \& {Swank}, J.~H. 1993, \aaps, 97, 355

\bibitem[{{Breivik} {et~al.}(2020){Breivik}, {Coughlin}, {Zevin}, {Rodriguez},
  {Kremer}, {Ye}, {Andrews}, {Kurkowski}, {Digman}, {Larson}, \&
  {Rasio}}]{Breivik2020}
{Breivik}, K., {Coughlin}, S., {Zevin}, M., {et~al.} 2020, \apj, 898, 71

\bibitem[{{Cannizzo} {et~al.}(1982){Cannizzo}, {Ghosh}, \&
  {Wheeler}}]{Cannizzo1982}
{Cannizzo}, J.~K., {Ghosh}, P., \& {Wheeler}, J.~C. 1982, \apjl, 260, L83

\bibitem[{{Cannizzo} {et~al.}(1988){Cannizzo}, {Shafter}, \&
  {Wheeler}}]{Cannizzo1988}
{Cannizzo}, J.~K., {Shafter}, A.~W., \& {Wheeler}, J.~C. 1988, \apj, 333, 227

\bibitem[{{Corral-Santana} {et~al.}(2016){Corral-Santana}, {Casares},
  {Mu{\~n}oz-Darias}, {Bauer}, {Mart{\'\i}nez-Pais}, \&
  {Russell}}]{Corral-Santana2016}
{Corral-Santana}, J.~M., {Casares}, J., {Mu{\~n}oz-Darias}, T., {et~al.} 2016,
  \aap, 587, A61

\bibitem[{{de Kool}(1990)}]{deKool1990}
{de Kool}, M. 1990, \apj, 358, 189

\bibitem[{{Degenaar} {et~al.}(2018){Degenaar}, {Ballantyne}, {Belloni},
  {Chakraborty}, {Chen}, {Ji}, {Kretschmar}, {Kuulkers}, {Li}, {Maccarone},
  {Malzac}, {Zhang}, \& {Zhang}}]{Degenaar2018}
{Degenaar}, N., {Ballantyne}, D.~R., {Belloni}, T., {et~al.} 2018, \ssr, 214,
  15

\bibitem[{{Dubus} {et~al.}(2001){Dubus}, {Hameury}, \& {Lasota}}]{Dubus2001}
{Dubus}, G., {Hameury}, J.~M., \& {Lasota}, J.~P. 2001, \aap, 373, 251

\bibitem[{{Dubus} {et~al.}(1999){Dubus}, {Lasota}, {Hameury}, \&
  {Charles}}]{Dubus1999}
{Dubus}, G., {Lasota}, J.-P., {Hameury}, J.-M., \& {Charles}, P. 1999, \mnras,
  303, 139

\bibitem[{{El-Badry} {et~al.}(2022){El-Badry}, {Seeburger}, {Jayasinghe},
  {Rix}, {Almada}, {Conroy}, {Price-Whelan}, \& {Burdge}}]{ElBadry2022}
{El-Badry}, K., {Seeburger}, R., {Jayasinghe}, T., {et~al.} 2022, \mnras, 512,
  5620

\bibitem[{{Ertl} {et~al.}(2020){Ertl}, {Woosley}, {Sukhbold}, \&
  {Janka}}]{Ertl2020}
{Ertl}, T., {Woosley}, S.~E., {Sukhbold}, T., \& {Janka}, H.~T. 2020, \apj,
  890, 51

\bibitem[{{Farr} {et~al.}(2011){Farr}, {Sravan}, {Cantrell}, {Kreidberg},
  {Bailyn}, {Mandel}, \& {Kalogera}}]{Farr2011}
{Farr}, W.~M., {Sravan}, N., {Cantrell}, A., {et~al.} 2011, \apj, 741, 103

\bibitem[{{Fragos} {et~al.}(2019){Fragos}, {Andrews}, {Ramirez-Ruiz}, {Meynet},
  {Kalogera}, {Taam}, \& {Zezas}}]{Fragos2019}
{Fragos}, T., {Andrews}, J.~J., {Ramirez-Ruiz}, E., {et~al.} 2019, \apjl, 883,
  L45

\bibitem[{{Fragos} {et~al.}(2023){Fragos}, {Andrews}, {Bavera}, {Berry},
  {Coughlin}, {Dotter}, {Giri}, {Kalogera}, {Katsaggelos}, {Kovlakas}, \&
  et~al.}]{Fragos2022}
{Fragos}, T., {Andrews}, J.~J., {Bavera}, S.~S., {et~al.} 2023, \apjs, 264, 45

\bibitem[{{Frank} {et~al.}(1992){Frank}, {King}, \& {Raine}}]{Frank1992}
{Frank}, J., {King}, A., \& {Raine}, D. 1992, {Accretion power in
  astrophysics.}, Vol.~21

\bibitem[{{Fryer} {et~al.}(2012){Fryer}, {Belczynski}, {Wiktorowicz},
  {Dominik}, {Kalogera}, \& {Holz}}]{Fryer2012}
{Fryer}, C.~L., {Belczynski}, K., {Wiktorowicz}, G., {et~al.} 2012, \apj, 749,
  91

\bibitem[{{Fryer} {et~al.}(2022){Fryer}, {Olejak}, \& {Belczynski}}]{Fryer2022}
{Fryer}, C.~L., {Olejak}, A., \& {Belczynski}, K. 2022, \apj, 931, 94

\bibitem[{{Gallegos-Garcia} {et~al.}(2021){Gallegos-Garcia}, {Berry},
  {Marchant}, \& {Kalogera}}]{GallegosGarcia2021}
{Gallegos-Garcia}, M., {Berry}, C. P.~L., {Marchant}, P., \& {Kalogera}, V.
  2021, \apj, 922, 110

\bibitem[{{Giacobbo} \& {Mapelli}(2018)}]{Giacobbo2018}
{Giacobbo}, N., \& {Mapelli}, M. 2018, \mnras, 480, 2011

\bibitem[{{Giacobbo} \& {Mapelli}(2019)}]{Giacobbo2019}
---. 2019, \mnras, 482, 2234

\bibitem[{{Hobbs} {et~al.}(2005){Hobbs}, {Lorimer}, {Lyne}, \&
  {Kramer}}]{Hobbs2005}
{Hobbs}, G., {Lorimer}, D.~R., {Lyne}, A.~G., \& {Kramer}, M. 2005, \mnras,
  360, 974

\bibitem[{{Hunter}(2007)}]{matplotlib}
{Hunter}, J.~D. 2007, Computing in Science and Engineering, 9, 90

\bibitem[{{Hurley} {et~al.}(2002){Hurley}, {Tout}, \& {Pols}}]{Hurley2002}
{Hurley}, J.~R., {Tout}, C.~A., \& {Pols}, O.~R. 2002, \mnras, 329, 897

\bibitem[{{Ivanova}(2006)}]{Ivanova2006}
{Ivanova}, N. 2006, \apjl, 653, L137

\bibitem[{{Ivanova} \& {Taam}(2003)}]{Ivanova2003}
{Ivanova}, N., \& {Taam}, R.~E. 2003, \apj, 599, 516

\bibitem[{{Jayasinghe} {et~al.}(2021){Jayasinghe}, {Stanek}, {Thompson},
  {Kochanek}, {Rowan}, {Vallely}, {Strassmeier}, {Weber}, {Hinkle}, {Hambsch},
  {Martin}, {Prieto}, {Pessi}, {Huber}, {Auchettl}, {Lopez}, {Ilyin},
  {Badenes}, {Howard}, {Isaacson}, \& {Murphy}}]{Jayasinghe2021}
{Jayasinghe}, T., {Stanek}, K.~Z., {Thompson}, T.~A., {et~al.} 2021, \mnras,
  504, 2577

\bibitem[{{Jonker} {et~al.}(2021){Jonker}, {Kaur}, {Stone}, \&
  {Torres}}]{Jonker2021}
{Jonker}, P.~G., {Kaur}, K., {Stone}, N., \& {Torres}, M. A.~P. 2021, \apj,
  921, 131

\bibitem[{{Kalogera} \& {Baym}(1996)}]{Kalogera1996}
{Kalogera}, V., \& {Baym}, G. 1996, \apjl, 470, L61

\bibitem[{{King} {et~al.}(1996){King}, {Kolb}, \& {Burderi}}]{King1996}
{King}, A.~R., {Kolb}, U., \& {Burderi}, L. 1996, \apjl, 464, L127

\bibitem[{{King} {et~al.}(1997){King}, {Kolb}, \& {Szuszkiewicz}}]{King1997a}
{King}, A.~R., {Kolb}, U., \& {Szuszkiewicz}, E. 1997, \apj, 488, 89

\bibitem[{{King} \& {Ritter}(1998)}]{King1998}
{King}, A.~R., \& {Ritter}, H. 1998, \mnras, 293, L42

\bibitem[{{Knevitt} {et~al.}(2014){Knevitt}, {Wynn}, {Vaughan}, \&
  {Watson}}]{Knevitt2014}
{Knevitt}, G., {Wynn}, G.~A., {Vaughan}, S., \& {Watson}, M.~G. 2014, \mnras,
  437, 3087

\bibitem[{{Knigge} {et~al.}(2011){Knigge}, {Baraffe}, \&
  {Patterson}}]{Knigge2011}
{Knigge}, C., {Baraffe}, I., \& {Patterson}, J. 2011, \apjs, 194, 28

\bibitem[{{Kolb} \& {Ritter}(1990)}]{Kolb1990}
{Kolb}, U., \& {Ritter}, H. 1990, \aap, 236, 385

\bibitem[{{Kreidberg} {et~al.}(2012){Kreidberg}, {Bailyn}, {Farr}, \&
  {Kalogera}}]{Kreidberg2012}
{Kreidberg}, L., {Bailyn}, C.~D., {Farr}, W.~M., \& {Kalogera}, V. 2012, \apj,
  757, 36

\bibitem[{{Lasota}(2001)}]{Lasota2001}
{Lasota}, J.-P. 2001, \nar, 45, 449

\bibitem[{{Lasota}(2016)}]{Lasota2016}
---. 2016, Astrophysics and Space Science Library, Vol. 440, {Black Hole
  Accretion Discs}, ed. C.~{Bambi}, 1

\bibitem[{{Lasota} {et~al.}(2008){Lasota}, {Dubus}, \& {Kruk}}]{Lasota2008}
{Lasota}, J.~P., {Dubus}, G., \& {Kruk}, K. 2008, \aap, 486, 523

\bibitem[{{Lattimer}(2019)}]{Lattimer2019}
{Lattimer}, J.~M. 2019, Universe, 5, 159

\bibitem[{{Levine} {et~al.}(1996){Levine}, {Bradt}, {Cui}, {Jernigan},
  {Morgan}, {Remillard}, {Shirey}, \& {Smith}}]{Levine1996}
{Levine}, A.~M., {Bradt}, H., {Cui}, W., {et~al.} 1996, \apjl, 469, L33

\bibitem[{{Liu} {et~al.}(2021){Liu}, {Wei}, {Xue}, \& {Sun}}]{Liu2021}
{Liu}, T., {Wei}, Y.-F., {Xue}, L., \& {Sun}, M.-Y. 2021, \apj, 908, 106

\bibitem[{{Mandel} \& {M{\"u}ller}(2020)}]{Mandel2020}
{Mandel}, I., \& {M{\"u}ller}, B. 2020, \mnras, 499, 3214

\bibitem[{{Margalit} \& {Metzger}(2017)}]{Margalit2017}
{Margalit}, B., \& {Metzger}, B.~D. 2017, \apjl, 850, L19

\bibitem[{{McMillan}(2011)}]{McMillan2011}
{McMillan}, P.~J. 2011, \mnras, 414, 2446

\bibitem[{{M{\"u}ller} \& {Serot}(1996)}]{Muller1996}
{M{\"u}ller}, H., \& {Serot}, B.~D. 1996, \nphysa, 606, 508

\bibitem[{{{\"O}zel} \& {Freire}(2016)}]{ozel2016}
{{\"O}zel}, F., \& {Freire}, P. 2016, \araa, 54, 401

\bibitem[{{{\"O}zel} {et~al.}(2010){{\"O}zel}, {Psaltis}, {Narayan}, \&
  {McClintock}}]{Ozel2010}
{{\"O}zel}, F., {Psaltis}, D., {Narayan}, R., \& {McClintock}, J.~E. 2010,
  \apj, 725, 1918

\bibitem[{{Patton} {et~al.}(2022){Patton}, {Sukhbold}, \&
  {Eldridge}}]{Patton2022}
{Patton}, R.~A., {Sukhbold}, T., \& {Eldridge}, J.~J. 2022, \mnras, 511, 903

\bibitem[{{Paxton} {et~al.}(2011){Paxton}, {Bildsten}, {Dotter}, {Herwig},
  {Lesaffre}, \& {Timmes}}]{Paxton2011}
{Paxton}, B., {Bildsten}, L., {Dotter}, A., {et~al.} 2011, \apjs, 192, 3

\bibitem[{{Paxton} {et~al.}(2013){Paxton}, {Cantiello}, {Arras}, {Bildsten},
  {Brown}, {Dotter}, {Mankovich}, {Montgomery}, {Stello}, {Timmes}, \&
  {Townsend}}]{Paxton2013}
{Paxton}, B., {Cantiello}, M., {Arras}, P., {et~al.} 2013, \apjs, 208, 4

\bibitem[{{Paxton} {et~al.}(2015){Paxton}, {Marchant}, {Schwab}, {Bauer},
  {Bildsten}, {Cantiello}, {Dessart}, {Farmer}, {Hu}, {Langer}, {Townsend},
  {Townsley}, \& {Timmes}}]{Paxton2015}
{Paxton}, B., {Marchant}, P., {Schwab}, J., {et~al.} 2015, \apjs, 220, 15

\bibitem[{{Paxton} {et~al.}(2018){Paxton}, {Schwab}, {Bauer}, {Bildsten},
  {Blinnikov}, {Duffell}, {Farmer}, {Goldberg}, {Marchant}, {Sorokina},
  {Thoul}, {Townsend}, \& {Timmes}}]{Paxton2018}
{Paxton}, B., {Schwab}, J., {Bauer}, E.~B., {et~al.} 2018, \apjs, 234, 34

\bibitem[{{Paxton} {et~al.}(2019){Paxton}, {Smolec}, {Schwab}, {Gautschy},
  {Bildsten}, {Cantiello}, {Dotter}, {Farmer}, {Goldberg}, {Jermyn}, {Kanbur},
  {Marchant}, {Thoul}, {Townsend}, {Wolf}, {Zhang}, \& {Timmes}}]{Paxton2019}
{Paxton}, B., {Smolec}, R., {Schwab}, J., {et~al.} 2019, \apjs, 243, 10

\bibitem[{{Price-Whelan} {et~al.}(2018){Price-Whelan}, {Sip{\H{o}}cz},
  {G{\"u}nther}, {Lim}, {Crawford}, {Conseil}, {Shupe}, {Craig}, {Dencheva},
  {Ginsburg}, {VanderPlas}, {Bradley}, {P{\'e}rez-Su{\'a}rez}, {de Val-Borro},
  {Paper Contributors}, {Aldcroft}, {Cruz}, {Robitaille}, {Tollerud},
  {Coordination Committee}, {Ardelean}, {Babej}, {Bach}, {Bachetti}, {Bakanov},
  {Bamford}, {Barentsen}, {Barmby}, {Baumbach}, {Berry}, {Biscani}, {Boquien},
  {Bostroem}, {Bouma}, {Brammer}, {Bray}, {Breytenbach}, {Buddelmeijer},
  {Burke}, {Calderone}, {Cano Rodr{\'\i}guez}, {Cara}, {Cardoso}, {Cheedella},
  {Copin}, {Corrales}, {Crichton}, {D{\textquoteright}Avella}, {Deil},
  {Depagne}, {Dietrich}, {Donath}, {Droettboom}, {Earl}, {Erben}, {Fabbro},
  {Ferreira}, {Finethy}, {Fox}, {Garrison}, {Gibbons}, {Goldstein}, {Gommers},
  {Greco}, {Greenfield}, {Groener}, {Grollier}, {Hagen}, {Hirst}, {Homeier},
  {Horton}, {Hosseinzadeh}, {Hu}, {Hunkeler}, {Ivezi{\'c}}, {Jain}, {Jenness},
  {Kanarek}, {Kendrew}, {Kern}, {Kerzendorf}, {Khvalko}, {King}, {Kirkby},
  {Kulkarni}, {Kumar}, {Lee}, {Lenz}, {Littlefair}, {Ma}, {Macleod},
  {Mastropietro}, {McCully}, {Montagnac}, {Morris}, {Mueller}, {Mumford},
  {Muna}, {Murphy}, {Nelson}, {Nguyen}, {Ninan}, {N{\"o}the}, {Ogaz}, {Oh},
  {Parejko}, {Parley}, {Pascual}, {Patil}, {Patil}, {Plunkett}, {Prochaska},
  {Rastogi}, {Reddy Janga}, {Sabater}, {Sakurikar}, {Seifert}, {Sherbert},
  {Sherwood-Taylor}, {Shih}, {Sick}, {Silbiger}, {Singanamalla}, {Singer},
  {Sladen}, {Sooley}, {Sornarajah}, {Streicher}, {Teuben}, {Thomas},
  {Tremblay}, {Turner}, {Terr{\'o}n}, {van Kerkwijk}, {de la Vega}, {Watkins},
  {Weaver}, {Whitmore}, {Woillez}, {Zabalza}, \& {Contributors}}]{astropy:2018}
{Price-Whelan}, A.~M., {Sip{\H{o}}cz}, B.~M., {G{\"u}nther}, H.~M., {et~al.}
  2018, \aj, 156, 123

\bibitem[{{Raaijmakers} {et~al.}(2021){Raaijmakers}, {Greif}, {Hebeler},
  {Hinderer}, {Nissanke}, {Schwenk}, {Riley}, {Watts}, {Lattimer}, \&
  {Ho}}]{Raaijmakers2021}
{Raaijmakers}, G., {Greif}, S.~K., {Hebeler}, K., {et~al.} 2021, \apjl, 918,
  L29

\bibitem[{{Ray} {et~al.}(2019){Ray}, {Arzoumanian}, {Ballantyne}, {Bozzo},
  {Brandt}, {Brenneman}, {Chakrabarty}, {Christophersen}, {DeRosa}, {Feroci},
  {Gendreau}, {Goldstein}, {Hartmann}, {Hernanz}, {Jenke}, {Kara}, {Maccarone},
  {McDonald}, {Martindale}, {Nowak}, {Phlips}, {Remillard}, {Schanne},
  {Stevens}, {Tomsick}, {Watts}, {Wilson-Hodge}, {Wolff}, {Wood}, {Zane},
  {Ajello}, {Alston}, {Altamirano}, {Antoniou}, {Arur}, {Ashton}, {Auchettl},
  {Ayres}, {Bachetti}, {Balokovic}, {Baring}, {Baykal}, {Begelman}, {Bhat},
  {Bogdanov}, {Briggs}, {Bulbul}, {Bult}, {Burns}, {Cackett}, {Campana},
  {Caspi}, {Cavecchi}, {Chenevez}, {Cherry}, {Corbet}, {Corcoran}, {Corsi},
  {Degenaar}, {Drake}, {Eikenberry}, {Enoto}, {Fragile}, {Fuerst}, {Gandhi},
  {Garcia}, {Goldstein}, {Gonzalez}, {Grefenstette}, {Grinberg}, {Grossan},
  {Guillot}, {Guver}, {Haggard}, {Heinke}, {Heinz}, {Hemphill}, {Homan}, {Hui},
  {Huppenkothen}, {Ingram}, {Irwin}, {Jaisawal}, {Jaodand}, {Kalemci},
  {Kaplan}, {Keek}, {Kennea}, {Kerr}, {van der Klis}, {Kocevski}, {Koss},
  {Kowalski}, {Lai}, {Lamb}, {Laycock}, {Lazio}, {Lazzati}, {Longcope},
  {Loewenstein}, {Maitra}, {Majid}, {Maksym}, {Malacaria}, {Margutti},
  {Martindale}, {McHardy}, {Meyer}, {Middleton}, {Miller}, {Miller}, {Motta},
  {Neilsen}, {Nelson}, {Noble}, {O'Brien}, {Osborne}, {Osten}, {Ozel},
  {Palliyaguru}, {Pasham}, {Patruno}, {Pelassa}, {Petropoulou}, {Pilia},
  {Pohl}, {Pooley}, {Prescod-Weinstein}, {Psaltis}, {Raaijmakers}, {Reynolds},
  {Riley}, {Salvesen}, {Santangelo}, {Scaringi}, {Schanne}, {Schnittman},
  {Smith}, {Smith}, {Snios}, {Steiner}, {Steiner}, {Stella}, {Strohmayer},
  {Sun}, {Tauris}, {Taylor}, {Tohuvavohu}, {Vacchi}, {Vasilopoulos},
  {Veledina}, {Walsh}, {Weinberg}, {Wilkins}, {Willingale}, {Wilms}, {Winter},
  {Wolff}, {in 't Zand}, {Zezas}, {Zhang}, \& {Zoghbi}}]{STROBEX}
{Ray}, P., {Arzoumanian}, Z., {Ballantyne}, D., {et~al.} 2019, in Bulletin of
  the American Astronomical Society, Vol.~51, 231

\bibitem[{{Remillard} \& {McClintock}(2006)}]{Remillard2006}
{Remillard}, R.~A., \& {McClintock}, J.~E. 2006, \araa, 44, 49

\bibitem[{{Rhoades} \& {Ruffini}(1974)}]{Rhoades1974}
{Rhoades}, C.~E., \& {Ruffini}, R. 1974, \prl, 32, 324

\bibitem[{{Robin} {et~al.}(2003){Robin}, {Reyl{\'e}}, {Derri{\`e}re}, \&
  {Picaud}}]{Robin2003}
{Robin}, A.~C., {Reyl{\'e}}, C., {Derri{\`e}re}, S., \& {Picaud}, S. 2003,
  \aap, 409, 523

\bibitem[{{Robitaille} {et~al.}(2013){Robitaille}, {Tollerud}, {Greenfield},
  {Droettboom}, {Bray}, {Aldcroft}, {Davis}, {Ginsburg}, {Price-Whelan},
  {Kerzendorf}, {Conley}, {Crighton}, {Barbary}, {Muna}, {Ferguson},
  {Grollier}, {Parikh}, {Nair}, {Unther}, {Deil}, {Woillez}, {Conseil},
  {Kramer}, {Turner}, {Singer}, {Fox}, {Weaver}, {Zabalza}, {Edwards}, {Azalee
  Bostroem}, {Burke}, {Casey}, {Crawford}, {Dencheva}, {Ely}, {Jenness},
  {Labrie}, {Lim}, {Pierfederici}, {Pontzen}, {Ptak}, {Refsdal}, {Servillat},
  \& {Streicher}}]{astropy:2013}
{Robitaille}, T.~P., {Tollerud}, E.~J., {Greenfield}, P., {et~al.} 2013, \aap,
  558, A33

\bibitem[{{Sana} {et~al.}(2012){Sana}, {de Mink}, {de Koter}, {Langer},
  {Evans}, {Gieles}, {Gosset}, {Izzard}, {Le Bouquin}, \&
  {Schneider}}]{Sana2012}
{Sana}, H., {de Mink}, S.~E., {de Koter}, A., {et~al.} 2012, Science, 337, 444

\bibitem[{{Santoliquido} {et~al.}(2021){Santoliquido}, {Mapelli}, {Giacobbo},
  {Bouffanais}, \& {Artale}}]{Santoliquido2021}
{Santoliquido}, F., {Mapelli}, M., {Giacobbo}, N., {Bouffanais}, Y., \&
  {Artale}, M.~C. 2021, \mnras, 502, 4877

\bibitem[{{Shao} {et~al.}(2020){Shao}, {Tang}, {Jiang}, \& {Fan}}]{Shao2020}
{Shao}, D.-S., {Tang}, S.-P., {Jiang}, J.-L., \& {Fan}, Y.-Z. 2020, \prd, 102,
  063006

\bibitem[{{Shenar} {et~al.}(2022){Shenar}, {Sana}, {Mahy}, {El-Badry},
  {Marchant}, {Langer}, {Hawcroft}, {Fabry}, {Sen}, {Almeida}, \&
  et~al.}]{Shenar2022}
{Shenar}, T., {Sana}, H., {Mahy}, L., {et~al.} 2022, Nature Astronomy, 6, 1085

\bibitem[{{The pandas development team}(2020)}]{reback2020pandas}
{The pandas development team}. 2020, pandas-dev/pandas: Pandas, v1.1.1,
  Zenodo, doi:10.5281/zenodo.3509134.
\newblock \url{https://doi.org/10.5281/zenodo.3509134}

\bibitem[{{Thompson} {et~al.}(2019){Thompson}, {Kochanek}, {Stanek}, {Badenes},
  {Post}, {Jayasinghe}, {Latham}, {Bieryla}, {Esquerdo}, {Berlind}, {Calkins},
  {Tayar}, {Lindegren}, {Johnson}, {Holoien}, {Auchettl}, \&
  {Covey}}]{Thompson2019}
{Thompson}, T.~A., {Kochanek}, C.~S., {Stanek}, K.~Z., {et~al.} 2019, Science,
  366, 637

\bibitem[{{van der Walt} {et~al.}(2011){van der Walt}, {Colbert}, \&
  {Varoquaux}}]{numpy}
{van der Walt}, S., {Colbert}, S.~C., \& {Varoquaux}, G. 2011, Computing in
  Science and Engineering, 13, 22

\bibitem[{{Virtanen} {et~al.}(2020){Virtanen}, {Gommers}, {Oliphant},
  {Haberland}, {Reddy}, {Cournapeau}, {Burovski}, {Peterson}, {Weckesser},
  {Bright}, \& et~al.}]{2020NatMe..17..261V}
{Virtanen}, P., {Gommers}, R., {Oliphant}, T.~E., {et~al.} 2020, Nature
  Methods, 17, 261

\bibitem[{{Webbink}(1984)}]{Webbink1984}
{Webbink}, R.~F. 1984, \apj, 277, 355

\bibitem[{{Wen} {et~al.}(2006){Wen}, {Levine}, {Corbet}, \& {Bradt}}]{Wen2006}
{Wen}, L., {Levine}, A.~M., {Corbet}, R. H.~D., \& {Bradt}, H.~V. 2006, \apjs,
  163, 372

\bibitem[{{W}es {M}c{K}inney(2010)}]{mckinney-proc-scipy-2010}
{W}es {M}c{K}inney. 2010, in {P}roceedings of the 9th {P}ython in {S}cience
  {C}onference, ed. {S}t\'efan van~der {W}alt \& {J}arrod {M}illman, 56--61

\bibitem[{{Zevin} {et~al.}(2020){Zevin}, {Spera}, {Berry}, \&
  {Kalogera}}]{Zevin2020}
{Zevin}, M., {Spera}, M., {Berry}, C. P.~L., \& {Kalogera}, V. 2020, \apjl,
  899, L1

\bibitem[{{Zevin} {et~al.}(2021){Zevin}, {Bavera}, {Berry}, {Kalogera},
  {Fragos}, {Marchant}, {Rodriguez}, {Antonini}, {Holz}, \&
  {Pankow}}]{Zevin2021}
{Zevin}, M., {Bavera}, S.~S., {Berry}, C. P.~L., {et~al.} 2021, \apj, 910, 152

\bibitem[{Zhang {et~al.}(2017)Zhang, Feroci, Santangelo, Dong, Feng, Lu,
  Nandra, Wang, Zhang, Bozzo, Brandt, Rosa, Gou, Hernanz, van~der Klis, Li,
  Liu, Orleanski, Pareschi, Pohl, Poutanen, Qu, Schanne, Stella, Uttley, Watts,
  Xu, Yu, in~’t Zand, Zane, Alvarez, Amati, Baldini, Bambi, Basso,
  Bhattacharyya, Bellazzini, Belloni, Bellutti, Bianchi, Brez, Bursa, Burwitz,
  Budtz-J{\o}rgensen, Caiazzo, Campana, Cao, Casella, Chen, Chen, Chen, Chen,
  Chen, Chen, Civitani, Zelati, Cui, Cui, Dai, Monte, de~Martino, Cosimo,
  Diebold, Dovciak, Donnarumma, Doroshenko, Esposito, Evangelista, Favre,
  Friedrich, Fuschino, Galvez, Gao, Ge, Gevin, Goetz, Han, Heyl, Horak, Hu,
  Huang, Huang, Hudec, Huppenkothen, Israel, Ingram, Karas, Karelin, Jenke, Ji,
  Korpela, Kunneriath, Labanti, Li, Li, Li, Liang, Limousin, Lin, Ling, Liu,
  Liu, Liu, Lu, Lund, Lai, Luo, Luo, Ma, Mahmoodifar, Marisaldi, Martindale,
  Meidinger, Men, Michalska, Mignani, Minuti, Motta, Muleri, Neilsen,
  Orlandini, Pan, Patruno, Perinati, Picciotto, Piemonte, Pinchera, Rachevski,
  Rapisarda, Rea, Rossi, Rubini, Sala, Shu, Sgro, Shen, Soffitta, Song,
  Spandre, Stratta, Strohmayer, Sun, Svoboda, Tagliaferri, Tenzer, Hong,
  Taverna, Torok, Turolla, Vacchi, Wang, Walton, Wang, Wang, Wang, Wang, Weng,
  Wilms, Winter, Wu, Wu, Xiong, Xu, Xue, Yan, Yang, Yang, Yang, Yuan, Yuan,
  Yuan, Zampa, Zampa, Zdziarski, Zhang, Zhang, Zhang, Zhang, Zhang, Zhang,
  Zheng, Zhou, \& Zhou}]{extp}
Zhang, S.~N., Feroci, M., Santangelo, A., {et~al.} 2017, in Space Telescopes
  and Instrumentation 2016: Ultraviolet to Gamma Ray, ed. J.-W.~A. den Herder,
  T.~Takahashi, \& M.~Bautz, Vol. 9905, International Society for Optics and
  Photonics (SPIE), 99051Q.
\newblock \url{https://doi.org/10.1117/12.2232034}

\end{thebibliography}
\end{document}